\documentclass[a4paper,11pt]{article}
\usepackage[utf8]{inputenc}
\pdfoutput=1 

\usepackage{jheppub} 

\usepackage[T1]{fontenc} 
\usepackage{graphicx}
\usepackage[makeroom]{cancel}
\usepackage{makecell}
\usepackage{bm}
\usepackage{amssymb}
\usepackage{amsmath}
\usepackage{epsfig}
\usepackage{hyperref}
\usepackage{hhline}
\usepackage{multirow}
\usepackage[normalem]{ulem}
\usepackage{subfig}
\usepackage{tikz}
\usepackage{epstopdf}
\usetikzlibrary{snakes}
\usepackage{slashed}
\usepackage{float}
\allowdisplaybreaks

 \definecolor{jd}{rgb}{0.858, 0.188, 0.478}

\def\lapp{\mathrel{\rlap{\raise.5ex\hbox{$<$}}
                    {\lower.5ex\hbox{$\sim$}}}}
\def\gapp{\mathrel{\rlap{\raise.5ex\hbox{$>$}}
                    {\lower.5ex\hbox{$\sim$}}}}

{
{

\newcommand{\bmt}{\begin{pmatrix}}
\newcommand{\emt}{\end{pmatrix}}
\newcommand{\ba}{\begin{array}{c}}
\newcommand{\ea}{\end{array}}
\newcommand{\be}{\begin{equation}}
\newcommand{\ee}{\end{equation}}
\newcommand{\bea}{\begin{eqnarray}}
\newcommand{\eea}{\end{eqnarray}}

\newcommand{\bi}{\begin{itemize}}
\newcommand{\ei}{\end{itemize}}

\newcommand{\baz}{\begin{array}{cc}}

\newcommand{\mathsym}[1]{{}}

\newcommand{\bt}{\begin{tabular}}
\newcommand{\et}{\end{tabular}}

\newcommand{\benu}{\begin{enumerate}}
\newcommand{\eenu}{\end{enumerate}}

\def\mH{m_{H^0}}
\def\mA{m_{A^0}}
\def\mHp{m_{H^\pm}}
\def\mph{m_\phi}
\def\lc{\lambda_c}
\def\lph{\lambda_{\phi h}}
\def\ll{\lambda_{L}}
\def\Oph{\Omega_\phi h^2}
\def\OH{\Omega_{H^0} h^2}
\def\ODM{\Omega_{DM}h^2}

\newcommand{\bav}{\begin{array}{cccc}}

\title{Two component dark matter with inert Higgs doublet: neutrino mass, high scale validity and collider searches}
\author[a]{Subhaditya Bhattacharya,}
\emailAdd{subhab@iitg.ac.in}
\author[a]{Purusottam Ghosh,}
\emailAdd{pghoshiitg@gmail.com}
\author[b]{Abhijit Kumar Saha,}
\emailAdd{aks@prl.res.in}
\author[a]{Arunansu Sil}
\emailAdd{asil@iitg.ac.in}
\affiliation[a]{Department of Physics, Indian Institute of Technology Guwahati, North Guwahati, Assam- 781039, India}
\affiliation[b]{Theoretical Physics Division, Physical Research Laboratory, Ahmedabad 380009, India}

\abstract{The idea of this work is to investigate the constraints on the dark matter (DM) allowed parameter space from 
 high scale validity (absolute stability of Higgs vacuum and perturbativity) in presence of multi particle dark sector and 
 heavy right handed neutrinos to address correct neutrino mass. We illustrate a simple two component DM model, consisting of one inert 
 $SU(2)_L$ scalar doublet and a scalar singlet, both stabilised by additional $\mathcal{Z}_2 \times \mathcal{Z}^{'}_2$ symmetry, which also aid to vacuum stability. 
 We demonstrate DM-DM interaction helps achieving a large allowed parameter space for both the DM components by evading direct search bound. High scale validity puts further constraints on the model, for example, on the mass splitting between the charged and neutral component of inert doublet, 
 which has important implication to its leptonic signature(s) at the Large Hadron Collider (LHC).
 }


\begin{document}

\maketitle
\flushbottom

\setcounter{footnote}{0}
\renewcommand*{\thefootnote}{\arabic{footnote}}
\section{Introduction}
\label{sec:intro}
Discovery of the `Higgs' boson at Large Hadron Collider (LHC) in 2012 \cite{Chatrchyan:2012xdj,Aad:2012tfa} strongly validates the Standard Model (SM) of particle
physics as the fundamental governing theory of Strong, Weak and Electromagnetic interactions. However many unresolved issues persist. For example, the electroweak (EW) vacuum
turns out to be {\it metastable} \cite{Alekhin:2012py,Buttazzo:2013uya,Isidori:2001bm,Anchordoqui:2012fq,Tang:2013bz,Ellis:2009tp,EliasMiro:2011aa} with the
present measured value of Higgs mass ($m_h \sim 125.09$ GeV \cite{PhysRevD.98.030001}) and top quark mass ($m_t \sim 173.1$ GeV \cite{PhysRevD.98.030001}). 
Large uncertainty in the measured value of $m_t$ can even make EW vacuum unstable, questioning the existence of our universe. It is well known that inclusion of additional scalars in the theory
can help stabilizing EW vacuum by compensating the negative contributions of fermionic couplings in renormalization group (RG) running of Higgs quartic coupling $\lambda_H$ \cite{EliasMiro:2012ay,Lebedev:2012zw}. 
This motivates us to look for extended scalar sector.

On the other hand, the existence of dark matter (DM) in the Universe
 is convincingly supported from the observations of various experiments around the globe, for example, WMAP \cite{Spergel:2006hy} and 
Planck \cite{Aghanim:2018eyx}. Extensions of SM to accommodate DM is therefore inevitable. The simplest of its kind is Weakly
Interacting Massive Particle (WIMP) \cite{Roszkowski:2017nbc} and most economical is the presence of a singlet scalar dark matter 
\cite{Silveira:1985rk,McDonald:1993ex,Cline:2013gha,Guo:2010hq,Feng:2014vea,Bhattacharya:2016qsg,Casas:2017jjg,Bhattacharya:2017fid,Bhattacharya:2019mmy}
with Higgs portal interaction with SM. Non observation of DM in direct search experiments like LUX \cite{Akerib:2016vxi}, XENON 1T \cite{Akerib:2016vxi,Aprile:2017iyp,Messina:2018fmz}, 
Panda X \cite{Zhang:2018xdp} put a stringent bound on this model (with mass below 1 TeV getting disallowed) due to its prediction of large spin independent (SI) direct search cross section. As an alternative,
multi-component DM scenarios \cite{Bhattacharya:2013hva,Bian:2013wna,Esch:2014jpa,Karam:2015jta,Karam:2016rsz,Ahmed:2017dbb,Herrero-Garcia:2018qnz,Poulin:2018kap,Aoki:2018gjf,Aoki:2017eqn,Bhattacharya:2016ysw,Bhattacharya:2018cgx,Biswas:2013nn,Borah:2019aeq,Chakraborti:2018lso,Chakraborti:2018aae,Barman:2018esi,DuttaBanik:2016jzv,Bhattacharya:2018cqu,YaserAyazi:2018lrv} are proposed where DM-DM interactions (see for example, \cite{Bhattacharya:2016ysw,Bhattacharya:2018cgx}) play an important role to 
evade direct search bound.Such an attempt was made in \cite{Bhattacharya:2016ysw}, where
two singlet scalars serve as two DM components, that satisfy required DM constraints. However, the framework is unable to accommodate the low mass region corresponding to both the DM's masses, simultaneously below 500 GeV, to satisfy DM direct search 
bounds. 

Singlet DM fields (even in multipartite framework consisting of two singlet DM candidates \cite{Bhattacharya:2016ysw}) also
have limited colliders search possibilities, due to small interaction (Higgs portal) with visible sector. 
The search strategy for such DM is therefore mainly limited to mono-$X$ signature with missing energy, 
where $X$ can be jet, $W$, $Z$ or Higgs \cite{Abdallah:2015ter,Abercrombie:2015wmb,Han:2016gyy}. 
Such signal arises out of initial state radiation (ISR) and suffers from huge SM background. 
Therefore, searching such singlet DM candidates at collider turn out to be difficult, even more due to direct search constraint limiting the DM 
mass to a higher value. Higher multiplet in dark sector, being equipped with charge components have better possibilities of getting unravelled in 
future collider search experiments, but tighter constraints arise from dark sector. 
The simplest of its kind is to assume an inert Higgs $SU(2)_L$ doublet (IDM). However, due to gauge coupling, a single component 
IDM is severely constrained and is not allowed between DM mass within $80-550$ GeV, referred as the {\it desert region} for under abundance. 
However, it can produce hadronically quiet single and two lepton signatures with missing energy at LHC \cite{Gustafsson:2012aj,Bhardwaj:2019mts,Chakraborti:2018aae}. Therefore, it is ideal to study a multipartite DM framework, involving an inert doublet to satisfy DM constraints as well 
as to have interesting phenomenological consequences.

Physics beyond the SM (BSM) is also strongly motivated by the presence of tiny neutrino mass ($\sim$ eV). In order to explain 
this tiny neutrino masses, type-I seesaw model is proposed where right handed SM singlet fermions are introduced with their 
Majorana mass terms. One can then have SM gauge invariant dimension five effective operator of the form $\sim \frac{1}
{\Lambda}~LLHH$, where $L$ represents SM lepton doublet, $H$ represents SM Higgs doublet and $\Lambda$ is 
the scale representative of RH neutrino mass scale. Different seesaw models other than type-I are also invented. They all necessarily 
predict BSM physics with important and interesting phenomenological consequences. Among these, type-I seesaw is the 
simplest possibility \cite{Mohapatra:1979ia,Schechter:1980gr}, which indicates the presence of heavy right handed (RH) 
neutrinos to yield correct light neutrino mass through additional Yukawa coupling with SM Higgs. This in 
turn may alter the Higgs vacuum stability condition at high scale (larger than the RH neutrino masses) with 
large neutrino Yukawa coupling \cite{Gabrielli:2013hma,Chen:2012faa,Rodejohann:2012px,Rose:2015fua,Lindner:2015qva,Chakrabortty:2012np,Coriano:2014mpa,Ng:2015eia,Bonilla:2015kna,Khan:2012zw,Garg:2017iva,Chakrabarty:2015yia,Ghosh:2017pxl}. Inclusion of an inert Higgs 
doublet along with right handed neutrinos can also generate the light neutrino mass radiatively \cite{Ma:2006km}. 
In that case, the usual neutrino Yukawa coupling involving the SM Higgs doublet can be forbidden with 
the choice of appropriate discrete symmetry, while a new Yukawa interaction involving the inert Higgs doublet is present. However in such a scenario (with radiative light neutrino mass generation), the EW vacuum stability would 
not be much different from that of the SM as the running of the Higgs quartic coupling remains unaffected 
by the presence of this new Yukawa interaction.

Our model under scrutiny addresses some of the interesting features in view of above discussion. We address a multipartite dark sector consisting of a $SU(2)_L$ doublet scalar (the IDM) and a scalar singlet, both stabilized by additional $\mathcal{Z}_2 \times \mathcal{Z}^{'}_2$ symmetry (for an earlier effort, see \cite{Biswas:2013nn}) and provide a two component DM set up. The presence of DM-DM interactions enlarge 
the available parameter space significantly (by reviving the below 500 GeV mass region for both 
the DM candidates), while the inert DM can also produce leptonic collider signature at LHC.
We augment the model with heavy RH neutrinos to address the light neutrino masses. Though the set-up 
with inert Higgs doublet and RH neutrinos opens up both the possibilities of generating light neutrinos 
masses, either through type-I seesaw or via radiative generation, we consider here the neutrino mass 
generation through type-I seesaw and accordingly choose the appropriate discrete charges of the fields so 
as to forbid the Yukawa interaction involving SM lepton doublet, RH neutrinos and inert Higgs doublet. The primary reason behind such a choice is to involve the study of the stability of the EW vacuum with large neutrino  
Yukawa couplings (involving SM Higgs doublet) as mentioned before. Then the parameter space obtained from DM phenomenology will be further 
constrained from vacuum stability criteria. In brief, we want to study a multipartite DM scenario which would be 
adversely affected by the non-zero light neutrino mass in terms of EW vacuum stability.
Note that while the neutrino Yukawa
coupling involving RH neutrinos and SM Higgs tends to destabilze the EW vacuum, the presence of the additional scalars in the set-up 
tends to stabilize it \cite{Gabrielli:2013hma,Chen:2012faa,Rodejohann:2012px,Rose:2015fua,Lindner:2015qva,Chakrabortty:2012np,Coriano:2014mpa,Ng:2015eia,Bonilla:2015kna,Khan:2012zw,Garg:2017iva,Chakrabarty:2015yia,Ghosh:2017pxl}. So, we take up an interesting exercise of validating the model from DM constraints, neutrino masses and high scale validity (absolute stability of the Higgs vacuum and perturbativity). This analysis provides some important conclusions, which are phenomenologically viable at LHC.  

Let us finally discuss the plan of the paper. In section \ref{sec:model}, we discuss the model construct in details. Section \ref{sec:constraints} presents possible theoretical and experimental
constraints on the model parameters. Then in sections \ref{sec:single-component},\ref{sec:2compDM} and \ref{sec:RD-ps} subsequently, we discuss DM phenomenology. Indirect search constraints are discussed in \ref{sec:indirect}. In section \ref{sec:vs}, we
investigate the high scale validity of the model. Section \ref{sec:collider} summarises collider signature(s) in context of the proposed set up. 
Finally we conclude in section \ref{sec:conclusion}. Tree level unitarity condition is elaborated in Appendix \ref{treeuni}. The high scale stability condition on the single component DM frameworks in presence of RH Neutrino is chalked out in Appendix \ref{sec:HSV_single} for the sake of comparison. All the constraints together on the 
model parameter space along with different choices of RH neutrino mass and Yukawa couplings are listed in two tables in Appendix \ref{sec:Allrange}. 

\section{The Model}
\label{sec:model}
The model is intended to capture the phenomenology of two already established DM frameworks involving that of a singlet scalar and that of an inert 
scalar doublet together with right handed neutrinos to address neutrino mass under the same umbrella. Therefore, we extend SM by an inert doublet scalar $(\Phi)$ 
and a real scalar singlet $(\phi)$ and include three RH Majorona neutrinos $N_i (i=1,2,3)$ in the set up. The lightest neutral scalar mode of the IDM and $\phi$ are the DM candidates 
provided an appropriate symmetry in addition to that of SM stabilizes both of them. This is minimally\footnote{Introduction of a single $\mathcal{Z}_2$ could lead to a decay of one of the DM candidates, as seen from an allowed term $\Phi^\dagger H \phi$ in that case.} possible by introducing an additional 
 $\mathcal{Z}_2 \times \mathcal{Z}^\prime_2$ discrete symmetry under which all SM fields along with the right handed neutrinos transform trivially and the other additional fields transform non-trivially as  
 tabulated in the Table \ref{tab1:charges}. We also note the charges of SM Higgs ($H$) explicitly in Table \ref{tab1:charges}, as it will be required to form the scalar potential of the model. 
 Note here, that charges of the two DM candidates ($\Phi$ and $\phi$) are complementary, $i.e.$ odd under either $\mathcal{Z}_2$ or $\mathcal{Z}^\prime_2$ for their stability. 
\begin{table}[htb!]
\begin{center}
\begin{tabular}{|p{4.5cm}|p{8cm}|}
 \hline
  ~BSM and SM Higgs Fields  &  $\hspace{0.8cm} SU(3)_C \times SU(2)_L \times U(1)_Y\times \mathcal{Z}_2 \times \mathcal{Z}^\prime_2$ $\equiv \mathcal{G}$ \\
 \hline
 \hspace{0.1cm}$\Phi \equiv \left(\begin{matrix}
 H^+  \\  \frac{1}{\sqrt{2}}(H^0+iA^0) 
\end{matrix}\right)$ & \hspace{1cm} 1 \hspace{1.3cm} 2 \hspace{1.1cm} +1 \hspace{0.8cm} - \hspace{0.45cm} +\\
\hline
 \hspace{2.0cm}$\phi$ & \hspace{1cm} 1 \hspace{1.3cm} 1 \hspace{1.31cm} 0 \hspace{0.8cm} +\hspace{0.7cm}-\\
 \hline
 \hspace{1.0cm}$N_i~(i=1,2,3)$ & \hspace{1cm} 1 \hspace{1.3cm} 1 \hspace{1.29cm} 0 \hspace{0.8cm} +\hspace{0.7cm}+\\
 \hline
  \hspace{0.1cm}$H \equiv \left(\begin{matrix}
 w^+  \\  \frac{1}{\sqrt{2}}(h+v+iz) 
\end{matrix}\right)$ & \hspace{1cm} 1 \hspace{1.3cm} 2 \hspace{1.1cm} +1 \hspace{0.72cm} + \hspace{0.4cm} +\\
\hline
 \end{tabular}
\end{center}
\caption{Charge assignments of the BSM fields assumed in the model under $ \mathcal{G}$ as well as that of SM Higgs. The $U(1)_Y$ hypercharge is chosen as $Q=T_3+Y/2$.}
  \label{tab1:charges}
\end{table} 
We also point out that the $U(1)_Y$ hypercharge assignment of $\Phi$ is identical to SM doublet $H$. Therefore the only $SU(2)_L \times U(1)_Y$ invariant terms are $H^\dagger H, ~\Phi^\dagger \Phi,~ H^\dagger \Phi$ and its conjugate. 

\noindent The scalar Lagrangian reads as :

\bea
\mathcal{L}_{scalar}&=& |D^\mu{H}|^2+|D^\mu{\Phi}|^2+ \frac{1}{2} (\partial^\mu{\phi})^2-V(H,\Phi,\phi),\\
{\rm where}~ D^\mu&=&\partial^\mu-ig_2 \frac{\sigma^a}{2} {W^{a}}^\mu-ig_1 \frac{Y} {2} B^\mu \nonumber
\eea
and $g_2,g_1$ denote $SU(2)_L$ and $U(1)_Y$ coupling respectively. 

\begin{figure}[h]
 $$
  \includegraphics[width=8.5cm,height=6.5cm]{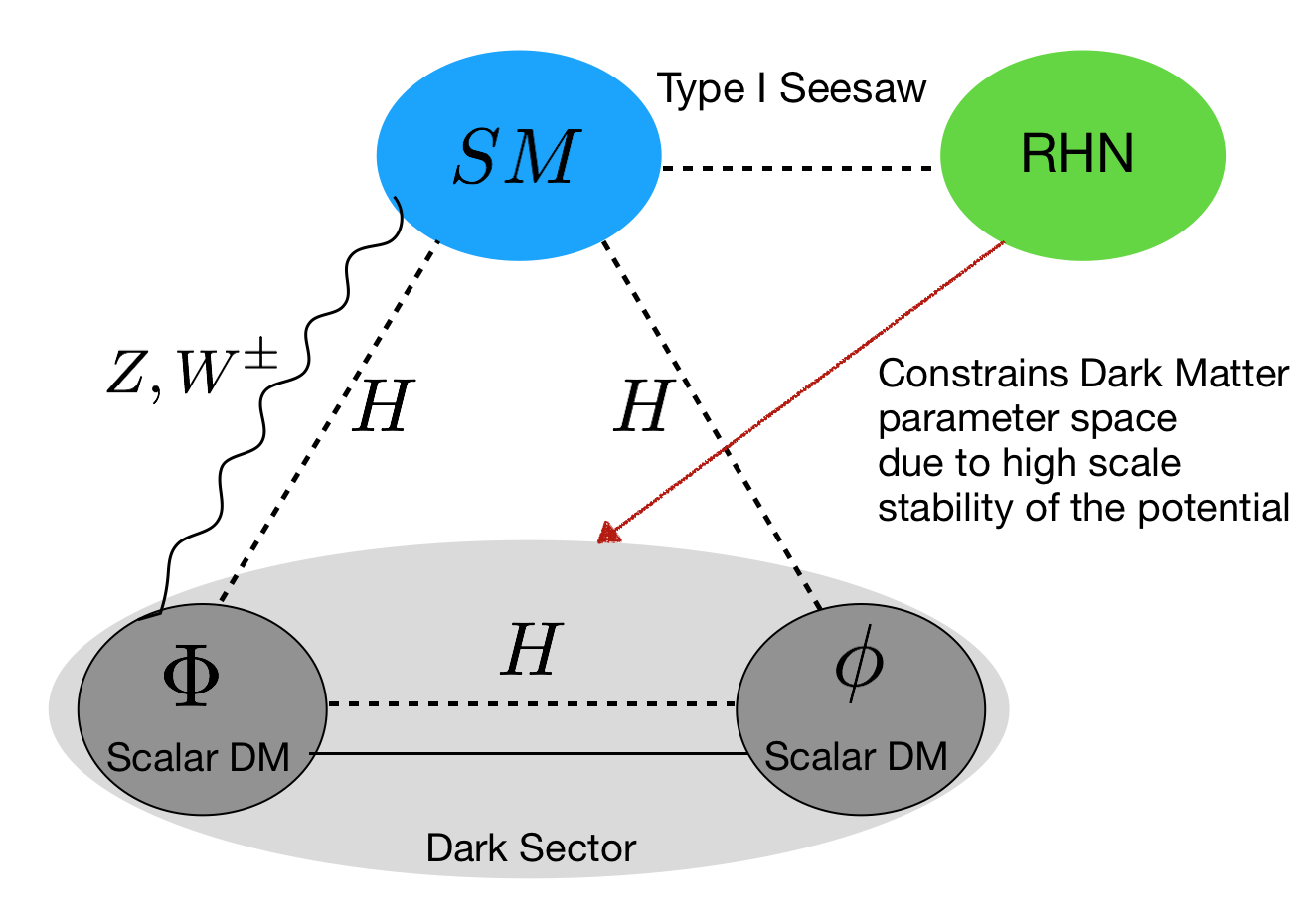} 
 $$
  \caption{A schematic diagram illustrating the different sectors of the model and their connection to SM. The dotted lines represent Higgs portal coupling, wavy line indicate gauge coupling, 
  while the thin solid line indicates direct DM-DM coupling through $\frac{\lambda_c}{2}\phi^2(\Phi^\dagger \Phi)$ term.}
  \label{fig:cartoon}
 \end{figure}

\noindent The most relevant renormalizable scalar potential in this case is given by,
\bea
V(H,\Phi,\phi) &=& - \mu_H^2(H^\dagger H)+\lambda_H (H^\dagger H)^2 + V(H,\Phi) + V(H,\phi) + V(\Phi,\phi) ,
\label{eq:totP}
\eea
where,
\bea\label{pot:pot}
V(H,\Phi)&=& \mu_\Phi^2(\Phi^\dagger \Phi)+\lambda_\Phi (\Phi^\dagger \Phi)^2 +\lambda_1 (H^\dagger H)(\Phi^\dagger \Phi) \nonumber \\
&&+\lambda_2 (H^\dagger \Phi)(\Phi^\dagger H)+\frac{\lambda_3}{2}[(H^\dagger \Phi)^2+h.c.] , \\ 
V(H,\phi)&=& \frac{1}{2} \mu_{\phi}^2 \phi^2 + \frac{\lambda_\phi}{4!}{\phi}^4 + \frac{1}{2}\lph \phi^2(H^\dagger H), \\
 V(\Phi,\phi)&=&\frac{\lc}{2} (\phi^2 )\Big(\Phi^\dagger \Phi\Big) .
\eea
The Lagrangian involving right handed neutrinos can be written as,
\bea
\mathcal{L^\nu}=-{(Y_{\nu})}_{ij} \bar{l}_{L_i} \widetilde{H} N_j - \frac{1}{2} M_{N_{ij}} \overline{N_i^C} N_j,\label{FLag}
\eea
where $\tilde{H}=i\sigma_2 H^*$. We have considered three generations of RH neutrinos with $\{i,j\}=1,2,3$, which can acquire Majorana masses and can possess Yukawa interactions with SM lepton doublet $l_{L}$. 
Note here, that the charge assignment of the $N$ fields then aid us to obtain neutrino masses through standard Seesaw-I \cite{Mohapatra:1979ia,Schechter:1980gr} 
mechanism (as detailed later), while it also prohibits the operator like $\bar{l_{L}}\Phi N$ due to $ \mathcal{Z}_2 $ charge assignment, 
and hence discards the possibility of generating the light neutrino mass radiatively. The ingredients and interactions of the model set up is described in the cartoon as in Fig.~\ref{fig:cartoon}.

After spontaneous symmetry breaking, SM Higgs doublet acquires non-zero vacuum expectation value (VEV) as $H=(0~ \frac{v+h}{\sqrt{2}})^T$ with $v$=246 GeV.
Also note that neither of the added scalars acquire VEV to preserve $\mathcal{Z}_2 {\times \mathcal{Z}^\prime}_2$ and act as DM components. After minimizing the potential $V(H,\Phi,\phi)$ along
different field directions, one can obtain the following relations between the physical masses and the couplings involved:
\bea
&\mu_H^2 ~=~ \frac{m_h^2}{2} , ~~~~ \mu_\Phi^2~=~m_{H^0}^2-\lambda_L v^2 ,~~~ \lambda_3~=~\frac{1}{v^2}(m_{H^0}^2-m_{A^0}^2), \nonumber \\
&\lambda_2~=~\frac{1}{v^2}(m_{H^0}^2+m_{A^0}^2-2 m_{H^\pm}^2)~~\rm and~ ~~ \lambda_1~=~2\lambda_L-\frac{2}{v^2}(m_{H^0}^2-m_{H^\pm}^2)~,
\label{eq:mass-coupling}
\eea
where $\lambda_L=\frac{1}{2}(\lambda_1+\lambda_2+\lambda_3)$ and $m_h,m_{H^0},m_{A^0}$ are the mass eigenvalues of SM-like neutral
scalar found at LHC ($m_h=125.09$ GeV), heavy or light additional neutral scalar and the CP-odd neutral scalar respectively. $m_{H^\pm}$ denotes the mass of charged scalar eigenstate(s).
The mass for $\phi$ DM will be rescaled as $m_\phi^2=\mu_\phi^2+\frac{1}{2} \lambda_{\phi h}v^2$.
The independent parameters of the model, those are used to evaluate the DM, neutrino mass constraints are as follows: 
\bea
\textrm{ Parameters}: ~~~ \{m_{H^0},m_{A^0},m_{H^\pm}, m_{\phi}, \lambda_L, \lambda_{\phi h}, \lambda_\Phi, \lambda_\phi, Y_{\nu_{ij}}, M_{N_{ij}}\}.
\label{eq:parameters}
\eea
\section{Theoretical and Experimental constraints }
\label{sec:constraints}
We would like to address possible theoretical and experimental constraints on model parameters here. \\~\\
\noindent $\bullet$ \textbf{ Stability:} In oder to get the potential bounded from below, the quartic couplings of the potential $V(H,\Phi,\phi)$ must have to satisfy following co-positivity conditions (\textrm{CPC}) as given by \cite{Kannike:2012pe,Chakrabortty:2013mha},
\begin{align}
\textrm{CPC\{1,2,3\}}: \lambda_{H}(\mu) \geq 0 , ~~~~\lambda_{\Phi}(\mu) \geq 0,~~~~~\lambda_{\phi}(\mu) \geq 0, \nonumber \\
\textrm{CPC\{4,5\}}: \Big(\lambda_1 (\mu) + \lambda_2 (\mu) \pm \lambda_3 (\mu) \Big) + \sqrt{\lambda_H (\mu) \lambda_\Phi (\mu)} \geq 0, \nonumber \\   
 \textrm{CPC\{6,7\}}:\lambda_1 (\mu) + 2 \sqrt{\lambda_H (\mu) \lambda_\Phi (\mu)} \geq 0 , ~~~
 \lambda_{\phi h} (\mu) +  \sqrt{ \frac{2}{3}\lambda_H (\mu) \lambda_\Phi (\mu)} \geq 0 ,\nonumber\\
  \textrm{CPC8}:\lambda_c(\mu)+\sqrt{\frac{2}{3}\lambda_\Phi (\mu) \lambda_\phi(\mu)} \geq 0~,
 \label{eq:stability}
\end{align}
 where CPC$\{i\}$ denotes $i^{\rm th}$ copositivity condition and $\mu$ is the running scale. The above conditions show that the model offers to choose even negative $\lambda_{1,2,3, \phi h}$ satisfying the above conditions. However, as demonstrated in 
Eqn.~\ref{eq:parameters}, we use $\lambda_L$ and physical masses to be the parameters. Therefore, if we choose a specific mass hierarchy as: $m_{H^\pm} \ge m_{A^0} \ge m_{H^0}$ with 
positive $\lambda_L$, we are actually using $\lambda_1$ to be positive while $\lambda_{2,3}$ negative abiding by the above conditions (see Eqn. \ref{eq:mass-coupling}). The conditions CPC$\{i\}$ as in 
Eqn.~\ref{eq:stability}, will be used later in demonstrating stability of the potential in Sec.~\ref{sec:vs}.
\\~\\
\noindent $\bullet$\textbf{ Perturbativity:} 
In oder to maintain perturbativity, the quartic couplings of the scalar potential $V(H,\Phi,\phi)$, gauge couplings ($g_{i=1,2,3}$) and neutrino Yukawa coupling $Y_\nu$ should obey:
\begin{align}
 |\lambda_{H} (\mu)| < 4 \pi,~~~ |\lambda_{\Phi} (\mu)| < 4 \pi,~~~|\lambda_{\phi} (\mu)| < 4 \pi, \nonumber \\
 |\lambda_{c} (\mu)| < 4 \pi,~~~ |\lambda_{\phi h} (\mu)| < 4 \pi, \nonumber \\
 |\lambda_{1} (\mu)| < 4 \pi,~~~|\lambda_{2} (\mu)| < 4 \pi,~~~|\lambda_{3} (\mu)| < 4 \pi,\nonumber \\
 |g_{i=1,2,3}| <\sqrt{4\pi}~~ \textrm{and}~~~~~ \big|\textrm{Tr}\Big[Y_{\nu}^\dagger(\mu) Y_{\nu}(\mu)\Big]\big| < 4 \pi ~ .
  \label{eq:perturbativity}
\end{align}
~\\
\noindent$\bullet$\textbf{ Tree Level Unitarity:} 
Next we turn to the constraints imposed by tree level unitarity of the theory, coming from all possible $2 \to 2$ scattering amplitudes as detailed in Appendix~\ref{treeuni} follows as \cite{Horejsi:2005da,Bhattacharyya:2015nca}:
\begin{align}\label{eq:unitarity}
  |\lambda_H|<4\pi,~~~|\lambda_\Phi|< 4 \pi, \nonumber \\
 |\lc|<8\pi,~~~|\lph|< 8 \pi,\nonumber \\
 |\lambda_1|< 8 \pi,~~~|\lambda_1+2(\lambda_2+\lambda_3)|< 8 \pi \nonumber \\
  |\lambda_1+\lambda_2+\lambda_3| < 8\pi, ~~|\lambda_1-\lambda_2-\lambda_3| < 8\pi,\nonumber \\ 
 |(\lambda_\Phi + \lambda_H ) \pm \sqrt{(\lambda_2+\lambda_3)^2 + (\lambda_H-\lambda_\Phi)^2}| < 8 \pi, \nonumber \\
 ~\text{and}~~~ |x_{1,2,3}| < 16 \pi ~.
\end{align}
where $x_{1,2,3}$ be the roots of the cubic equation as detailed in Appendix~\ref{treeuni}. \\~\\
\noindent
$\bullet$\textbf{ Electroweak precision parameters:} There exists an additional $SU(2)_L$ doublet ($\Phi$) in our model in addition to a gauge singlet scalar ($\phi$). 

\begin{figure}[htb!]
 $$
  \includegraphics[height=6.0cm]{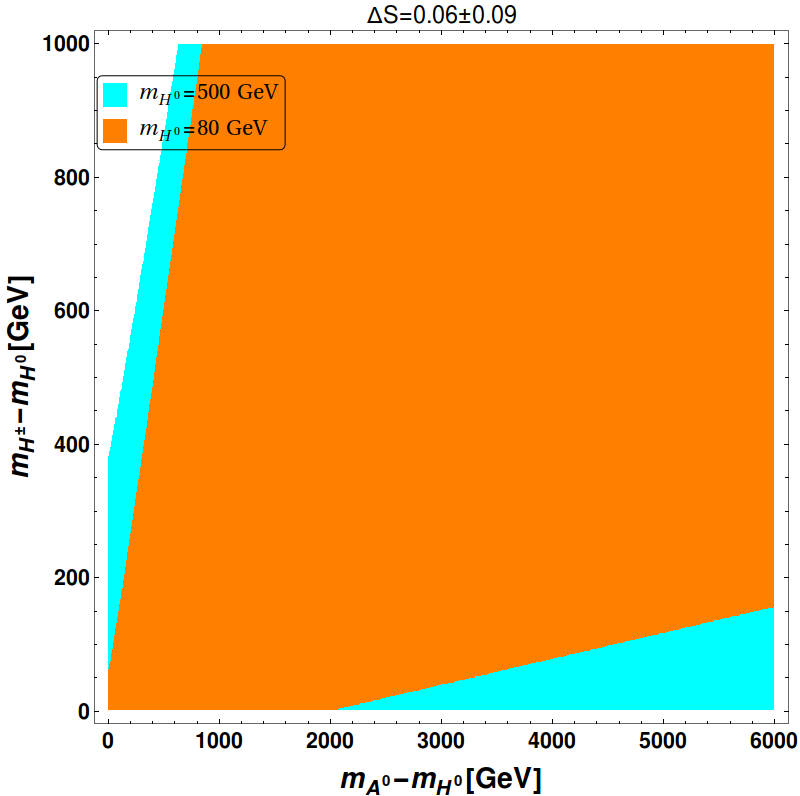} ~~~~~
   \includegraphics[height=6.0cm]{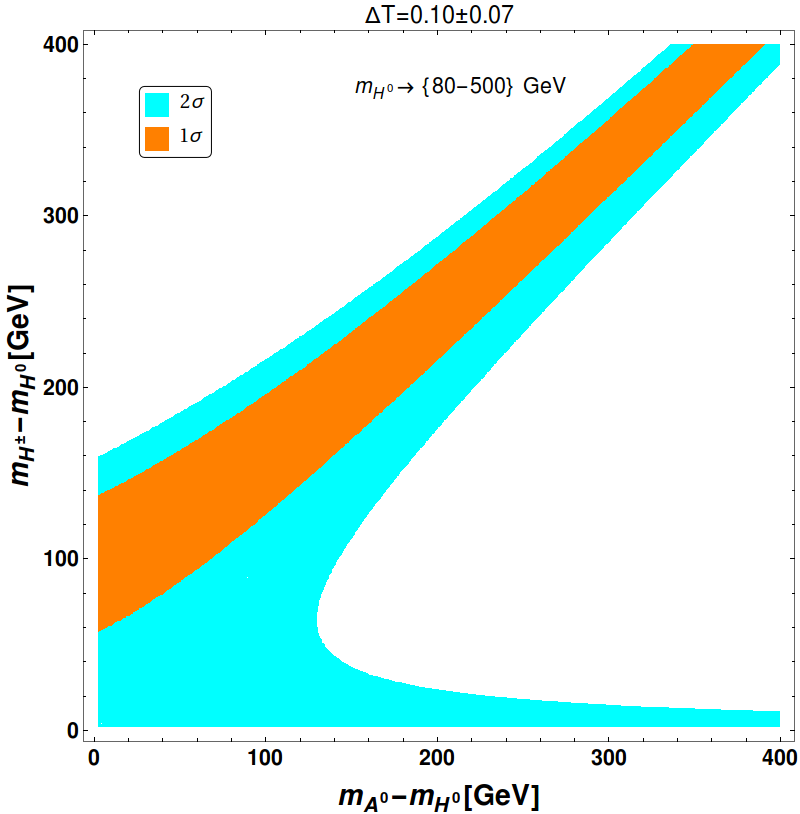} 
  $$
  \caption{Constraints from $\Delta S$ (left) and $\Delta T$ (right) in $m_{A^0}-m_{H^0}$ and $m_{H^\pm}-m_{H^0}$ plane. For $\Delta S$, we have taken 1$\sigma$ limit for two different choices of $\mH=\{80,500\}$ GeV. 
  For $\Delta T$ scan, we show both 1$\sigma$ and 2$\sigma$ limits for a range of $\mH=\{80-500\}$ GeV.}
  \label{fig:STParameter}
 \end{figure}

As the vev of $\Phi$ is zero, it does not alter the SM predictions of electroweak $\rho$ parameter \cite{Peskin:1991sw}. However IDM, being an $SU(2)_L$
doublet makes a decent contribution to $S$ and $T$ parameters \cite{Arhrib:2012ia,Barbieri:2006dq} which we will identify as $\Delta S$ and $\Delta T$. 
The experimental bound from the global electroweak fit results on $\Delta S$ and $\Delta T$ using $\Delta U=0$ are given by:
\begin{align}
 \Delta S\vert_{\Delta U=0}=0.06\pm 0.09, ~~ \Delta T|_{\Delta U=0}=0.1\pm 0.07,
 \label{eq:STU}
\end{align}
at 1$\sigma$ level with correlation coefficient 0.91 \cite{Baak:2014ora}. 

 We show the constraint from $\Delta S$ and $\Delta T$ on the model parameter space in Fig. \ref{fig:STParameter} using the standard formula as presented 
in \cite{Arhrib:2012ia,Barbieri:2006dq}. In left plot, we scan 1$\sigma$ fluctuation on $\Delta S$ in $m_{A^0}-m_{H^0}$ versus $m_{H^\pm}-m_{H^0}$ plane for two different values of $\mH=\{80,500\}$ GeV. 
We see that for smaller $\mH$, the constraint is larger. In right panel, we show 1$\sigma$ 
and 2$\sigma$ limits from $\Delta T$ in $m_{A^0}-m_{H^0}$ versus $m_{H^\pm}-m_{H^0}$ plane for a range of IDM mass $\mH=\{80-500\}$ GeV. 
We can clearly see, that $\Delta T$ constrains the mass splitting much more than $\Delta S$.  \\~\\
\noindent
$\bullet$\textbf{ Higgs invisible decay:} Whenever the DM particles are lighter than half of 
the SM Higgs mass, the Higgs can decay to DM and therefore it will contribute to Higgs invisible decay. Therefore, in such circumstances, we have to employ the bound on the invisible decay width of the 125.09 GeV 
Higgs as \cite{PhysRevD.98.030001}:
\bea
Br(h \to \textrm{Inv}) &<& 0.24 \nonumber \\
 \frac{\Gamma (h \to \textrm{Inv})}{\Gamma (h \to \textrm{SM})+\Gamma (h \to \textrm{Inv})}&<& 0.24 ~.
\eea
where 
\bea
\Gamma (h \to \textrm{Inv})&=&\Gamma (h \to H^0~H^0)+\Gamma (h \to \phi~\phi), ~\textrm{when}~ m_\phi,m_{H_0}< m_h/2 \sim 62.5~\textrm{GeV}; \nonumber 
\eea
and $\Gamma (h \to \textrm{SM})= 4.2 $ MeV \cite{PhysRevD.98.030001}. In this analysis, we have mostly focused in the region where $m_\phi,m_{H_0}> m_h/2$, actually larger than $W$ mass, i.e. $m_\phi,m_{H_0} \geq m_W$, so that the above constraint is not applicable.\\~\\
\noindent
$\bullet$\textbf{ Collider search constraints:}
Experimental searches for additional charged scalars and pseudoscalar in LEP and LHC provide bound on IDM mass parameters and coupling coefficients of IDM with SM particles. \\
\textbf{(i) Bounds from LEP: } The observed decay widths of $Z$ and $W$ bosons from LEP data restrict the decay of gauge bosons to the additional scalars and therefore provide a bound on IDM mass parameters
as $m_{A^0}+m_{H^0}> m_Z,~ 2~m_{H^\pm}>m_Z$ and  $m_{H^\pm}+m_{H^0,A^0}>m_W$. In addition, neutralino searches at LEP-II, provides a lower limit on the pseudoscalar Higgs ($m_{A^0}$) to 
100 GeV when $m_{H^0} < m_{A^0}$ \cite{ Lundstrom:2008ai}. The chargino search at LEP-II limits indicate a bound on the charged Higgs to $m_{H^\pm }> 70$ GeV \cite{Pierce:2007ut}.  \\
 \textbf{(ii) Bounds from LHC: }Due to the presence of SM Higgs and IDM interaction, charged scalars $H^\pm$ take part
 into the decay of SM Higgs to diphoton. Thus it contributes to Higgs to diphoton signal strength $\mu_{\gamma\gamma}$ which is defined as \cite{Cao:2007rm,Djouadi:2005gj,Swiezewska:2012eh,Krawczyk:2013jta}
 
 \begin{align}
  \mu_{\gamma\gamma}=\frac{\sigma(gg\rightarrow h\rightarrow \gamma\gamma) }{\sigma(gg\rightarrow h\rightarrow \gamma\gamma)_{\rm SM}}\simeq \frac{{\rm Br}(h\rightarrow\gamma\gamma)_{\rm IDM}}{{\rm Br}(h\rightarrow\gamma\gamma)_{\rm SM}}.
 \end{align}
 Now when IDM particles are heavier than $m_h/2$, one can further
 write
 \begin{align}
  \frac{{\rm Br}(h\rightarrow\gamma\gamma)_{\rm IDM}}{{\rm Br}(h\rightarrow\gamma\gamma)_{\rm SM}}=\frac{\Gamma(h\rightarrow\gamma\gamma)_{\rm IDM}}{\Gamma(h\rightarrow\gamma\gamma)_{\rm SM}}.
 \end{align}
 The analytic expression of $\Gamma(h\rightarrow\gamma\gamma)_{\rm IDM}$ can be obtained as \cite{Cao:2007rm,Djouadi:2005gj,Swiezewska:2012eh,Krawczyk:2013jta}:
 \begin{align}
  \Gamma(h\rightarrow\gamma\gamma)_{\rm IDM}=\Big|\mathcal{A}_{\rm SM}+\frac{\alpha_e m_h^{3/2}}{16\pi^{3/2}}\frac{\lambda_1 v}{m_{H^\pm}^2}F\Big(\frac{m_h^2}{4 m_{H^\pm}^2}\Big)\Big|^2,
 \end{align}
 where $\mathcal{A}_{\rm SM}$ represents pure SM contribution (see \cite{Cao:2007rm,Djouadi:2005gj,Swiezewska:2012eh,Krawczyk:2013jta}). And  $F(x)=-[x-f(x)]x^{-2}$ where
 \begin{align}
  f(x)= 
\begin{cases}
   (\sin^{-1} x)^2,&  x\leq 1\\
    -\frac{1}{4}\Big[\textrm{ln}\frac{1+\sqrt{1-x^{-1}}}{1-\sqrt{1-x^{-1}}}-i\pi\Big]^2              & x>1  .
\end{cases}
 \end{align}
 \begin{figure}[htb!]
 $$
  \includegraphics[height=6.0cm,width=8cm]{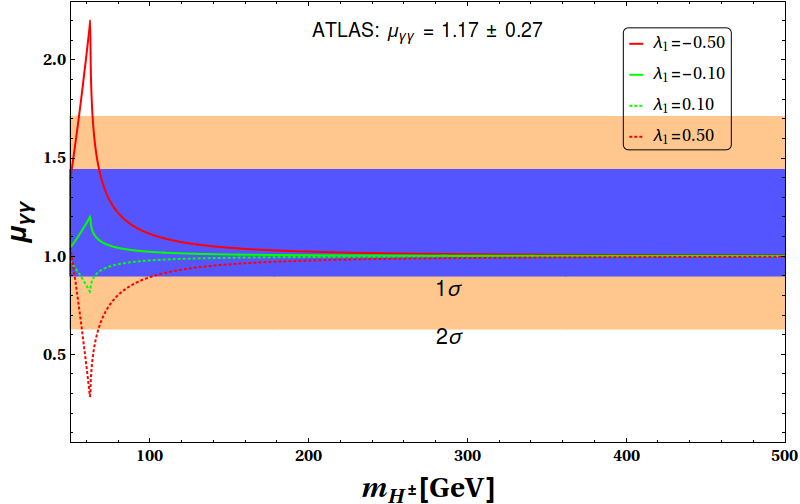} ~~~~~
  $$
  \caption{$\mu_{\gamma\gamma}$ as function of $m_{H^\pm}$ for different values of $\lambda_1$ as defined in the inset.
   $1\sigma$ and $2\sigma$ limits of $\mu_{\gamma\gamma}$ from ATLAS are also shown in blue and orange colours for comparison purpose.}
  \label{fig:muGamma}
 \end{figure}
Therefore it turns out that IDM contribution to $\mu_{\gamma\gamma}$ is function of both mass of the charged Higgs ($m_{H^\pm}$) and 
the coefficient of the trilinear coupling $hH^+H^-$ {\it i.e.} $\lambda_1$. The measured value of $\mu_{\gamma\gamma}$ are given by
$\mu_{\gamma\gamma}= 1.17\pm 0.27$ from ATLAS \cite{Aad:2014eha} and $\mu_{\gamma\gamma}=1.14^{+ 0.26}_{-0.23}$ from
 CMS \cite{Khachatryan:2014ira}.
 In Fig. \ref{fig:muGamma}, we show the variation of $\mu_{\gamma\gamma}$ as function of $m_{H^\pm}$ for different values of 
 $\lambda_1$. We also present the experimental limits on $\mu_{\gamma\gamma}$ from ATLAS in Fig. \ref{fig:muGamma}. Excepting for the resonance at $m_{h/2}$, we see that our choice of $\lambda_1$ is consistent with experimental bound. 
 The larger is $m_{H^\pm}$ GeV, $\mu_{\gamma\gamma} \to 1$ i.e. approches to SM value. Also, we see that $\lambda_1>0$ diminishes  $\mu_{\gamma\gamma}$, while $\lambda_1<0$ tends to enhance it. In this analysis, 
 we mostly consider $m_{H^\pm}>m_{h/2}$ and positive $\lambda_1$ within correct experimental limit.\\~\\
\noindent
$\bullet$ \textbf{Relic Density of DM:} The PLANCK experiment \cite{Aghanim:2018eyx} provides the observed amount of relic abundance 
\begin{align}
 0.1166 \leq \Omega_{\rm DM}h^2  \leq 0.1206 ~.
\end{align}
Furthermore, strong constraints exist from direct DM search experiments. In our analysis we will
consider the most recent bound on direct detection cross section provided by XENON 1T~\cite{Aprile:2017iyp}. Relic density and direct search allowed parameter space of the model will be evaluated in details. \\~\\
%
\noindent
$\bullet$ \textbf{Neutrino observables:} The parameters associated to the neutrino sector should satisfy the bounds provided by different ongoing neutrino experiments. Limit on sum of light neutrino masses
$\sum m_{\nu_i} \leq 0.12$ eV as provided by PLANCK data \cite{Vagnozzi:2017ovm,Aghanim:2018eyx} is incorporated. The present values of neutrino mass hierarchies and mixing angle
can be found in \cite{deSalas:2017kay,Esteban:2016qun}. Due to the presence of RH neutrino in the set up, the constraint from lepton flavor violating decay (LFV) 
(dominantly from $\mu\rightarrow e\gamma$ ) will be applicable \cite{Ilakovac:1994kj,Tommasini:1995ii,Dinh:2012bp}. LFV constraint can be successfully evaded for 
$M_N \gtrsim 10^{3.5}$ GeV \cite{Bambhaniya:2016rbb,Ghosh:2017fmr} even for neutrino Yukawa coupling of $\mathcal{O}(1)$. It is important to note that in our model, the IDM does not interact with
SM leptons and thus plays no role in LFV.
\section{Single component DM frameworks involving $\phi ~\textrm{or}~ H^0$}
\label{sec:single-component}
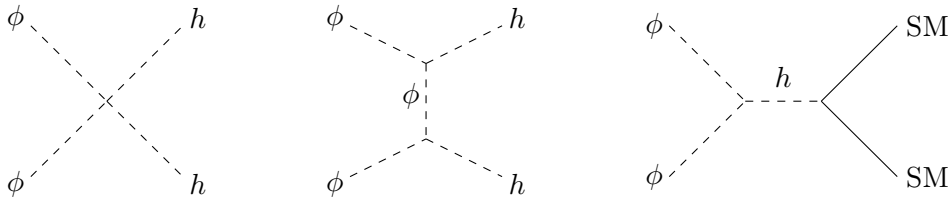
\begin{figure}[htb!]
 \begin{center}
    \begin{tikzpicture}[line width=0.4 pt, scale=1.0]
        \draw[dashed] (-6,1)--(-5,0);
	\draw[dashed] (-6,-1)--(-5,0);
	\draw[dashed] (-5,0)--(-4,1);
	\draw[dashed] (-5,0)--(-4,-1);
	\node at (-6.2,1.1) {$\phi$};
	\node at (-6.2,-1.1) {$\phi$};
	\node at (-3.8,1.1) {$h$};
	\node at (-3.8,-1.1) {$h$};
	\draw[dashed] (-1.8,1.0)--(-0.8,0.5);
	\draw[dashed] (-1.8,-1.0)--(-0.8,-0.5);
	\draw[dashed] (-0.8,0.5)--(-0.8,-0.5);
	\draw[dashed] (-0.8,0.5)--(0.2,1.0);
	\draw[dashed] (-0.8,-0.5)--(0.2,-1.0);
	\node at (-2.0,1.1) {$\phi$};
	\node at (-2.0,-1.1) {$\phi$};
	\node at (-1.0,0.07) {$\phi$};
	\node at (0.4,1.1) {$h$};
	\node at (0.4,-1.1) {$h$};
	%
        \draw[dashed] (2.4,1)--(3.4,0);
	\draw[dashed] (2.4,-1)--(3.4,0);
	\draw[dashed] (3.4,0)--(4.4,0);
	\draw[solid] (4.4,0)--(5.4,1);
	\draw[solid] (4.4,0)--(5.4,-1);
	\node  at (2.2,-1) {$\phi$};
	\node at (2.2,1) {$\phi$};
	\node [above] at (3.9,0.05) {$h$};
	\node at (5.8,1.0){{\rm SM}};
	\node at (5.8,-1.0) {{\rm SM}};
     \end{tikzpicture}
 \end{center}
\caption{Annihilation processes of real scalar singlet DM ($\phi$) to SM particles. SM in the last graph stands for $W^\pm, Z,h$ and SM fermions. }
\label{Feyn_ann_SS}
 \end{figure}

The model inherits two DM candidates: inert DM $H^0$ and singlet scalar DM $\phi$. 
Both the DM components have been studied extensively in literature as individual candidates to satisfy relic density and direct search bounds. 
Let us first revisit the single component frameworks for these two cases here. The relic density of scalar singlet ($\phi$) is obtained via thermal freeze out through annihilation to SM through the Feynman graphs shown in 
Fig.~\ref{Feyn_ann_SS}. The direct search constraint for $\phi$ comes from the $t$- channel Higgs portal interaction (turning the last graph of Fig.~\ref{Feyn_ann_SS} upside down). 
The relevant parameters of the model are \cite{McDonald:1993ex,Feng:2014vea,Bhattacharya:2017fid}:
\bea
\phi \textrm{~as single~component DM}: \{m_\phi, \lambda_{\phi h}\}.
\eea

 \begin{figure}[htb!]
 $$
  \includegraphics[height=4.8cm]{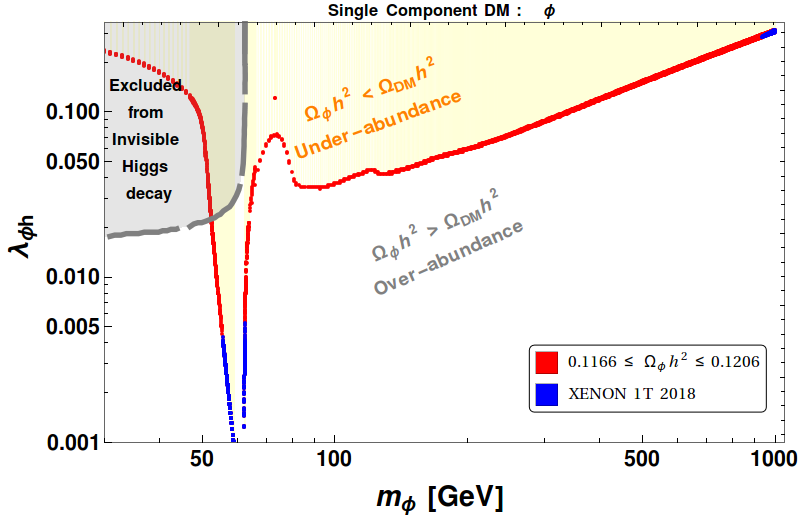} ~
   \includegraphics[height=4.8cm]{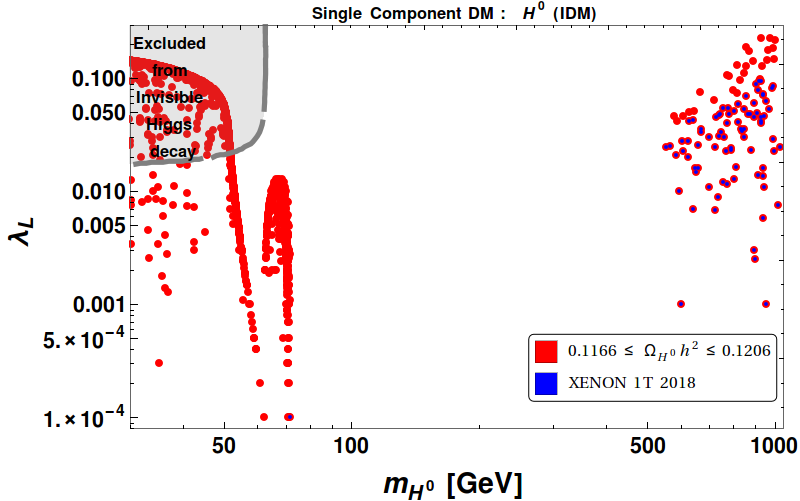} 
  $$
  \caption{Relic density (red dots) and direct search/XENON1T (blue dots) allowed parameter space of the single component DM; scalar singlet ($\phi$) on left 
  (in $m_\phi-\lambda_{\phi h}$ plane) and IDM ($H^0$) on right (in $m_{H^0}-\lambda_L$ plane). 
  For the right hand side plot, we have used: $0 \le m_{A^0}-m_{H^0}\le 200$ GeV, $1 \le m_{H^\pm}-m_{H^0}\le 400$ GeV, and $\lambda_\Phi=0.001$.}
  \label{fig:single-comp}
 \end{figure}
The allowed parameter space of $\phi$ is depicted in left hand side (LHS) of Fig.~\ref{fig:single-comp} in $m_\phi-\lambda_{\phi h}$ plane by the red dots. Direct search allowed parameter space from XENON1T data  
\cite{Aprile:2017iyp,Messina:2018fmz} using spin-independent DM-nucleon scattering cross section is shown by the blue dots. We therefore see that the model can only survive either in the Higgs resonance region 
($\sim m_h/2$) or at a very heavy mass $\gtrsim$ 900 GeV. The under abundant (shown in yellow) and over abundant regions are also indicated. \\~\\
\begin{figure}[htb!]
 \begin{center}
    \begin{tikzpicture}[line width=0.4 pt, scale=1.0]
        \draw[dashed] (-6,1)--(-5,0);
	\draw[dashed] (-6,-1)--(-5,0);
	\draw[dashed] (-5,0)--(-4,1);
	\draw[dashed] (-5,0)--(-4,-1);
	\node at (-6.2,1.1) {$H^0$};
	\node at (-6.2,-1.1) {$H^0$};
	\node at (-3.8,1.1) {$h$};
	\node at (-3.8,-1.1) {$h$};
	\draw[dashed] (-1.8,1.0)--(-0.8,0.5);
	\draw[dashed] (-1.8,-1.0)--(-0.8,-0.5);
	\draw[dashed] (-0.8,0.5)--(-0.8,-0.5);
	\draw[dashed] (-0.8,0.5)--(0.2,1.0);
	\draw[dashed] (-0.8,-0.5)--(0.2,-1.0);
	\node at (-2.1,1.1) {$H^0$};
	\node at (-2.1,-1.1) {$H^0$};
	\node at (-1.2,0.07) {$H^0$};
	\node at (0.4,1.1) {$h$};
	\node at (0.4,-1.1) {$h$};
        \draw[dashed] (2.4,1)--(3.4,0);
	\draw[dashed] (2.4,-1)--(3.4,0);
	\draw[dashed] (3.4,0)--(4.4,0);
	\draw[solid] (4.4,0)--(5.4,1);
	\draw[solid] (4.4,0)--(5.4,-1);
	\node  at (2.1,-1.1) {$H^0$};
	\node at (2.1,1.1) {$H^0$};
	\node [above] at (3.9,0.05) {$h$};
	\node at (5.8,1.0){{\rm SM}};
	\node at (5.8,-1.0) {{\rm SM}};
     \end{tikzpicture}
 \end{center}
 \begin{center}
    \begin{tikzpicture}[line width=0.4 pt, scale=1.0]
        \draw[dashed] (-6,1)--(-5,0);
	\draw[dashed] (-6,-1)--(-5,0);
	\draw[snake] (-5,0)--(-4,1);
	\draw[snake] (-5,0)--(-4,-1);
	\node at (-6.2,1.1) {$H^0$};
	\node at (-6.2,-1.1) {$H^0$};
	\node at (-3.4,1.1) {$W^+(Z)$};
	\node at (-3.4,-1.1) {$W^-(Z)$};
	\draw[dashed] (-1.8,1.0)--(-0.8,0.5);
	\draw[dashed] (-1.8,-1.0)--(-0.8,-0.5);
	\draw[dashed] (-0.8,0.5)--(-0.8,-0.5);
	\draw[snake] (-0.8,0.5)--(0.2,1.0);
	\draw[snake] (-0.8,-0.5)--(0.2,-1.0);
	\node at (-2.1,1.1) {$H^0$};
	\node at (-2.1,-1.1) {$H^0$};
	\node at (-0.1,0.07) {$H^{\pm}(A^0)$};
	\node at (0.8,1.2) {$W^{\pm}(Z)$};
	\node at (0.8,-1.2) {$W^{\mp}(Z)$};
        \draw[dashed] (2.4,1)--(3.4,0);
	\draw[dashed] (2.4,-1)--(3.4,0);
	\draw[dashed] (3.4,0)--(4.4,0);
	\draw[snake] (4.4,0)--(5.4,1);
	\draw[snake] (4.4,0)--(5.4,-1);
	\node  at (2.1,-1.1) {$H^0$};
	\node at (2.1,1.1) {$H^0$};
	\node [above] at (3.9,0.05) {$h$};
	\node at (6.0,1.2){$W^+(Z)$};
	\node at (6.0,-1.2){$W^-(Z)$};
     \end{tikzpicture}
 \end{center}
\caption{Annihilation processes of IDM ($H^0$) to SM particles. SM in the top right graph stands for $W^\pm, Z,h$ and SM fermions. }
\label{Feyn-ann-ID}
 \end{figure}
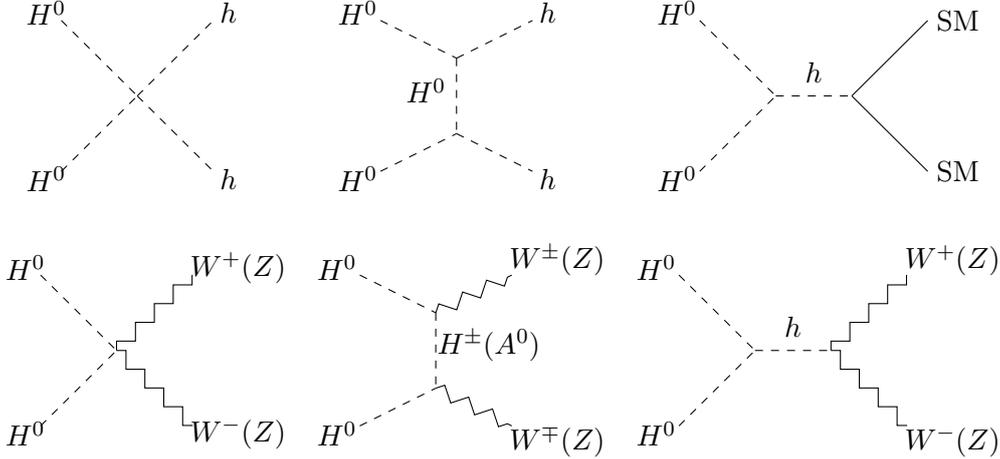
\begin{figure}[htb!]
 \begin{center}
    \begin{tikzpicture}[line width=0.4 pt, scale=1.0]
        \draw[dashed] (-5.0,1)--(-4.0,0);
	\draw[dashed] (-5.0,-1)--(-4.0,0);
	\draw[snake] (-4.0,0)--(-2.4,0);
	\draw[solid] (-2.4,0)--(-1.4,1);
	\draw[solid] (-2.4,0)--(-1.4,-1);
	\node  at (-5.3,1.2) {$H^0$};
	\node at (-5.7,-1.2) {$A^0(H^\pm)$};
	\node [above] at (-3.2,0.05) {$Z(W^\pm)$};
	\node at (-1.0,1.2){{\rm SM}};
	\node at (-1.0,-1.2) {{\rm SM}};
         \draw[dashed] (0.8,1)--(1.8,0);
	\draw[dashed] (0.8,-1)--(1.8,0);
	\draw[snake] (1.8,0)--(2.8,1);
	\draw[snake] (1.8,0)--(2.8,-1);
	\node at (0.6,1.2) {$H^0$};
	\node at (0.5,-1.2) {$H^\pm$};
	\node at (3.4,1.2) {$Z,A$};
	\node at (3.4,-1.2) {$W^{\pm}$};
     \end{tikzpicture}
 \end{center}
   \begin{center}
    ~~~~~~~~~~ ~~ \begin{tikzpicture}[line width=0.4 pt, scale=1.0]
       \draw[dashed] (-1.8,1.0)--(-0.8,0.5);
	\draw[dashed] (-1.8,-1.0)--(-0.8,-0.5);
	\draw[dashed] (-0.8,0.5)--(-0.8,-0.5);
	\draw[snake] (-0.8,0.5)--(0.2,1.0);
	\draw[snake] (-0.8,-0.5)--(0.2,-1.0);
	\node at (-2.1,1.1) {$H^0$};
	\node at (-2.1,-1.1) {$A^0$};
	\node at (0.4,0.07) {$H^0,A^0 (H^\pm)$};
	\node at (1.0,1.2) {$h,Z (W^\pm)$};
	\node at (1.0,-1.2) {$Z,h(W^\pm)$};
        \draw[dashed] (4.0,1.0)--(5.0,0.5);
	\draw[dashed] (4.0,-1.0)--(5.0,-0.5);
	\draw[dashed] (5.0,0.5)--(5,-0.5);
	\draw[snake] (5,0.5)--(6,1.0);
	\draw[snake] (5,-0.5)--(6,-1.0);
	\node at (3.8,1.2) {$H^0$};
	\node at (3.8,-1.2) {$H^\pm$};
	\node at (5.8,0.07) {$H^{\pm}(A^0)$};
	\node at (7.1,1.2) {$W^{\pm}(Z,h,A)$};
	\node at (7.1,-1.2) {$h,Z,A(W^\pm)$};
     \end{tikzpicture}
 \end{center}
\caption{Co-annihilation processes of IDM ($H^0$) with $A^0$ and $H^\pm$ to SM particles. SM in the top left graph stands for $W^\pm, Z,h$ and SM fermions in suitable combination. }
\label{Feyn-coann-ID}
 \end{figure}
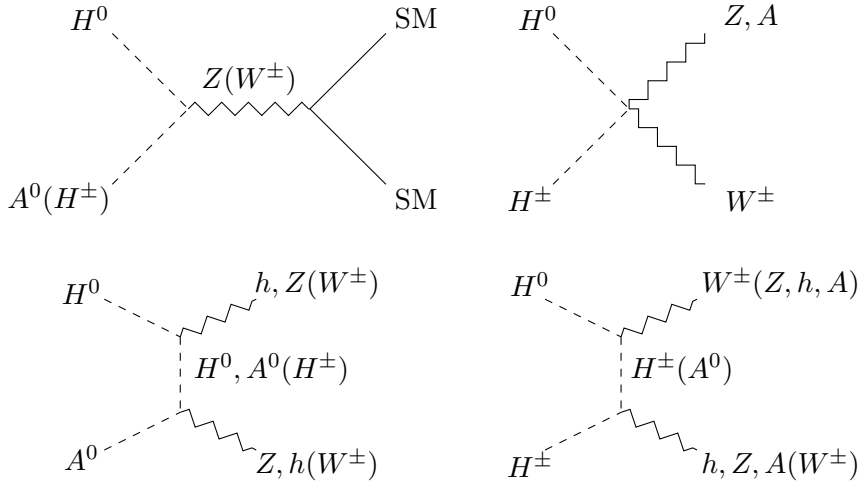
IDM ($H^0$) as a single component DM have annihilation and co-annihilation channels for freeze-out due to both gauge and Higgs portal interactions as shown by the Feynman graphs in 
Figs.~\ref{Feyn-ann-ID} and \ref{Feyn-coann-ID}.
The parameters, which govern the IDM phenomenology are \cite{LopezHonorez:2006gr}:
\bea
H^0 ~\textrm{as~single~component~DM:}~\{m_{H^0},m_{A^0},m_{H^\pm}, \lambda_L\}.
\eea
Relic density allowed parameter space for single component IDM is shown in right hand side (RHS) of Fig.~\ref{fig:single-comp} in $m_{H^0}-\lambda_L$ plane by red dots. Direct search (XENON1T data) allowed points are shown by blue dots.  
The scan is obtained using $1\le m_{A^0}-m_{H^0}\le 200$ GeV, $1\le m_{H^\pm}-m_{H^0}\le 400$ GeV with self coupling $\lambda_\Phi=0.001$ kept constant. 
Here we see again that allowed region from relic density and direct search constraint for IDM lies either in the small mass region $m_{H^0}\lesssim m_W\sim 80$ GeV or in the heavy mass region $m_{H^0}\gtrsim 550$ GeV. It is 
a well known result coming essentially due to too much annihilation and co-annihilation of IDM to SM through gauge interactions \cite{LopezHonorez:2006gr}. The disallowed region $80~\textrm{GeV}<m_{H^0}<550$ GeV is often called 
desert region, which is obviously under abundant. Another important point of single component IDM is that the direct search allowed 
parameter space beyond resonance ($m_{H^0}\gtrsim550$ GeV) have significant co-annihilation dependence with $m_{H^\pm}-m_{H^0}\lesssim 10$ GeV and $m_{A^0}-m_{H^0} \lesssim 10$ GeV. 

\section{Two component DM set-up with $\phi ~\textrm{and}~ H^0$}
\label{sec:2compDM}
\subsection{Coupled Boltzmann Equations and Direct search}
\label{subsec:cbeq}
In presence of two DM components ($\phi ~\textrm{and}~ H^0$) DM-DM conversion plays a crucial role. The heavier 
DM can annihilate to the lighter component and thus contribute to the freeze-out of heavier DM. The conversion processes are shown in Fig.~\ref{Feyn_Conver}, which shows that they are 
dictated by four point contact interactions as well as by Higgs portal coupling. It is clear that $H^\pm,A^0$ are not really DM, but belongs to dark sector, hence annihilation to them is broadly classified within DM-DM conversion. More importantly none of them contribute to direct search. 
The two component DM set up therefore requires following parameters for analysis:
\begin{figure}[htb!]
 \begin{center}
    \begin{tikzpicture}[line width=0.4 pt, scale=1.0]
        \draw[dashed] (-5.0,1)--(-4.0,0);
	\draw[dashed] (-5.0,-1)--(-4.0,0);
	\draw[dashed] (-4.0,0)--(-2.4,0);
	\draw[dashed] (-2.4,0)--(-1.4,1);
	\draw[dashed] (-2.4,0)--(-1.4,-1);
	\node  at (-5.3,1.2) {$\phi$};
	\node at (-5.3,-1.2) {$\phi$};
	\node [above] at (-3.2,0.05) {$h$};
	\node at (-0.3,1.2){{$H^0/~A^0/~H^+$}};
	\node at (0,-1.2) {{$H^0(A^0)/~A^0/ ~H^-$}};
         \draw[dashed] (3,1)--(4,0);
	\draw[dashed] (3,-1)--(4,0);
	\draw[dashed] (4,0)--(5,1);
	\draw[dashed] (4,0)--(5,-1);
	\node at (2.7,1.2) {$\phi$};
	\node at (2.7,-1.2) {$\phi$};
	\node at (6.2,1.2) {$H^0/~A^0/~H^+$};
	\node at (6.2,-1.2) {$H^0/~A^0/ ~H^-$};
     \end{tikzpicture}
 \end{center}
\caption{DM $\rightarrow$ DM conversion processes in a model with $\phi ~\textrm{and}~ H^0$. We have assumed $m_{\phi}>m_{H^0},m_{H^\pm},m_{A^0}$ here. The reverse processes occur with reverse hierarchy.}
\label{Feyn_Conver}
 \end{figure}
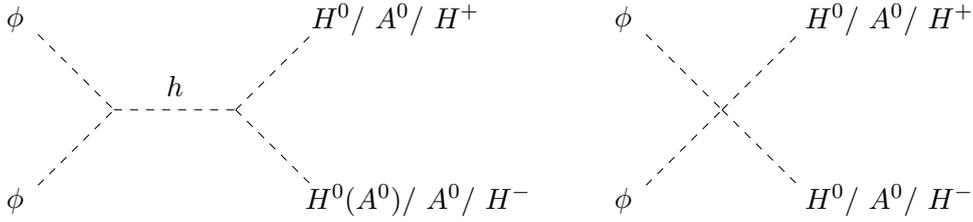
\bea
\textrm{Two~component~DM}:~\{m_{H^0},m_{A^0},m_{H^\pm}, m_{\phi}, \lambda_L, \lambda_{\phi h} \}.
\eea
There are two self interacting quartic couplings present in the model; namely $\lambda_\Phi$ and $\lambda_\phi$,
which do not play an important role in DM analysis, but appear in vacuum stability constraint that we discuss later.

When $m_{\phi}>m_{H^0},m_{H^\pm},m_{A^0}$, then $\phi$ can annihilate to all possible IDM components. 
 The dominant $s$- wave DM-DM conversion cross-sections ($\sigma v $) of $\phi$ in non-relativistic approximation are given by: 
\bea
\nonumber
{(\sigma v)}_{\phi~\phi \rightarrow H^0~H^0}&=&\frac{1}{64 \pi {m_{\phi}}^2}\Big[\lc + \frac{2\lambda_L \lph v^2}{(4 {m_{\phi}}^2-m_h^2)}\Big]^2\sqrt{1-\frac{4 m_{H^0}^2}{4 m_{\phi}^2}}~ \Theta(m_{\phi}-m_{H^0})\\
\nonumber
{(\sigma v)}_{\phi~\phi \rightarrow A^0~A^0}&=&\frac{1}{64 \pi m_{\phi}^2}\Big[\lc + \frac{2\lambda_S \lph v^2}{(4 {m_{\phi}}^2-m_h^2)}\Big]^2\sqrt{1-\frac{4 m_{A^0}^2}{4 m_{\phi}^2}}~ \Theta(m_{\phi}-m_{A^0})\\
{(\sigma v)}_{\phi~\phi \rightarrow H^+~H^-}&=&\frac{1}{32 \pi m_{\phi}^2}\Big[\lc + \frac{\lambda_1 \lph v^2}{(4 {m_{\phi}}^2-m_h^2)}\Big]^2\sqrt{1-\frac{4 m_{H^\pm}^2}{4m_{\phi}^2}}~ \Theta(m_{\phi}-m_{H^\pm}),
\label{eq:conv}
\eea
where $\lambda_S = \frac{1}{2}(\lambda_1+\lambda_2-\lambda_3)$. On the other hand, when $m_{\phi}<m_{H^0}$, the conversion process will be as $H^0H^0 (A^0) \rightarrow \phi\phi$ or $H^{+}H^{-} \rightarrow \phi\phi$. The corresponding cross-sections can easily be gauged from Eqn.~\ref{eq:conv}.

The evolution of DM number density for both components ($\phi ~\textrm{and}~ H^0$) in early universe as a function of time is obtained by coupled Boltzmann equations (CBEQ) as described in Eqn.~\ref{eq:CBEQs2}:
\begin{align}
\frac{dn_{H^0}}{dt}+3 H n_{H^0}&= -\sum_X{\langle \sigma v\rangle}_{H^0~X \rightarrow SM ~SM} \Big(n_{H^0} n_{X}-n_{H^0}^{eq} n_{X}^{eq}\Big)~ \Theta(m_{H^0}+m_{X}-2 ~m_{SM}) \nonumber, \\
& -\sum_X{\langle \sigma v\rangle}_{H^0~X \rightarrow \phi \phi} \Big(n_{H^0} n_{X}-\frac{n_{H^0}^{eq} n_{X}^{eq}}{{n_{\phi}^{eq}}^2}{n_{\phi}}^2\Big)~ \Theta(m_{H^0}+m_{X}-2 ~m_{\phi}) \nonumber ,\\
& +\sum_{X,Y}{\langle \sigma v\rangle}_{\phi \phi \rightarrow X~Y} \Big({n_{\phi}}^2 -\frac{{n_{\phi}^{eq}}^2}{n_{X}^{eq} n_{Y}^{eq}} n_{X} n_{Y}\Big)~ \Theta(2 m_{\phi}-m_{X}+m_{Y}) \nonumber ;\\
\frac{dn_{\phi}}{dt}+3 H n_{\phi}&= -{\langle \sigma v\rangle}_{\phi \phi \rightarrow SM ~SM} \Big({n_{\phi}}^2-{n_{\phi}^{eq}}^2\Big) ~ \Theta(m_{\phi}- ~m_{SM})\nonumber, \\
& -\sum_{X,Y}{\langle \sigma v\rangle}_{\phi \phi \rightarrow X~Y} \Big({n_{\phi}}^2 -\frac{{n_{\phi}^{eq}}^2}{n_{X}^{eq} n_{Y}^{eq}} n_{X} n_{Y}\Big)~ \Theta(2 m_{\phi}-m_{X}+m_{Y})  \nonumber, \\
& +\sum_X{\langle \sigma v\rangle}_{H^0~X \rightarrow \phi \phi} \Big(n_{H^0} n_{X}-\frac{n_{H^0}^{eq} n_{X}^{eq}}{{n_{\phi}^{eq}}^2}{n_{\phi}}^2\Big)~ \Theta(m_{H^0}+m_{X}-2 ~m_{\phi}) \label{eq:CBEQs2}
\end{align}
where $ \{X,Y\}=\{{H^0},{A^0},{H^\pm}\}$. We can clearly spot DM-DM conversion contributions in second and third lines of each equation, which actually make the two equations `coupled'. 
The freeze-out of two component DM is therefore obtained by numerically solving the above CBEQ and yields relic density (for a detailed discussion see for example \cite{Bhattacharya:2016ysw}).  
The total relic density ($\Omega_{DM}$) will then have contributions from both DM components as:
\begin{align}
\Omega_{DM} h^2 = \Omega_{H^0} h^2 +\Omega_{\phi} h^2.
\end{align}

\begin{figure}[h]
\centering
 \includegraphics[height=4.0cm]{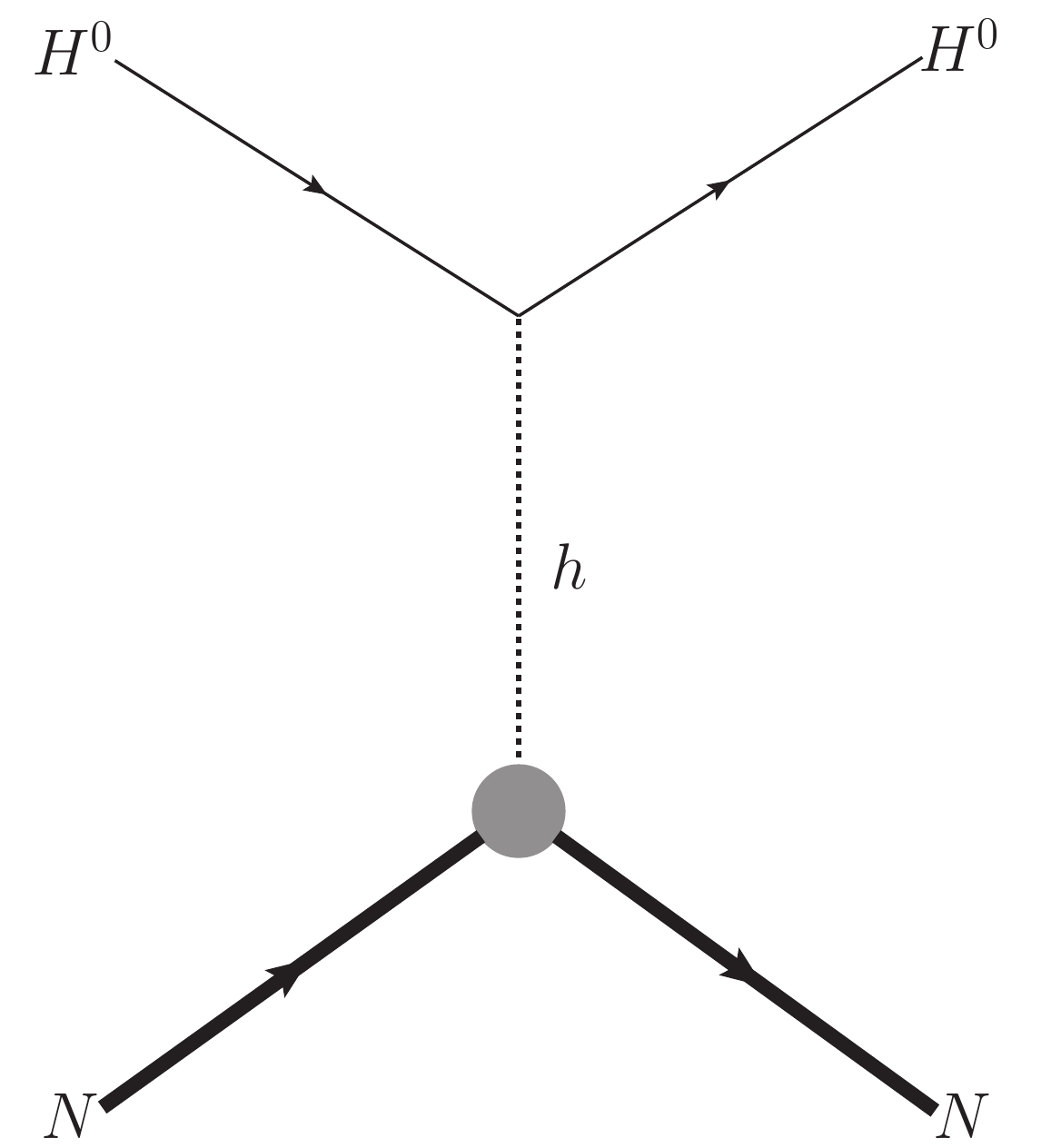} ~~~
 \includegraphics[height=4.0cm]{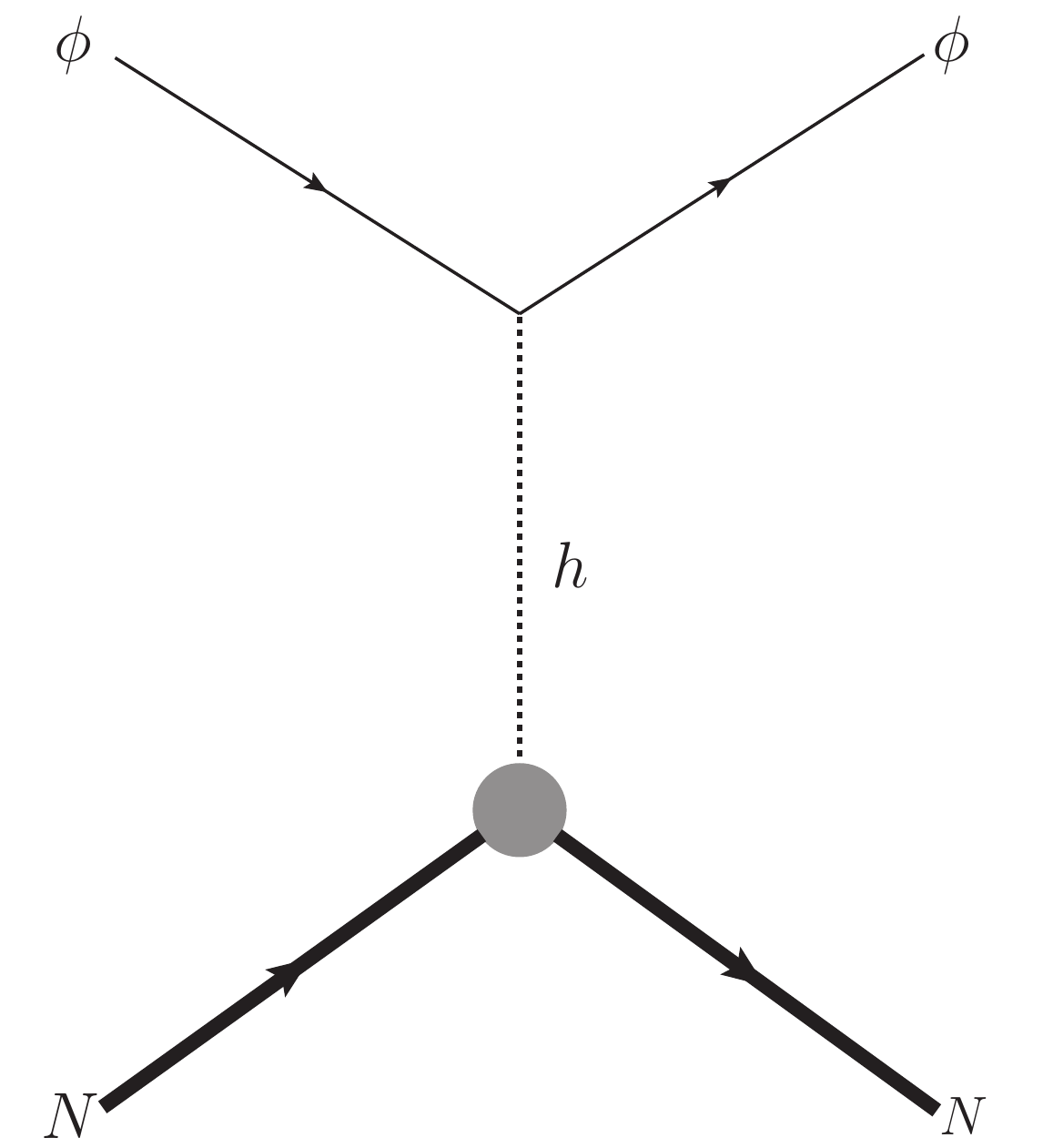}
 \caption{Spin independent direct detection processes for IDM (left) and scalar singlet DM (right).}
 \label{fig:FeynMDD}
\end{figure}

Now let us turn to direct search of two component DM set up. Both the DM candidates can be detected through the spin independent (SI) direct detection (DD) processes through $t$-channel 
Higgs mediation as depicted in Fig. \ref{fig:FeynMDD}.
\noindent The SI DD cross section for $H^0$ ($\sigma_{H^0}$) and for $\phi$ ($\sigma_{\phi}$) at tree level turn out to be 
\cite{Feng:2014vea,Bhattacharya:2016ysw}:
\begin{align}
 \sigma_{H^0}^{\rm eff,~tree}=\Big(\frac{\OH }{\ODM }\Big)\frac{\lambda_L^2f_N^2}{\pi}\frac{\mu_{H^0,N}^2m_N^2}{m_h^4m_{H_0}^2},~~~
 \sigma_{\phi}^{\rm eff,~tree}=\Big(\frac{\Oph}{\ODM }\Big)\frac{\lph^2f_N^2}{4\pi}\frac{\mu_{\phi,N}^2m_N^2}{m_h^4m_{\phi}^2},
 \end{align}
where $\mu_{\phi,N}=\frac{m_{\phi}~m_N}{m_{\phi}+m_N}$ and $\mu_{H^0,N}=\frac{m_{H^0}~m_N}{m_{H^0}+m_N}$
are the reduced masses with nucleon, $f_N=0.2837$ represents the form factor of nucleon \cite{Alarcon:2011zs,Alarcon:2012nr} and 
$m_N=0.939$ GeV represents nucleon mass. Importantly, the effective direct search cross-section 
for each individual component is modified by the fraction with which it is present in the universe\footnote{A more comprehensive bound on multipartite DM from direct search can be obtained from 
the recoil rate of the nucleus \cite{Herrero-Garcia:2017vrl,Ahmed:2017dbb}.}, 
given by $\frac{\OH}{\ODM }$ 
for $H^0$ and $\frac{\Oph}{\ODM }$ for $\phi$. It is pertinent to note here that next to leading order (NLO) 
corrections to SI direct detection cross section can turn dominant for some region of parameter space in 
case of IDM \cite{Klasen:2013btp,Abe:2015rja}. The NLO corrections arise mainly from one loop triangular and box diagrams mediated by gauge bosons and also from Higgs vertex corrections. In \cite{Abe:2015rja}, authors extend the analysis of
\cite{Klasen:2013btp} to include gluon contribution at two loop and also contribution of IDM self coupling. In our case, the self coupling of IDM can be considered to be sufficiently 
small (as it is not playing any significant role in DM phenomenology) and hence we can ignore the effect of it. As shown in \cite{Abe:2015rja}, in presence of loop effects, 
the coupling ($\lambda_L$) that enters into $\sigma_{H^0}^{\rm eff}$ gets replaced by 
$\lambda_L^{\rm eff}=\lambda_L+\delta\lambda_L$ where $\delta\lambda_L$
takes into account the loop contributions effectively. Following this, we rewrite 
$\sigma_{H^0}^{\rm eff}$ as
\begin{align}
\sigma_{H^0}^{\rm eff,~1~loop}=\sigma_{H^0}^{\rm eff,~tree}(1+\kappa),
\label{eq:NLO-DD}
\end{align}
where $\kappa$ is the enhancement factor to SI DD cross section of IDM candidate due to loop correction. It is shown in 
\cite{Klasen:2013btp}, $\kappa$ can take different values depending on $\lambda_L$. For smaller values of $\lambda_L$, 
$\kappa$ turns bigger; for example, with $0.001\lesssim\lambda_L < 0.01$ the correction turns out to be $100  \gtrsim \kappa > 10$; 
for $0.01 \lesssim \lambda_L < 0.1 \to 10  \gtrsim \kappa > 1$ and $\kappa \lesssim 1$ for $\lambda_L \gtrsim 0.1$ for single component IDM 
relic density allowed points. Here we follow similar scaling and take $\kappa\sim \frac{0.1}{\lambda_L}$ to accommodate 
NLO corrections for IDM direct search in our analysis.

To obtain relic density and tree level DD cross sections of both the DM candidates numerically, 
we have used {\tt MicrOmegas}~\cite{Barducci:2016pcb}. It is noteworthy, that version 4.3 of {\tt MicrOmegas} is capable of 
handling two component DM and we have cross-checked the solution from the code to match very closely to the 
numerical solution of CBEQ in Eqn. \ref{eq:CBEQs2}. 
For generating the model files compatible with {\tt MicrOmegas}, we have implemented the model in {\tt LanHEP} \cite{Semenov:2008jy}.

\subsection{Role of DM-DM conversion in relic density}
\label{sec:conversion}
We first study the variation of relic density with respect to DM mass and other relevant parameters to extract the importance 
of DM-DM conversion in this two-component set up before elaborating on the relic density and direct search allowed parameter space of the model. 
\begin{figure}[htb!]
 $$
 \includegraphics[height=4.6cm]{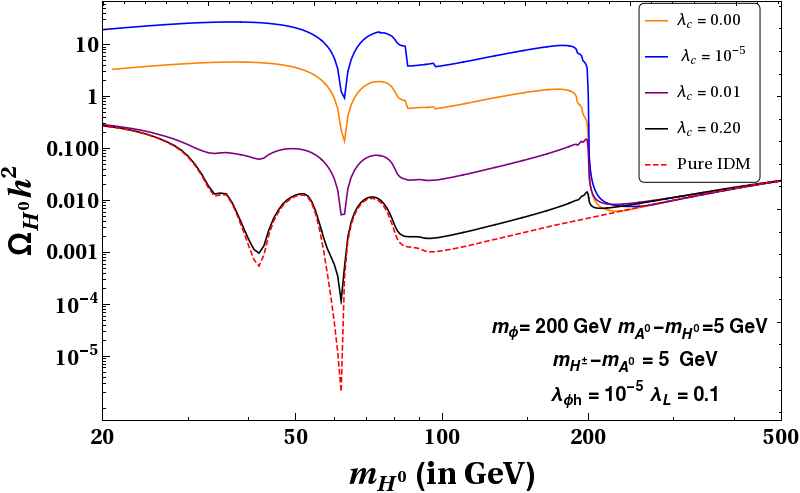} ~~
 \includegraphics[height=4.6cm]{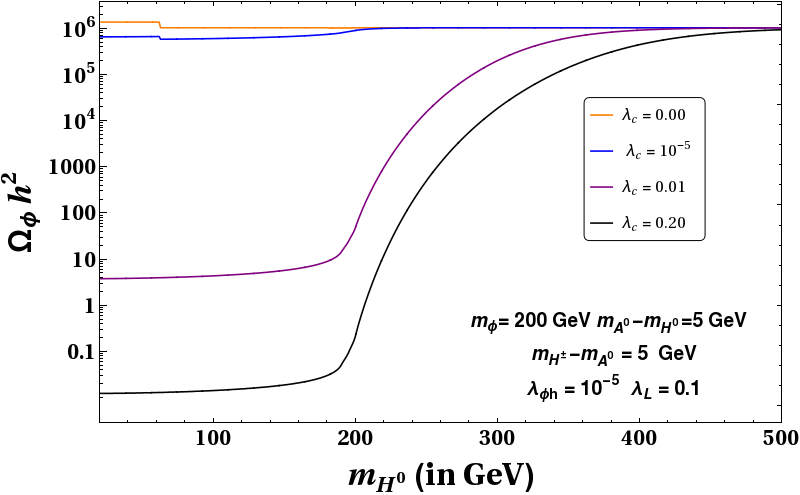}
 $$
 $$
 \includegraphics[height=4.6cm]{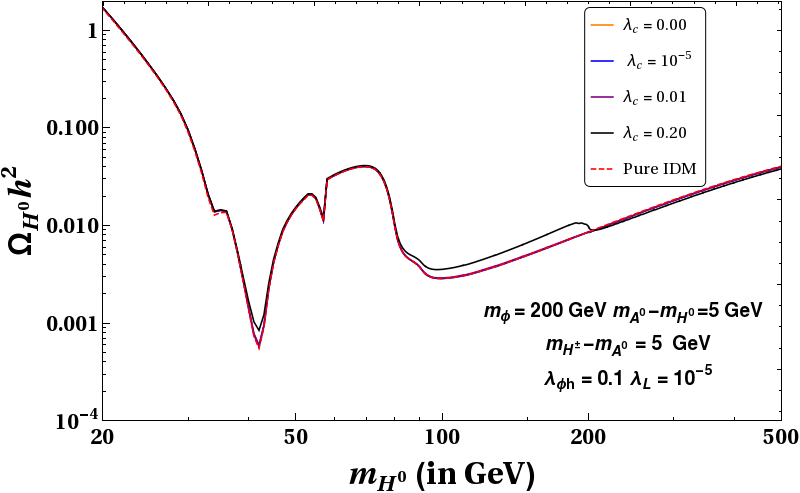} ~~
 \includegraphics[height=4.6cm]{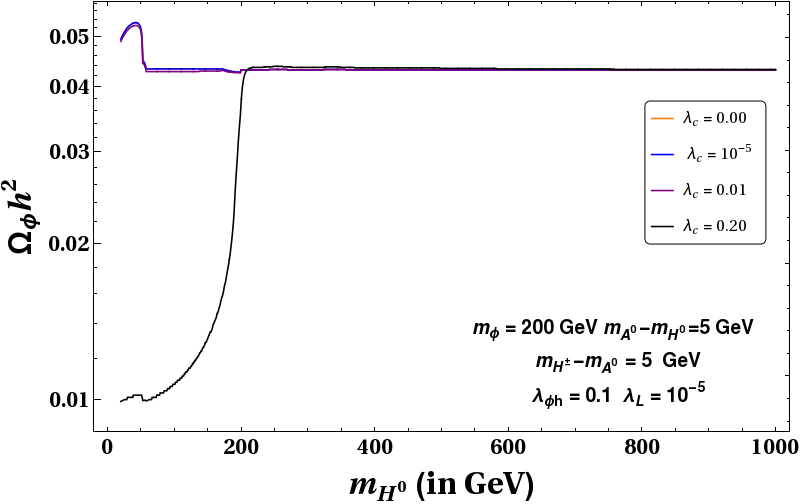}
 $$
 \caption{ Relic Density of IDM, $H^0$ (left panel) and scalar DM, $\phi$ (right panel) as a function of IDM mass $m_{H^0}$, with different choices of $\lc$. We illustrate two different combinations of DM-SM 
 couplings, in the top panel: $\{\lph=10^{-5}, ~\ll=0.1\}$ and in bottom panel: $\{\lph=0.1,~\ll=10^{-5}\}$. Other parameters kept fixed, are mentioned in the inset of each figure.}
 \label{fig:relic_mH0}
\end{figure}

In Fig. \ref{fig:relic_mH0} we plot relic densities of two DM candidates: $\Omega_{H^0} h^2$ in left panel and
$\Omega_{\phi} h^2$ in right panel figures as function of $m_{H^0}$ for different
 values of $\lambda_c$. We also keep $m_{\phi}$ fixed at 200 GeV here. 
 Other parameters are chosen as mentioned in the inset of individual plots. \\
 
 $\bullet$ Top left of Fig. \ref{fig:relic_mH0}: In this figure we have
 shown the variation of $\Omega_{H^0} h^2$ with $m_{H^0}$ for different 
 values of $\lambda_c=0,10^{-5},0.01,0.2$ by yellow, blue, purple and black solid lines respectively. 
 Pure IDM case (in a single component framework) is depicted by red dotted line, where also evidently $\lambda_c=0$. 
 It is important to note the other parameters kept fixed for this plot are: 
 $\lambda_{\phi h}=10^{-5}, \lambda_L=0.1, m_{H^\pm}-m_{A^0}=m_{A^0}-m_{H^0}=$ 5 GeV.
 We see that for $m_{H^0} < m_{\phi}$, relic density of $H^0$ changes significantly 
 with the variation of $\lambda_c$. We also see that $\lambda_c=0$ (in two component set-up) is way above the pure IDM case yielding a 
 large relic density. Now, with slight increase in $\lambda_c=10^{-5}$, relic density goes further up and then reduces significantly for 
 larger $\lambda_c=0.01,0.2$. The interesting point is that the case of $\lambda_c=0.2$ lies very close to 
the pure IDM case. It is therefore evident that the presence of the second DM component $\phi$ plays an important role in $\Omega_{H^0} h^2$
through the coupling $\lambda_c$. For $m_{H^0} < m_{\phi}$, relic density of the heavier component $\phi$ can easily inherit the annihilation to other DM components $\{X,Y\}=\{H^0,A^0,H^\pm\}$ in addition to SM as\cite{Kolb:1990vq,Bhattacharya:2016ysw}:
\bea
\Omega_{\phi} h^2 \simeq \frac{854.45 \times 10^{-13}x_f}{\sqrt{g_{*}} ({\langle \sigma v\rangle}_
{\phi \phi \rightarrow ~\rm SM~\rm SM}+{\langle \sigma v\rangle}_{\phi \phi \rightarrow XY}) } \simeq \frac{0.1~{\rm pb}}{{\langle \sigma v\rangle}_
{\phi \phi \rightarrow ~\rm SM~\rm SM}+{\langle \sigma v\rangle}_{\phi \phi \rightarrow XY}},
\label{eq:relic-phi}
\eea
where in the last step, we have used $g_{*}=106.7$ and $x_f=20$ \cite{Kolb:1990vq}.
 However it is difficult to envisage the relic density for the lighter DM component due to such DM-DM conversion.
 This can be understood from the CBEQ for the two component DM as in Eqn.~\ref{eq:CBEQs2}. Let us define the following notations first:
\begin{align}
\nonumber 
&\langle \sigma v \rangle_{H^0X \to SM}~n_{H^0}n_X=\mathcal{F}_{H^0}~; ~~\langle \sigma v \rangle_{\phi\phi \to XY}~n_{\phi}^2=\mathcal{F}_{\phi \Phi}~; \\
&\langle \sigma v \rangle_{\phi\phi \to SM}~n_{\phi}^2=\mathcal{F}_{\phi};~~~~~~\langle \sigma v \rangle_{H^0 X \to \phi\phi}~n_{H^0}n_X=\mathcal{F}_{ \Phi \phi}~.
\end{align}
Again $\{X,Y\}={H^0,A^0,H^\pm}$; as earlier. The CBEQ with above notation, turns out to be (assuming $m_\phi>m_{H^0}$):
\bea
\nonumber 
\frac{dn_{H^0}}{dt}+3 H n_{H^0}& \simeq -\mathcal{F}_{H^0}+\mathcal{F}_{\phi \Phi} ~;\\
\frac{dn_{\phi}}{dt}+3 H n_{\phi}& \simeq -\mathcal{F}_{\phi} - \mathcal{F}_{\phi \Phi}~.
\label{CBEQ-simple}
\eea
In Eqn.~\ref{CBEQ-simple}, we neglected the equilibrium number densities as they are tiny near freeze-out, where the dynamics is under study. 
With $\lambda_{c} = 0, ~\textrm{and}~\lambda_{\phi h}=10^{-5}$, annihilation cross-sections for $\phi$, (${\langle \sigma v\rangle}_{\phi~\phi \rightarrow {\rm SM} ~{\rm SM}}$
 and ${\langle \sigma v\rangle}_{\phi~\phi \rightarrow {\rm X} ~{\rm Y}}$) are very small. Hence $\phi$ freezes out early and the number density
 of $\phi$ turns out to be large since $n_{\phi} \propto 1/\langle \sigma v\rangle_{\phi}^{\rm eff}$ where
 $\langle \sigma v\rangle_{\phi}^{\rm eff}\simeq\Big({\langle \sigma v\rangle}_{\phi \phi \rightarrow ~\rm SM~\rm SM}+{\langle \sigma v\rangle}_{\phi \phi \rightarrow ~X~Y}\Big)$ following Eqn.~\ref{eq:relic-phi}. 
 Now, it is easy to appreciate that with $\lambda_{c} = 0, ~\textrm{and}~ \lambda_{\phi h}=10^{-5}$, annihilation of $\phi$ to SM is larger than conversion to other DM ($H^0$), i.e. 
 ${\langle \sigma v\rangle}_{\phi \phi \rightarrow ~\rm SM~\rm SM}\gg{\langle \sigma v\rangle}_{\phi \phi \rightarrow XY}$. 
However due to large $\phi$ abundance\footnote{$n_{\phi}$ can be calculated by using the formula $\Omega_{\phi}=\frac{n_{\phi} m_{\phi}}{\rho_c}$, where $\rho_c=1.05\times 10^{-5} h^2$ GeV cm$^{-3}$\cite{Kolb:1990vq}.} $(n_{\phi}\sim 0.1~ {\rm cm^{-3}})$, $\mathcal{F}_{\phi \Phi}$ becomes comparable with $\mathcal{F}_{H^0}$. As these two terms ($\mathcal{F}_{\phi \Phi}~\textrm{and}~\mathcal{F}_{H^0}$) appear in the evolution of 
$n_{H^0}$ (Eqn.~\ref{CBEQ-simple}) with opposite sign, it is quite evident that effective annihilation cross-section for $H^0$ becomes small
and hence $n_{H^0}$ after freeze out turns out to be much larger than the pure IDM case. 
Next let us consider non zero but small $\lambda_c$ ($= 10^{-5}$). Then ${\langle \sigma v\rangle}_{\phi \phi \rightarrow ~X~Y}$ increases
compared to the earlier case of $\lambda_c=0$. However due to smallness of the coupling $\lambda_c$, this does not make any significant change 
in the number density of $\phi$ and $n_{\phi} \sim 0.1 ~ {\rm cm^{-3}}$ remains the same (as
$\phi~\phi \to {\rm SM}~{\rm SM}$ still dominantly contributes to the total annihilation
cross section of $\phi$). Therefore, $\mathcal{F}_{\phi \Phi}$ increases and reduces the separation
with $\mathcal{F}_{H^0}$. 
Hence the effective annihilation cross-section for $H^0$ turns out to be even smaller than $\lambda_c = 0$ case. Therefore, for  $\lambda_c= 10^{-5}$
relic density increases further than that of $\lambda_c=0$. For larger value of $\lambda_c=0.01$, 
contribution from DM-DM conversion, ${\langle \sigma v\rangle}_{\phi \phi \rightarrow ~X~Y}$
significantly rises and therefore the number density of $n_\phi$ drops to {\bf $n_\phi \sim 10^{-5} ~ {\rm cm^{-3}}$}, 
and therefore $\mathcal{F}_{\phi \Phi}$ becomes much smaller than $\mathcal{F}_{H^0}$. 
This increases the effective annihilation for $H^0$ and reduces relic density. 
This trend continues for higher values of $\lambda_c$ and eventually leads to a vanishingly small $\mathcal{F}_{\phi \Phi}$ to
closely mimic the case of single component IDM. 
The case of $m_{H^0}>m_\phi$ can also be understood from the 
CBEQ in this limit:
\bea
\nonumber 
\frac{dn_{H^0}}{dt}+3 H n_{H^0}& \simeq -\mathcal{F}_{H^0}-\mathcal{F}_{ \Phi \phi} ~;\\
\frac{dn_{\phi}}{dt}+3 H n_{\phi}& \simeq -\mathcal{F}_\phi +\mathcal{F}_{\Phi\phi}~.
\label{CBEQ-simple-2}
\eea
For $m_{H^0}>m_\phi$, relic density of $H^0$ can be written simply as 
\bea
\Omega_{H^0} h^2 \simeq \frac{0.1~{\rm pb}}{{\langle \sigma v\rangle}_
{H^0X \rightarrow ~\rm SM~\rm SM}~+~{\langle \sigma v\rangle}_{ H^0 X \rightarrow \phi \phi}}.
\label{eq:relic-H}
\eea
The annihilation to SM (${\langle \sigma v\rangle}_{H^0X \rightarrow ~\rm SM~\rm SM}$) due to gauge coupling is much larger than the conversion cross-section 
${\langle \sigma v\rangle}_{ H^0 X \rightarrow \phi \phi}$ for all the choices of $\lambda_c$, so we do not find any distinction between all those cases.

$\bullet$ Top right of Fig. \ref{fig:relic_mH0}: In the top right panel, the same parameter space is used to show the variation of $\Omega_\phi h^2$ with respect to $m_{H^0}$. The 
dynamics is much simpler for $m_\phi>m_{H^0}$ (see BEQ.~\ref{CBEQ-simple}) where $\Omega_{\phi} h^2 \simeq \frac{0.1~{\rm pb}}{{\langle \sigma v\rangle}_
{\phi \phi \rightarrow ~\rm SM~\rm SM}~+~{\langle \sigma v\rangle}_{\phi \phi \rightarrow ~X~Y}}$. With larger $\lambda_c$, the conversion to other DM ($\mathcal{F}_{\phi \Phi}$) becomes larger and relic density drops accordingly. 
For $m_\phi<m_{H^0}$, we see from Eqn.~\ref{CBEQ-simple-2}, that there is a competition between $\mathcal{F}_{\phi} $ and $\mathcal{F}_{\Phi\phi}$. With increasing $m_{H^0}$, 
$\langle \sigma v \rangle_{H^0 X \to \phi\phi}$ decreases, therefore $\mathcal{F}_{ \Phi \phi}$ decreases and eventually it 
becomes vanishingly small for $m_{H^0} \gtrsim 400$ GeV. 
The equation for $n_{\phi}$ then becomes equivalent to the single component case of $\phi$ where $m_{H^0}$ is no more relevant for $\Omega_\phi h^2$. 

$\bullet$ The bottom panel figures of Fig.~\ref{fig:relic_mH0} essentially indicate that with larger $\lambda_{\phi h}$, 
annihilation of $\phi$ to SM becomes large, 
resulting a smaller $n_\phi$ after freeze out. Therefore $\mathcal{F}_{\phi \Phi}$ in Eqn. \ref{CBEQ-simple} turns insignificant. 
On the other hand, $\mathcal{F}_{ \Phi \phi}$ also becomes smaller than 
$\mathcal{F}_\phi$ in Eqn. \ref{CBEQ-simple-2}. Together, relic density of the lighter DM component is not affected by the presence 
of a heavy DM component. 

 Before we move on, let us summarise the outcome of Fig.~\ref{fig:relic_mH0}. We see here that relic density of lighter 
 DM component is affected by the heavier one, 
 when the annihilation cross-section to SM is tiny.  
\begin{figure}[htb!]
 \includegraphics[height=4.6cm]{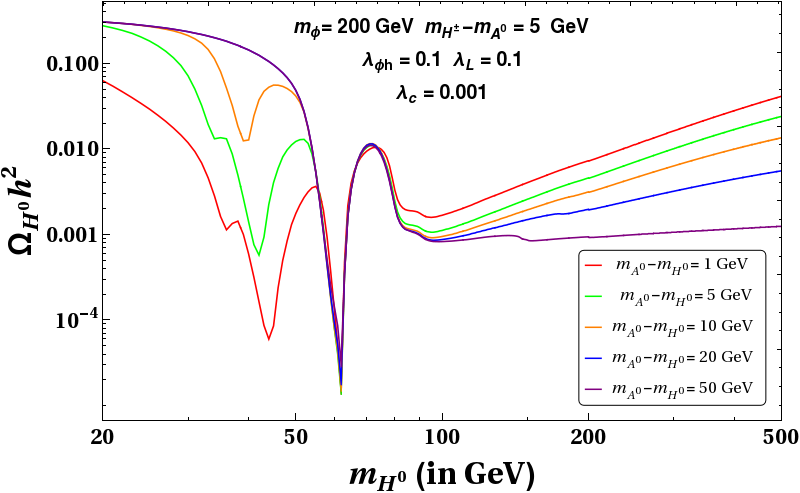}~~ 
 \includegraphics[height=4.6cm]{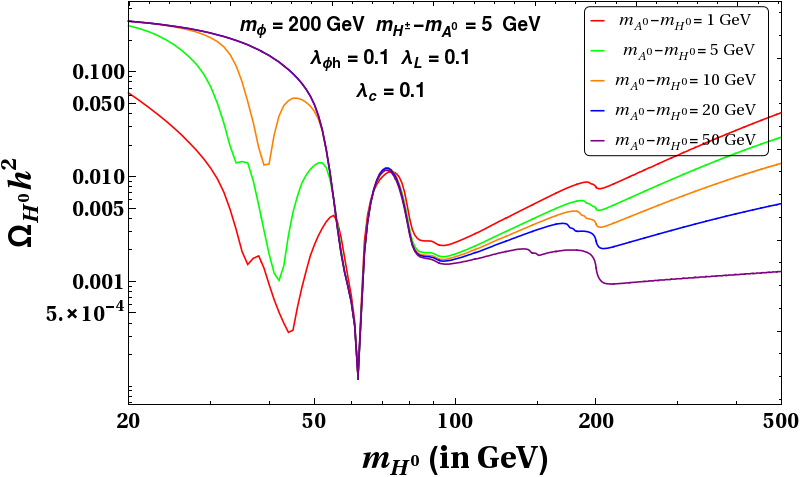}~~ 
 \caption{$\Omega_{H^0} h^2 ~\rm vs~ m_{H^0}$ for $\lc=0.001$ (left panel) and $\lc=0.1$ (right panel) with different choices of $m_{A^0}-m_{H^0}$ 
 which are depicted by different coloured lines in the figure. The other parameters kept fixed are mentioned in the inset of each figure. }
 \label{fig:IDM-coan}
\end{figure}

In Fig.~\ref{fig:IDM-coan}. 
we plot the variation of IDM relic density as function of DM mass $\mH$ for different choices of $\Delta m=m_{A^0}-m_{H^0}:\{1-50\}$ GeV shown by different coloured lines. The other parameters are kept 
fixed and mentioned explicitly in the figure insets. Let us first focus on the left panel plot. 
We notice that for $\mH < m_W$, with larger $\Delta m=\mA-\mH$, relic density becomes larger. 
Keeping in mind that in this case ($\mH<m_W$), co-annihilations  (see Fig.~\ref{Feyn-coann-ID}) 
dominate the total $\langle\sigma v \rangle$, this can be explained from the usual convention of co-annihilation cross-section being reduced by Boltzmann suppression ($\sim e^{-\Delta m/T}$) with larger splitting ($\Delta m$). 
However, for  $\mH>m_W$, the phenomena seems to be reversed, $i.e.,$ with larger mass splitting, relic 
density decreases.
 Note that for $\mH>m_W$, DM annihilations (in particular, annihilations into gauge bosons) 
become dominant. It turns out that the contribution of $\langle \sigma v \rangle _{H^0 H^0 \rightarrow 
W^{+} W^{-} {\rm or} ~ZZ}$ increases with larger $\Delta m$ (for the relevant cross-section, see \cite{Biswas:2013nn}). 
This happens due to presence of the destructive interference between the $s$ and $t$ channel contributions. With larger $\Delta m$, the destructive interference reduces and hence the annihilation cross section increases which in turn lowers the relic density of $H^0$ in this region.
The same feature prevails in the right panel figure. Here, for a larger $\lc$, 
the relic density of $H^0$ goes further down due to annihilation to $\phi$ beyond $\mH>\mph$.
\section{Relic density and Direct Search allowed parameter space}
\label{sec:RD-ps}
One of the important motivations of this analysis is to study interacting
multi-component DM phenomenology for DM mass lying between $80 \leq m_{DM} \leq 500 $ GeV for 
both the components in view of relic density ($0.1166 \leq \Omega_{DM} h^2 (\equiv \Omega_{H^0} h^2 + \Omega_{\phi} h^2) \leq 0.1206$~\cite{Aghanim:2018eyx}) and direct detection (XENON 1T~\cite{Aprile:2017iyp}) bounds. We
would like to recall that for the individual scenarios, none of the DM components satisfy relic and direct detection bounds simultaneously within
this mass range (see section \ref{sec:single-component}). Now in order to find a consistent parameter space in the set up, 
we perform a numerical scan of the relevant parameters within the specified ranges as mentioned below.
\begin{align}
80 ~\leq m_{H^0} \leq 500~{\rm GeV},~80 \leq~ m_{\phi} \leq 500~ \rm GeV, \nonumber\\
 0 ~\leq m_{A^0} - m_{H^0} \leq 200~{\rm GeV},~0 \leq~ m_{H^\pm} - m_{A^0} \leq 180~ \rm GeV, \nonumber\\
0.001 \leq \ll \leq 0.30 ,~~0.001 \leq \lph \leq 0.20~,~~0.01 \leq \lc \leq 1.00~.
\label{eq:}
\end{align}
We also note here that as $\ll$ and $\lph$ enters into direct search cross-sections for $H^0$ and $\phi$ DMs respectively, we keep those couplings in a moderate range, while $\lc$ 
governs DM-DM interactions, but do not directly enter into direct search bounds, therefore we choose a larger 
range (with natural values) for scanning $\lc$. $\lambda_\phi = 0.001$ and $\lambda_\Phi = 0.001$ are kept fixed throughout the 
analysis since those are not relevant for DM phenomenology.

 There are two possible mass hierarchies for the two-component DM set up relevant for phenomenological analysis: 
 (i) $\mph>\mH$ and (ii) $\mph \le \mH$, which we address separately below.  
 \subsubsection*{Case I: ~$\mph>\mH$ }
 Primarily, in such a scenario, the main physics arises due to annihilation of $\phi$ to $H^0$, on top of their individual annihilation to SM 
 to govern the freeze-out.
 \begin{figure}[htb!]
 \centering
 $$
 \includegraphics[height=5.0cm]{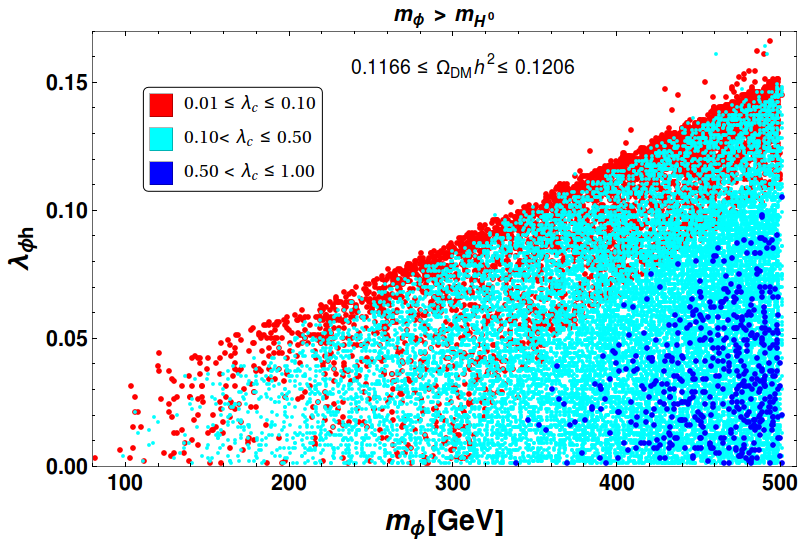} ~
 \includegraphics[height=5.0cm]{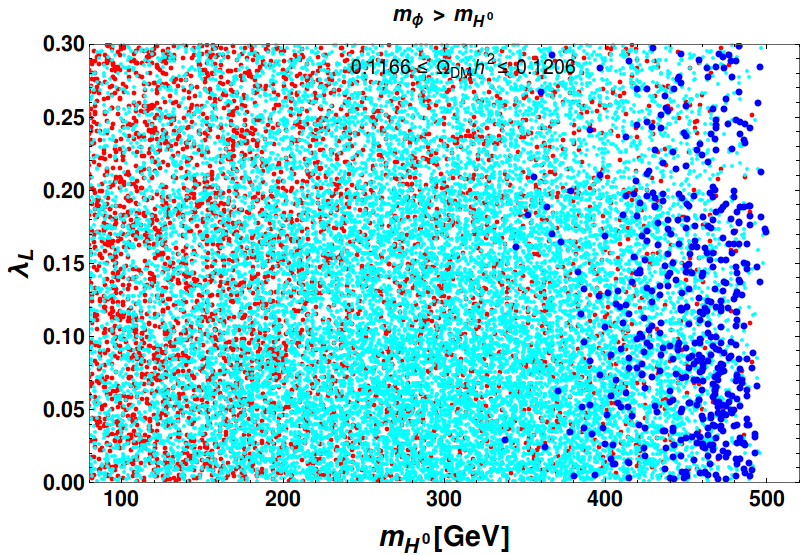}
 $$
 $$
 \includegraphics[height=5.0cm]{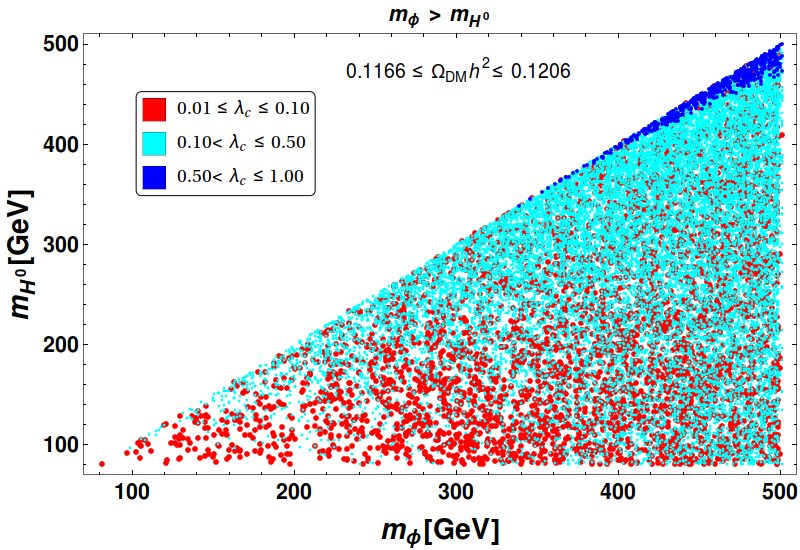} ~
 \includegraphics[height=5.0cm]{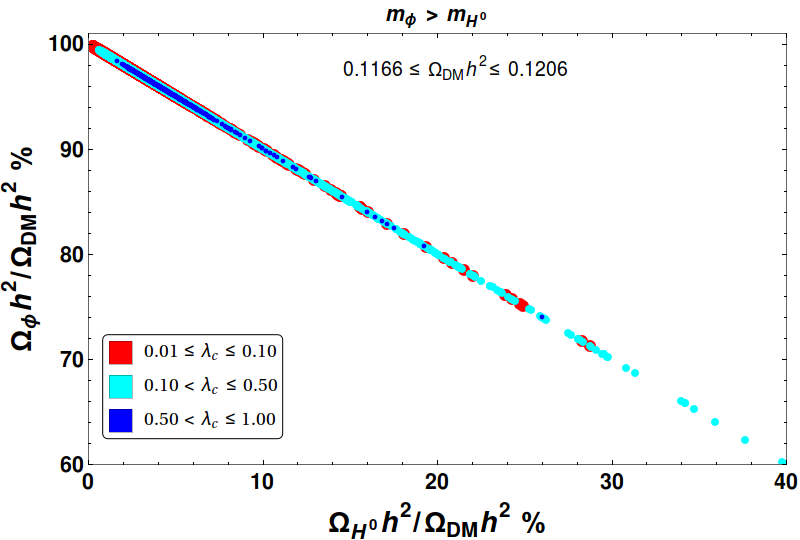} 
 $$
 \caption{Relic Density allowed parameter space is shown in $\mph-\lph$ plane (top left), $\mH-\ll$ plane (top right), 
 $\mph-\mH$ plane (bottom left) and $\Oph/\ODM (\%)-\OH/\ODM (\%)$ plane (bottom right) 
 for the mass hierarchy $\mph>\mH$.}
 \label{fig:lam_DM_mass}
\end{figure}

In Fig.~\ref{fig:lam_DM_mass}, we show relic density allowed parameter space for the model in terms of different relevant parameters.  
In top left panel of Fig.~\ref{fig:lam_DM_mass}, we have shown the relic density 
satisfied (considering contribution from both $\phi$ and $H^0$ components) points in $\mph-\lph$ plane for different ranges of $\lc$, as depicted
in the figure with different colour codes.
As seen from the plot, for a fixed $\mph$, there is a maximum $\lph$. 
All possible values less than the maximum $\lph$ is also allowed subject to different choices of $\lc$. The larger is $\lc$, the smaller is the 
required $\lph$ thanks to the conversion of $\phi \to H^0$ to yield relic density. It is also noted that for large $\lc \gtrsim 0.5$, as the DM-DM 
conversion is very high, the DM mass $\mph$ has to lie in the high mass region 
($\gtrsim 400$ GeV) to tame the annihilation cross-section within acceptable range. 
To summarise, this figure shows that due to the presence of second DM component,
much larger parameter space (actually the over-abundant regions of the single component framework, compare Fig.~\ref{fig:single-comp}) is allowed. 
Top right panel figure shows the relic density allowed parameter space in $\mH-\ll$ plane again for different ranges of $\lc$ as in the left plot. 
It naturally depicts that $\ll$ is insensitive to $\mH$ as the annihilation and co-annihilation cross-section of $H^0$ is mainly dictated 
by gauge interaction. However, we see a mild dependence on $\lc$, such that when $\lc \gtrsim 0.5$, the DM mass $\mH$ has to be heavy 
$\gtrsim 400$ GeV. This 
is because with large $\lc$, DM-DM conversion is large; to achieve relic density of correct order, $\mph$ requires to be large and the conversion can only be tamed down by 
phase space suppression, i.e. by choosing $H^0$ mass as close as possible to $\phi$ mass ($\mH \sim \mph$). Bottom left figure correlates the DM masses to obtain 
correct density within $\mph>\mH$. We see that for small $\lc$, particularly with higher $\mph$, large $\mH$ values are disfavoured 
in order to keep the DM-DM conversion in the right order. While for large $\lc$, mass degeneracy is required ($\mH \sim \mph$) to tame the DM-DM conversion. 
Bottom right figure shows the relative contribution of relic density of the two DM components. First of all, this shows that $\phi$ contributes with larger 
share of relic density, while the relic density of $H^0$ can at most be $40\%$ of the total.  This is natural from the perspective of IDM to be the lighter DM 
component, as we know it is always suppressed below 500 GeV (with a maximum contribution $40\%$) and 
the conversion channels do not alter IDM's annihilation for $m_\phi>m_{H^0}$. Although for $0.01<\lambda_c\lesssim 0.1$, individual relic abundance of IDM 
 could be large for very small (and fine tuned)  $\lambda_{\phi h}\sim 10^{-5}$, as shown in upper left of Fig. \ref{fig:relic_mH0}; with the increase of 
 $\lambda_{\phi h}$, this enhancement vanishes. Since we scan the parameter space
 where $\lambda_{\phi h}\gtrsim 0.001$, such effect can not be found in Fig. \ref{fig:lam_DM_mass}. For small $\lambda_c$, contribution from $H_0$ 
 is smaller, as relic density contribution from $\phi$ gets larger due to small DM-DM conversion. With larger $\lc\gtrsim 0.1$, the DM-DM conversion for 
 $\phi$ becomes larger and therefore the relic density of $\phi$ can easily span between $60-100 \%$. 
 With very high $\lc\sim 1$, DM-DM conversion becomes too large, therefore to keep relic density in the correct ballpark,
the DM mass ($\mph$) has to be heavy and almost degenerate with the heavier DM ($\mH \sim \mph$). $\lph$ in such cases, 
requires to be very small, which are validated by some dark blue 
points with $\Oph/\ODM \sim 60\%$.  
\begin{figure}[htb!]
 \centering
 $$
 \includegraphics[height=5.0cm]{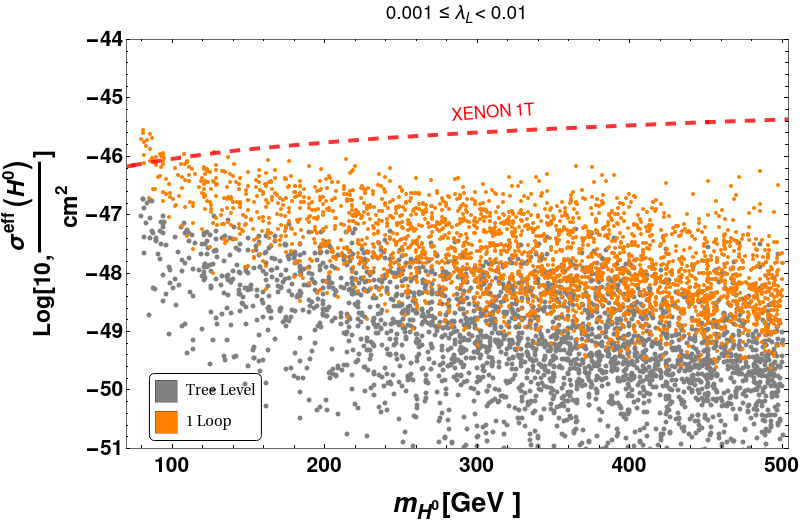} ~
 \includegraphics[height=5.0cm]{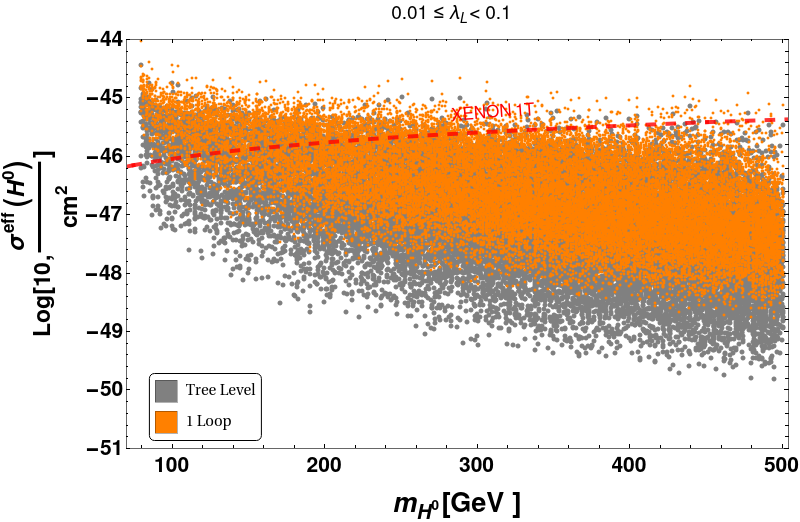}
 $$
 $$
 \includegraphics[height=5.0cm]{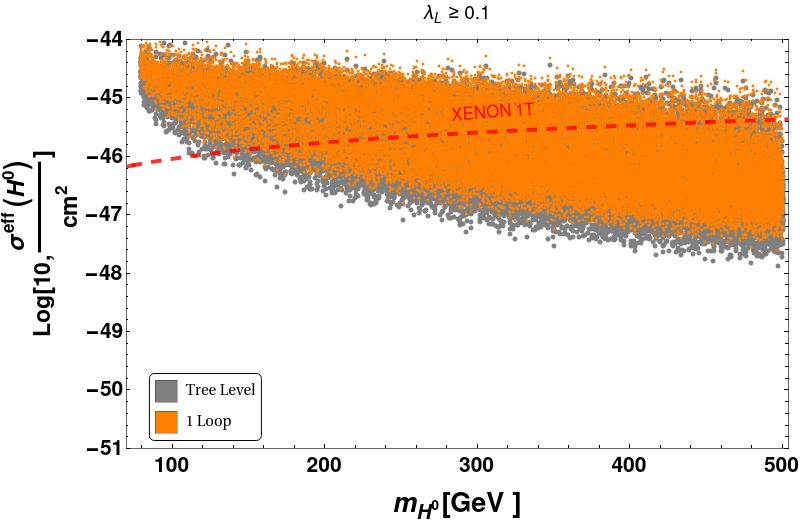}
 $$
 \caption{SI Direct detection cross section for IDM component is plotted against $m_{H^0}$ for (upper left) $0.001 \lesssim \lambda_L < 0.01$, (upper right)
 $0.01 \lesssim \lambda_L < 0.1$ and (bottom) $\lambda_L \gtrsim 0.1$. All the figures show $\sigma_{H^0}^{\rm eff}$ both at tree level (light black) and 
 at NLO level (orange). The XENON 2018 bound (red dotted line) is shown for comparison purpose.}
 \label{fig:DDNLO}
\end{figure}

Next in Fig. \ref{fig:DDNLO}, we show the SI direct detection
cross section for $H^0$ at relic density satisfied points in $\sigma^{\rm eff}_{H^0}-m_{H^0}$ plane. 
Direct search cross-section at tree level is shown by light black points 
and NLO effects as prescribed in Eq.~\ref{eq:NLO-DD}, is shown by orange points. For comparison, we also point out the most recent 
XENON 1T experimental bound in dotted red line.
Upper left panel of Fig. \ref{fig:DDNLO} shows that for $0.001 \lesssim \lambda_L < 0.01$, the one loop correction to $\sigma^{\rm eff}_{H^0}$ is really significant considering $\kappa \sim \frac{0.1}{\lambda_L}$ and rules out some of the DD bound satisfied points at tree level. Similar kind of feature can be found for $0.01 \lesssim\lambda_L < 0.1$ case (upper right panel figure) also. Once $\lambda_L \gtrsim 0.1$, as in the lower panel of 
Fig. \ref{fig:DDNLO}, $\sigma^{\rm eff}_{H^0}$ at NLO shows minor difference to tree level computation.

\begin{figure}[htb!]
\centering
$$
 \includegraphics[height=5.0cm]{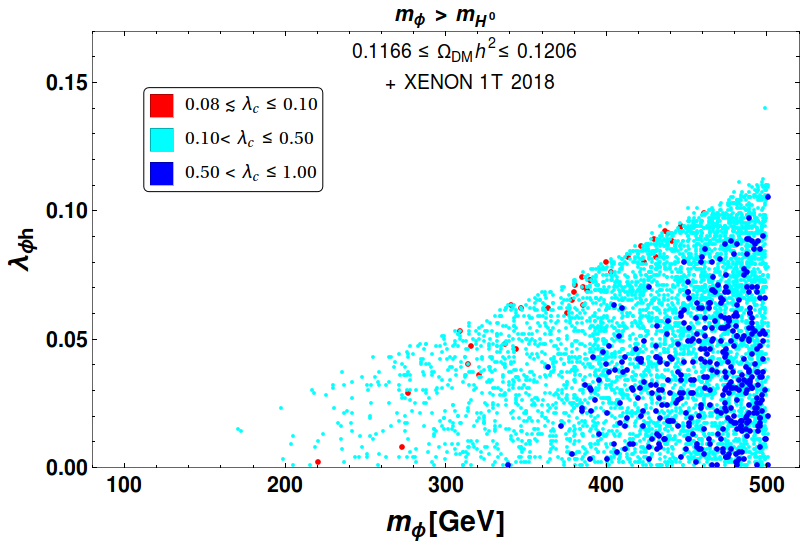} ~
 \includegraphics[height=5.0cm]{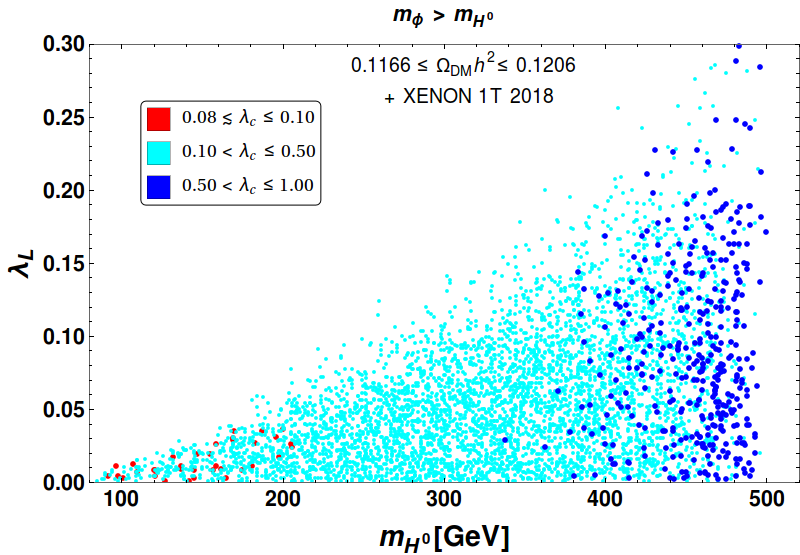}
 $$
 $$
 \includegraphics[height=5.0cm]{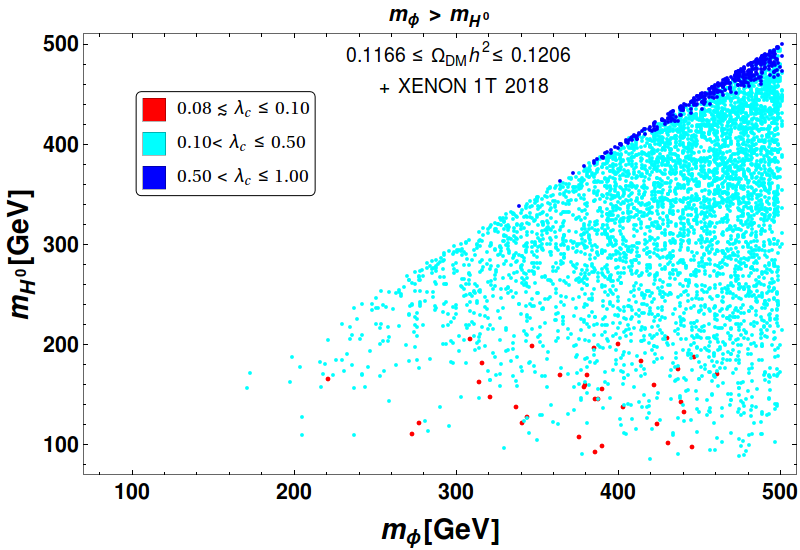}
 $$
 \caption{ Relic Density and direct detection (XENON 1T 2018) at NLO allowed parameter space is shown in: 
 $\mph-\lph$ plane (top left panel), $\mH-\ll$ plane (top right panel) and $\mph-\mH$ plane (bottom panel). The scans are performed for for the mass hierarchy $\mph>\mH$.}
 \label{fig:lam_DM_mass_A}
\end{figure}

Relic density allowed parameters space consistent with direct search constraints at NLO where
 both DMs $\phi$ and $H^0$ simultaneously satisfy XENON 1T 2018 \cite{Aprile:2018cxk} bound (for different ranges of $\lc$) are shown 
 next in Fig.~\ref{fig:lam_DM_mass_A}. This is illustrated in $\mph-\lph$ plane (top left panel), $\mH-\ll$ plane (top right panel) 
 and in $\mph-\mH$ plane (bottom panel) similar to Fig.~\ref{fig:lam_DM_mass}. 
 We have already mentioned that spin independent (SI) DM-nucleon cross-section depends on square of Higgs portal couplings of the 
 respective DM candidates, $\lph$ for $\phi$ and $\ll$ for $H^0$ scaled by a pre-factor of the relative number density $\Omega_i h^2/\Omega_{DM}h^2$ ($i=\phi,~H^0$). 
 Since in this two component scenario, the dominant contribution is coming from $\phi$ DM, the pre-factor $\Oph/\ODM \sim 1$. 
On the other hand, for $H^0$ the pre-factor is small, $\OH/\ODM < 1$, and will help $H^0$ reducing the effective direct search cross-section. 
Therefore, portal coupling $\lph$ is tightly constrained from XENON 1T bound to $\lph \lesssim 0.1$ for DM mass $m_\phi\lesssim 500$ GeV, as shown in top left panel of
Fig.~\ref{fig:lam_DM_mass_A} (compare it with the top left panel of Fig.~\ref{fig:lam_DM_mass}). 
Similarly in top right panel, the direct search allowed $\mH-\ll$ plane for $H^0$ shows that a large region corresponding to 
higher $\lambda_L$ is excluded as a function of $\mH$ (again, compare it with top right panel of Fig.~\ref{fig:lam_DM_mass}). 
A possible mass correlation after direct search bound are plotted in bottom panel of Fig.~\ref{fig:lam_DM_mass_A} in $\mph-\mH$ plane. 
The main outcomes from this figure are: (i) For $\lc \le 0.1$, small $\mH \sim 200$ GeV is favoured, (ii) for moderate values of $\lc\sim\{0.1-0.5\}$, there is no correlation 
and (iii) for large $\lc$, only degenerate mass scenario ($\mph \sim \mH$) with large $\mph\sim 400$ GeV is allowed.  

\subsubsection*{Case II: ~$\mph \leq \mH$ }
Naturally here the conversion of heavier $H^0$ to the lighter component $\phi$ will mainly dictate the relic density of DM components on top their annihilations to SM. 
\begin{figure}[htb!]
\centering
$$
 \includegraphics[height=5.0cm]{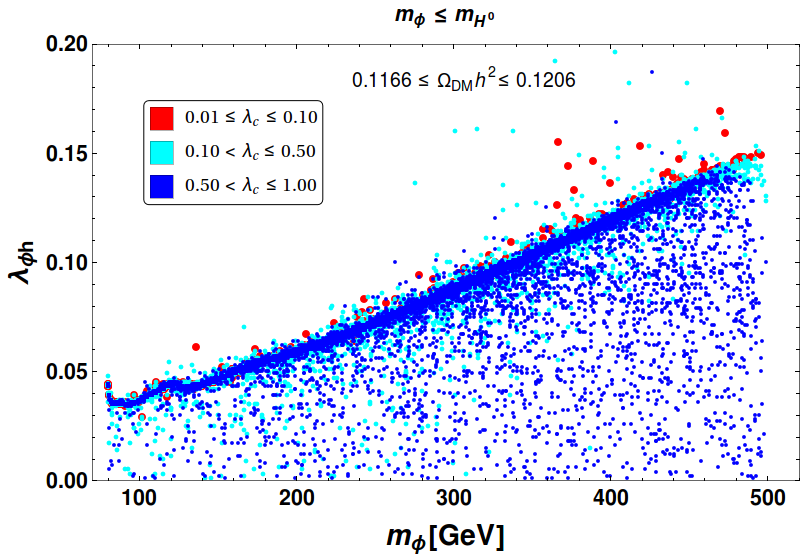} ~
 \includegraphics[height=5.0cm]{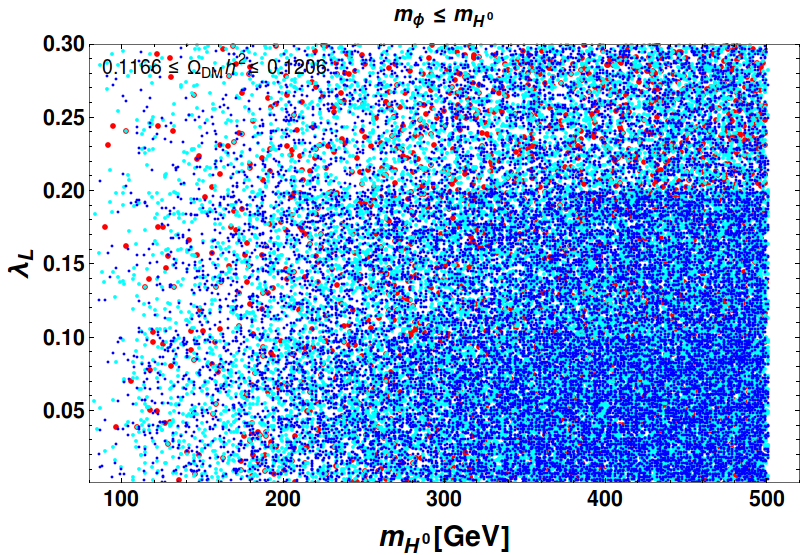}
 $$
 $$
 \includegraphics[height=5.0cm]{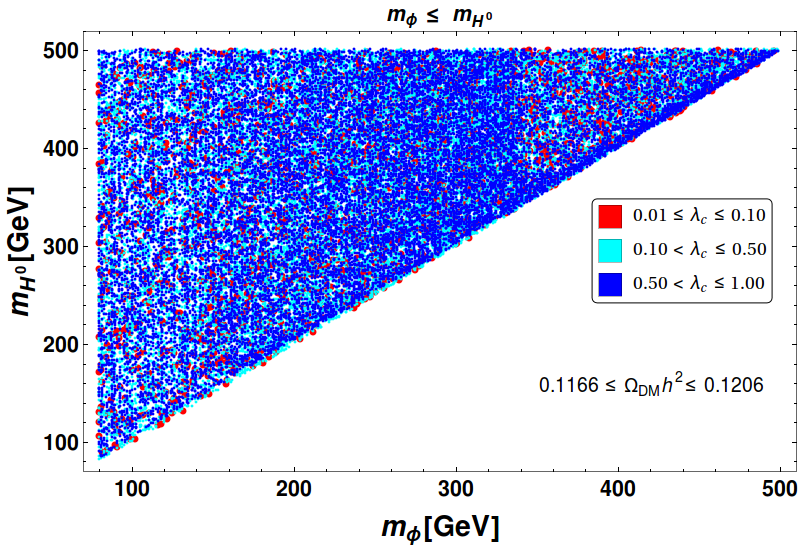} ~
 \includegraphics[height=5.0cm]{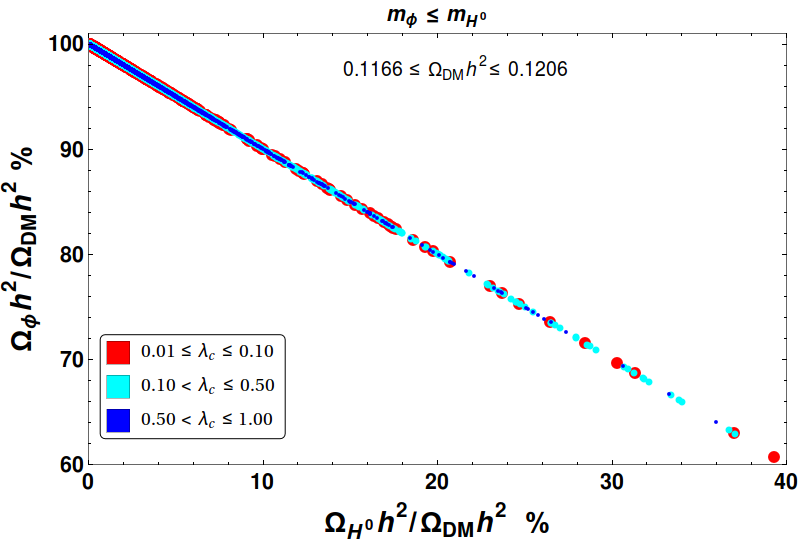} 
 $$
 \caption{Relic Density allowed parameter space is shown in $\mph-\lph$ plane (top left), $\mH-\ll$ plane (top right), 
 $\mph-\mH$ plane (bottom left) and $\Oph/\ODM (\%)-\OH/\ODM (\%)$ plane (bottom right) 
 for the mass hierarchy $\mph \leq \mH$.}
 \label{fig:lam_DM_mass_C}
 \end{figure}
 \begin{figure}[htb!]
\centering
$$
 \includegraphics[height=5.0cm]{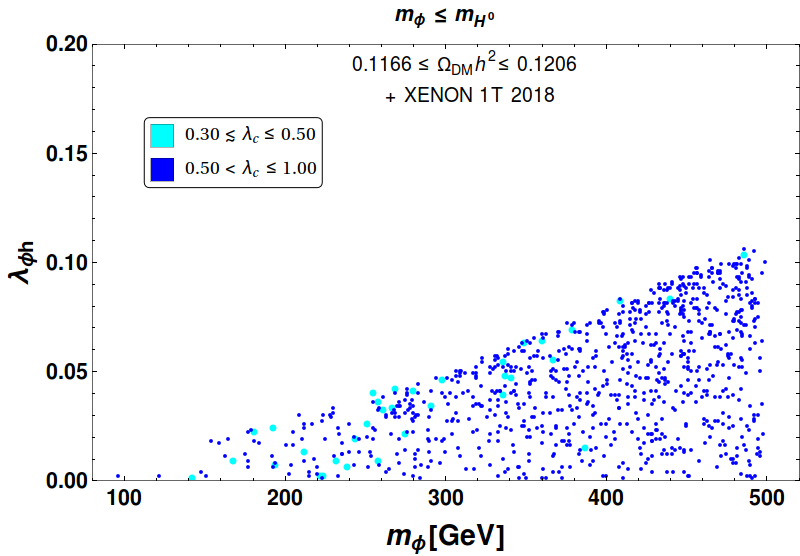} ~
 \includegraphics[height=5.0cm]{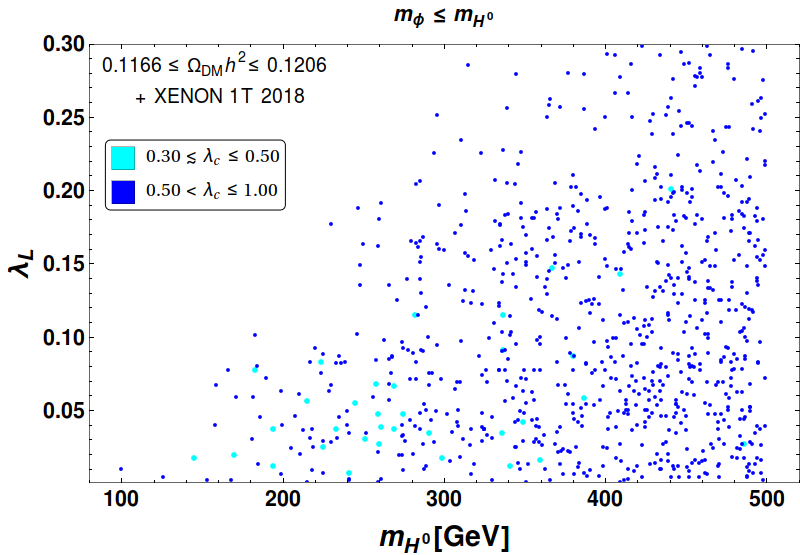}
 $$
 $$
 \includegraphics[height=5.0cm]{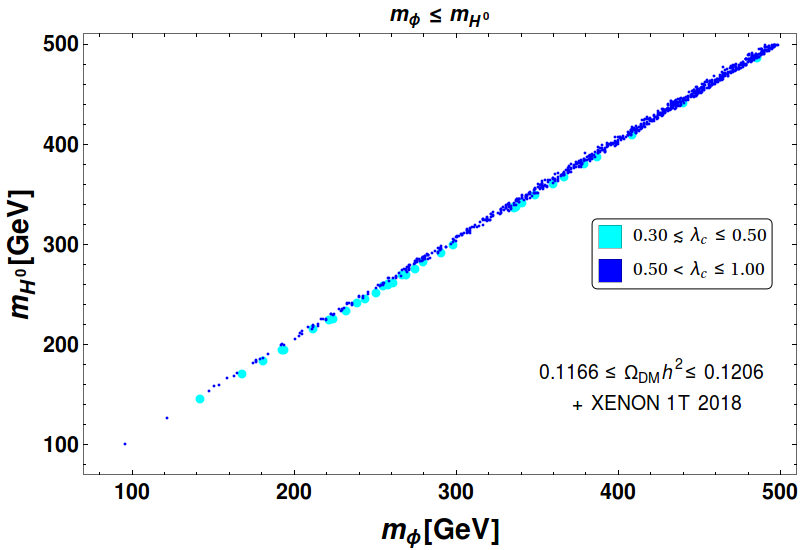} 
 $$
 \caption{Relic Density and direct detection (XENON 1T 2018) at NLO allowed parameter space is shown in: 
 $\mph-\lph$ plane (top left panel), $\mH-\ll$ plane (top right panel) and $\mph-\mH$ plane (bottom panel). The scans are performed for for the mass hierarchy $\mph \leq \mH$.}
 \label{fig:lam_DM_mass_D}
\end{figure}
 Relic density allowed parameter space for $\mph \leq \mH$ is shown in Fig.~\ref{fig:lam_DM_mass_C}. Again, this is illustrated in $\mph-\lph$ plane (top left), $\mH-\ll$ plane (top right), 
 $\mph-\mH$ plane (bottom left) and $\Oph/\ODM (\%)-\OH/\ODM (\%)$ plane (bottom right). Different ranges of $\lc$ are shown by the same colour code as in 
 Fig.~\ref{fig:lam_DM_mass}, \ref{fig:lam_DM_mass_A}. Let us first focus on the top left figure. It shows that for small values of $\lc$, relic density allowed 
 parameter space points lie in the vicinity of single component framework of $\phi$ (red points in figure). In absence of a lighter mode, the relic density of $\phi$ is essentially 
 governed by its annihilation to SM and due to small conversion cross-section the production of $\phi$ is also not large enough to change the conclusion. 
 However, the situation changes significantly with larger $\lc$ (cyan and blue points), where we see again that the overabundant region of the single component scenario
 is getting allowed by relic density. In order to understand this let us remind ourself of the CBEQ for $\mph<\mH$ as depicted in Eqn.~\ref{CBEQ-simple-2}. In particular, 
 the number density of $\phi$ is dictated by $\dot{n_\phi}+3Hn_\phi \simeq -\mathcal{F}_\phi+\mathcal{F}_{\Phi\phi}$. With larger $\lc$ and larger conversion, $\mathcal{F}_{\Phi\phi}$ increases
 to reduce the effective $\mathcal{F}_\phi$ that determines the relic of $\phi$. Therefore, to keep the relic density of $\phi$ to correct order, $\mathcal{F}_\phi$ has to increase. 
 Now recall, $\mathcal{F}_\phi=\langle \sigma v\rangle_{\phi\phi \to SM~SM}(n_\phi)^2 \sim 1/\langle \sigma v\rangle_{\phi\phi \to SM~SM}$ as 
 $n_\phi \sim 1/\langle \sigma v\rangle_{\phi\phi \to SM~SM}$. Therefore, to increase $\mathcal{F}_\phi$, one has to reduce the annihilation cross-section 
 $\langle \sigma v\rangle_{\phi\phi \to SM~SM}$. This is possible by reducing $\lph$ as we see here in the plot. Next let us discuss the top right figure. 
 This figure in $\mH-\ll$ plane essentially depicts that with larger $\lc$, larger $\mH$ is favoured to tame the DM conversion as well as annihilation cross-section
 to keep the relic within limit. The dependence however is not that much significant due to the presence of large number of co-annihilation channels which remain unaffected by $\lc$.  
 In the bottom left panel, mass correlation has been plotted and carries no information. Lastly, bottom right figure shows the relative relic density contributions of these 
 two components. It is well understood that an additional channel for annihilation of $H^0$ only reduces the possibility of bringing $\OH$ in the correct ballpark due to 
 already existing gauge mediated annihilation and co-annihilation channels. Therefore, for small $\lc$, it is still possible to get a contribution from $\OH \sim 40\%$, but 
 that becomes harder with large $\lc$, where the relic density contribution of $H^0$ is further limited to $\OH \sim 20\%$.
 
Direct search constraints from XENON 1T 2018 on the relic density allowed points are shown in Fig.~\ref{fig:lam_DM_mass_D}. 
To emphasise again, the demand of these plots are simultaneous satisfaction of XENON1T limit for both DM components. 
The main outcome of this plot is to see the absence of small $\lc$ points (red dots) upto $\lc \sim 0.30$. This is simply due to the fact that, with small $\lc$, the required $\lph$ is high enough 
for $\phi$ DM to be discarded by XENON1T data. The other important feature is that with larger $\lc$, larger DM masses are favoured. Lastly a very important conclusion comes from 
the bottom panel figure in the mass correlation plot. This shows, as only high $\lc$ region is allowed, the conversion of $H^0$ to $\phi$ still needs to 
be restricted and therefore the mass difference between the two DM components ($\mH-\mph$) has to be very very small. These features are all distinct from that of $\mph > \mH$ region. 

So far our discussion has been focused on DM mass region $m_{W^\pm} \leq m_{H^0}, m_{\phi} \leq 500$ GeV. But if $m_{H^0} < m_{W^\pm} ~\textrm{or}~ m_{\phi} < m_{W^\pm}$, while other DM mass is heavier than 
$m_{W^\pm}$, the only region available for lighter DM with mass $< m_{W^\pm}$ are the Higgs and $Z$ resonance regions: $m_{H^0}\sim~\frac{m_Z}{2},~\frac{m_h}{2}$ and $m_{\phi} \sim~\frac{m_h}{2}$. 
It is important to remind that the resonance regions are already available in absence of second DM component and therefore brings no new phenomenological outcome.

\section{Indirect searches}
\label{sec:indirect}
It is possible to probe DM signals in the context of different indirect detection 
experiments as well. The experiments look for astrophysical sources of SM particles produced
through DM annihilations or via DM decays. Amongst these final
states, photon and neutrinos, being neutral and stable can reach indirect detection
probes without getting affected much by intermediate regions. For WIMP
type DM, emitted photons lie in the gamma ray
regime that can be measured at space-based telescopes like the Fermi-LAT or ground-based
telescopes like MAGIC \cite{Ahnen:2016qkx}. The photon flux
in a specific energy range is written as 
\begin{align}
 \Phi_F=\frac{1}{4\pi}\frac{\langle\sigma v\rangle_{\rm ann}}{2m_{DM}^2}\int_{E_{\rm min}}^{E_{\rm max}}\frac{dN_\gamma}{dE_\gamma}dE_\gamma \int_{\rm LOS} dx \rho^2(r(b,l,x)),
 \label{eq:InPhoton}
\end{align}
where $r(b,l,x)$ is the distance of the DM halos from the galactic center. Galactic coordinates are defined by $b,l$ and 
$\rho(r)$ represents the DM density profile. Eq.(\ref{eq:InPhoton}) is further rewritten as
\begin{align}
 \Phi_F=\frac{1}{4\pi}\frac{\langle\sigma v\rangle_{\rm ann}}{2m_{DM}^2}\int_{E_{\rm min}}^{E_{\rm max}}\frac{dN_\gamma}{dE_\gamma}dE_\gamma\times J,
 \label{eq:PFlux}
\end{align}
where $J=\int dx \rho^2(r(b,l,x))$ is conventionally known as $J-$factor and LOS means line of sight. Now 
from the observed Gamma ray flux produced due to DM annihilations,  one can constrain the DM annihilation 
into different charged final states like $\mu^+\mu^-$, $\tau^+\tau^-$, $W^+,W^-$ and $b^+b^-$. It is pertinent 
to mention here that along with the continuous gamma ray spectrum, gamma ray lines originated from IDM annihilations are also of special interest \cite{Gustafsson:2007pc}. 
It has been shown in \cite{Garcia-Cely:2015khw} that such a feature, characteristic of inert doublet dark matter (IDM) model, can indeed be important for DM mass above TeV where 
Sommerfeld enhancement becomes relevant. Since we mostly focus on the intermediate mass range of IDM in this study ($\sim$ 80~-~500 GeV), aforementioned constraint is extremely subdued.

In the multicomponent dark matter set up as we have here, it is expected that the $\gamma$ ray flux will be the sum of the 
contributions from both $\phi$ and $H^0$. Then we can write the total $\gamma$ ray flux can be written as:
\begin{align}
 \Phi_F^T=\Phi_F^\phi+\Phi_F^{H^0},
\end{align}
where $\Phi_F^\phi$ and $\Phi_F^{H^0}$ are the individual contribution to $\Phi_F^T$ from $\phi$ and $H^0$. Following Eq. (\ref{eq:PFlux}), 
it would be natural to write
\begin{align}
 \frac{\langle\sigma v\rangle_{XX}^{\rm eff}}{m_{\rm R }^2}\rho_T^2=\frac{\langle\sigma v\rangle_{\phi\phi\rightarrow XX}}{m_\phi^2}\rho_\phi^2+\frac{\langle\sigma v\rangle_{H^0H^0\rightarrow XX}}{m_{H^0}^2}\rho_{H^0}^2,
 \label{eq:SigmaV1}
\end{align}
where $X,X$ denote the annihilation products and $m_R$ indicates some effective mass scale. One can further write the effective annihilation 
cross-section in Eq. (\ref{eq:SigmaV1}) considering $m_{\rm R}$ to be the reduced mass of the two component system as
\begin{align}
 \langle\sigma v\rangle_{XX}^{\rm eff}=\left(\frac{\Omega_{H^0}}{\Omega_T}\right)^2\frac{m_{\phi}^2}{(m_\phi+m_{H^0})^2}\langle\sigma v\rangle_{H^0 H^0\rightarrow X X}+\left(\frac{\Omega_{\phi}}{\Omega_T}\right)^2\frac{m_{H^0}^2}{(m_\phi+m_{H^0})^2}\langle\sigma v\rangle_{\phi\phi\rightarrow XX} .
\end{align}

\begin{figure}[htb!]
\centering
$$
 \includegraphics[height=4.7cm]{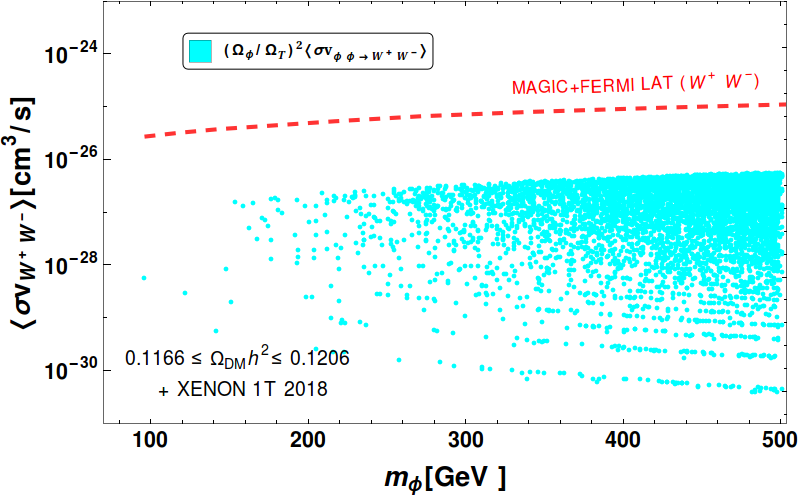} ~
 \includegraphics[height=4.7cm]{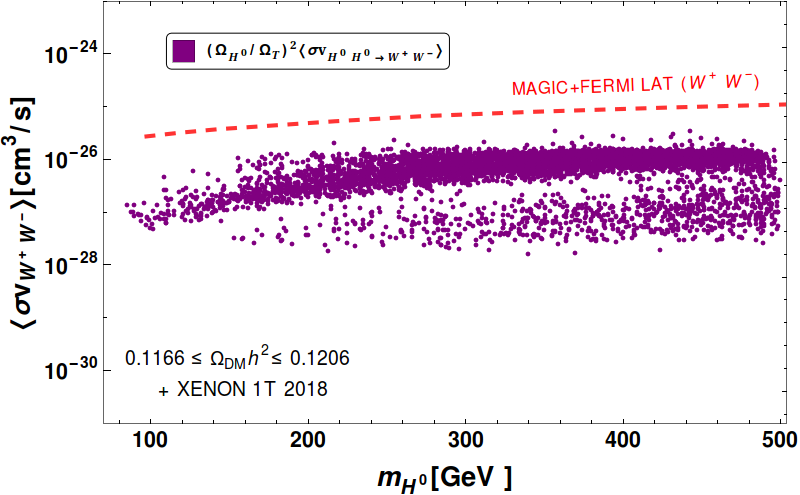}
 $$
 \caption{DM annihilations $\phi\phi \rightarrow W^{+} W^{-}$ (left) and $H^0 H^0 \rightarrow W^{+} W^{-}$ (right) normalised by 
 branching fraction of individual relic densities compared against the latest
indirect detection bounds of Fermi-LAT \cite{Ahnen:2016qkx}.}
 \label{fig:Ind12}
\end{figure}

\begin{figure}[htb!]
\centering
$$
 \includegraphics[height=7cm,width=10cm]{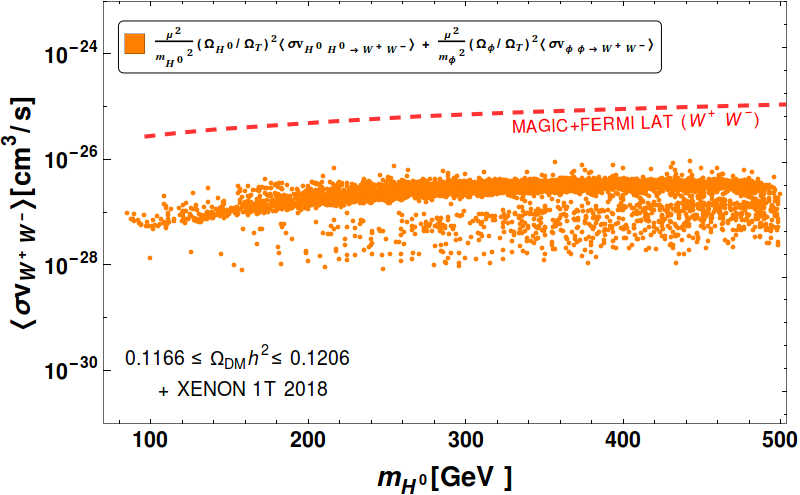} ~
 $$
  \caption{Effective DM annihilations to $ W^{+} W^{-} $ 
  as defined in Eq.(\ref{eq:SigmaV1}) in the two-component model for relic density and direct search allowed points compared against 
  the latest indirect detection bounds of Fermi-LAT \cite{Ahnen:2016qkx}.}
 \label{fig:Indeff}
\end{figure}

It turns out that for real singlet scalar dark matter indirect searches does not pose strong bound  
\cite{Cline:2013gha,Eiteneuer:2017hoh}; while for single component IDM candidate, the indirect search rules out the region of $m_{H^0}<400$ GeV due to
 large thermal average cross section of $H^0 H^0\rightarrow W^{+}W^{-}$ channel 
 \cite{Eiteneuer:2017hoh,Garcia-Cely:2015khw,Queiroz:2015utg,Garcia-Cely:2013zga}. Hence in our proposed two component DM set up
 we shall mostly focus on the $\langle\sigma v\rangle_{\rm{DM},\rm{DM} \rightarrow W^{+}W^{-}}$ and compare it with the indirect search experimental 
 bound from Magcic+Fermi Lat bound \cite{Ahnen:2016qkx}.
 In left panel of Fig. \ref{fig:Ind12}, we show the contribution of $\phi$ DM in $\langle\sigma v\rangle^{\rm eff}_{W^{+}W^{-}}$ for all relic and 
 SI DD cross section satisfied points and compare it with the limit coming from Fermi Lat experiment \cite{Ahnen:2016qkx}. 
 Similar comparison is drawn for IDM ($H^0$) in right panel of Fig. \ref{fig:Ind12}. Finally in Fig. \ref{fig:Indeff}, we show the order of magnitude
of total $\langle\sigma v\rangle^{\rm eff}_{W^{+}W^{-}}$ for all relic density and SI DD (at NLO) satisfied points 
and find it to lie well below the upper limit imposed by Fermi Lat experiment. The relative abundance of individual relic densities and 
ratio of reduced mass to the DM mass helps to evade the indirect search bound on $\langle \sigma v\rangle_{\rm{DM},\rm{DM}\rightarrow W^{+}W^{-}}$ 
successfully. We have also confirmed that the bounds on $\langle\sigma v\rangle^{\rm eff}$ for $b\bar{b}$, $\mu^{+}\mu^{-}$, $\tau^{+}\tau^{-}$ 
and other annihilation final products from Magic+Fermi Lat experiment \cite{Ahnen:2016qkx} are easily satisfied in our set up.

\section{Electroweak Vacuum stability and High Scale Perturbativity in presence of RH neutrino and DM}
\label{sec:vs}
One of the motivations of this work is to show that the presence of right handed neutrinos to 
yield correct neutrino masses in presence of multipartite DM. Although the neutrino sector considered here seems 
decoupled from the dark sector, is not completely true. The effect of the RH neutrinos alter the allowed DM parameter 
space when the model is validated at high scale. As already stated before, we employ type-I seesaw mechanism 
to generate the light neutrino mass, for which three RH neutrinos are included in the set up. 
\subsection{Neutrino Yukawa couplings}
We first describe the strategy in order to study their impact on RG evolution. For simplicity, the RH neutrino mass 
matrix $M_N$ is considered to be diagonal with degenerate entries, {\it{i.e.}} $M_{N_{i=1,2,3}}=M_R$. It is 
to be noted that in the RG equation, $\textrm{Tr}[Y_\nu^\dagger Y_\nu]$ enters. In order to extract the information on $Y_\nu$, we use the type-I seesaw formula for neutrino mass $m_\nu=Y_{\nu}^T Y_{\nu} \frac{v^2}{2M_R}$. 
Then, naively one would expect that large Yukawa couplings are possible with even heavier RH neutrino masses. 
For example with $M_R \sim 10^{14}$ GeV, $Y_{\nu}$ comes out to be $0.3$ in order to obtain $m_\nu\simeq 0.05$ eV. 
However, contrary to this, it can be shown that even with smaller $M_R$ one can achieve large values of $\textrm{Tr}[Y_\nu^\dagger Y_\nu]$, 
but effectively keeping $Y_{\nu}^T Y_{\nu}$ small, using a special flavor structure of $Y_{\nu}$ through Casas-Ibarra parametrization \cite{Rodejohann:2012px}. 
Note that our aim is to study the maximum effect coming from the right handed neutrino Yukawa, $i.e.$ with 
large $\textrm{Tr}[Y_\nu^\dagger Y_\nu]$, on EW vacuum stability. 
For this purpose, we use the parametrisation by \cite{Casas:2001sr} and write $Y_\nu$ as 
\begin{align}\label{casas}
    Y_\nu=\sqrt{2}\frac{\sqrt{M_R}}{v}  \mathcal{R}\sqrt{m_\nu^d} ~U^\dagger_{\textrm{PMNS}},
    \end{align}
where $m_\nu^d$ is the diagonal light neutrino mass matrix and $U_{\textrm{PMNS}}$ is the 
unitary matrix diagonalizing the neutrino mass matrix $m_{\nu}$ such that 
$m_{\nu} = U^*_{\textrm{PMNS}} m_\nu^d U^\dagger_{\textrm{PMNS}}$. Here  
$\mathcal{R}$ represents a complex orthogonal matrix which can be written as 
$\mathcal{R}=\mathcal{O}\textrm{exp}(i\mathcal{A})$ with $\mathcal{O}$ as real orthogonal matrix and $\mathcal{A}$
as real antisymmetric matrices. Using above parametrisation, then one gets,
 \begin{align}\label{trCa}     
\textrm{ Tr}[Y_\nu^\dagger Y_\nu] =  \frac{2M_R}{v^2}\textrm{Tr}\Big[\sqrt{m_\nu^d}e^{2i\mathcal{A}}\sqrt{m_\nu^d}\Big] . 
\end{align} 
Note that the real antisymmetric matrix $\mathcal{A}$ however does not appear in the seesaw expression for neutrino mass as $m_{\nu}=\frac{Y_\nu^TY_\nu v^2}{2M_R}$.
 Therefore with any suitable choice of $\mathcal{A}$, it would actually be possible to have relatively large Yukawa even with light $M_R$. 
\subsection{$\beta$ functions and RG running}  
To study the high scale validity of this multi-component DM model with neutrino mass (including perturbativity and vacuum stability), we need to consider the RG running of the associated couplings. The framework contains few additional mass scales: one extra 
scalar singlet, one inert doublet and three RH neutrinos with mass $M_{N_{i=1,2,3}}(= M_R)$. Hence in the study of RG running, different couplings will enter into different mass scales.
In DM phenomenology, we have considered the physical masses of DM sector particles within $\sim$ 500 GeV. 
Then for simplicity, we can safely ignore the small differences between the dark sector particle masses and identify the 
single additional mass scale as $m_{DM}$. On the other hand, RH neutrinos are considered to be heavier than the 
scalars present in the model ($M_{R}>m_{\rm{DM}}$). Hence, for running energy scale $\mu>m_{\rm{DM}}$, we need to consider the couplings associated to DM sector in addition to SM while while for $\mu>M_{N_{i=1,2,3}}$, the neutrino couplings will 
additionally enter into the scenario. Below we provide the one loop $\beta$ functions for the additional 
for the couplings involved in our model, when $\mu > M_{R}$.\\

\noindent{\bf $\beta$ functions of gauge couplings \cite{Branco:2011iw}: }
 \begin{align} 
\beta_{g_1} & =  
\frac{21}{5} g_{1}^{3}, \nonumber \\ 
\beta_{g_2} & =  -3 g_{2}^{3}, \nonumber \\ 
\beta_{g_3} & =  
-7 g_{3}^{3}. \nonumber \\
\end{align}
\noindent{\bf $\beta$ functions of Yukawa couplings \cite{Branco:2011iw,Pirogov:1998tj}:}
\begin{align} 
&\beta_{y_t} =  
\frac{3}{2} y_t^3 + y_t \Big( 3 y_t^2  -8 g_{3}^{2}  -\frac{17}{20} g_{1}^{2}  -\frac{9}{4} g_{2}^{2} + y_t \textrm{Tr}[Y_\nu^\dagger Y_\nu]\Big), \nonumber \\
&\beta_{\textrm{Tr}[Y_\nu^\dagger Y_\nu]} = 3 \textrm{Tr}[(Y_\nu^\dagger Y_\nu)^2]+\textrm{Tr}[Y_\nu^\dagger Y_\nu]\Big(-\frac{9}{10}g_1^2-\frac{9}{2}g_2^2 + 6 y_t^2 + 2 \textrm{Tr}[Y_\nu^\dagger Y_\nu]\Big).
\end{align}

\noindent{\bf $\beta$ functions of quartic scalar couplings \cite{Branco:2011iw}:}
 \begin{align} \label{eq:RGScalars}
\beta_{\lambda_H} & = \frac{27}{200} g_{1}^{4} +\frac{9}{20} g_{1}^{2} g_{2}^{2} +\frac{9}{8} g_{2}^{4} -\frac{9}{5} g_{1}^{2} \lambda_H -9 g_{2}^{2} \lambda_H +24 \lambda_{H}^{2}  +12 \lambda_H y_t^2 -6 y_t^4\nonumber \\ 
 & ~~~~~+2 \lambda_{1}^{2} +2 \lambda_{1} \lambda_{2} +\lambda_{2}^{2}+\lambda_{3}^{2} + \frac{1}{2}\lph^2 +4 \lambda_H {\rm  Tr}[Y_\nu^\dagger Y_\nu]-2 {\rm Tr}[(Y_\nu^\dagger Y_\nu)^2] \\
\beta_{\lambda_{3}} & =  
-\frac{9}{5} g_{1}^{2} \lambda_{3} -9 g_{2}^{2} \lambda_{3} +8 \lambda_{1} \lambda_{3} +12 \lambda_{2} \lambda_{3} +4 \lambda_{3} \lambda_H +4 \lambda_{3} \lambda_{\Phi} +6 \lambda_{3} y_t^2+ 2 \lambda_{3} \textrm{Tr}[Y_\nu^\dagger Y_\nu], 
\nonumber \\ 
\beta_{\lambda_{2}} & =  
+\frac{9}{5} g_{1}^{2} g_{2}^{2} -\frac{9}{5} g_{1}^{2} \lambda_{2} -9 g_{2}^{2} \lambda_{2} +8 \lambda_{1} \lambda_{2} +4 \lambda_{2}^{2} +8 \lambda_{3}^{2} +4 \lambda_{2} \lambda_H +4 \lambda_{2} \lambda_{\Phi}  +6 \lambda_{2} y_t^2 + 2 \lambda_{2} {\rm Tr}[Y_\nu^\dagger Y_\nu], \nonumber\\ 
\beta_{\lambda_{1}} & =  
\frac{27}{100} g_{1}^{4} -\frac{9}{10} g_{1}^{2} g_{2}^{2} +\frac{9}{4} g_{2}^{4} -\frac{9}{5} g_{1}^{2} \lambda_{1} -9 g_{2}^{2} \lambda_{1} +4 \lambda_{1}^{2} +2 \lambda_{2}^{2} +2 \lambda_{3}^{2} +12 \lambda_{1} \lambda_H  \nonumber \\
& ~~~+4 \lambda_{2} \lambda_H +12 \lambda_{1} \lambda_{\Phi} + 4 \lambda_{2} \lambda_{\Phi} + 6 \lambda_{1} y_t^2 + \lc \lph+2 \lambda_{1}{\rm Tr}[Y_\nu^\dagger Y_\nu], \nonumber\\ 
\beta_{\lambda_{\Phi}} & =  
24 \lambda_{\Phi}^{2}  + 2 \lambda_{1}^{2}  + 2 \lambda_{1} \lambda_{2}
-9 g_{2}^{2} \lambda_{\Phi}  + \frac{27}{200} g_{1}^{4}  + \frac{9}{20} g_{1}^{2} \Big(-4 \lambda_{\Phi}
+ g_{2}^{2}\Big) + \frac{9}{8} g_{2}^{4}  + \lambda_{2}^{2} + \lambda_{3}^{2}+\frac{1}{2}\lambda_c^2
, \nonumber \\
\beta_{\lph} & =  
-\frac{9}{10} g_{1}^{2} \lph -\frac{9}{2} g_{2}^{2} \lph +4 \lph^{2}+12 \lph \lambda_H +\lph \lambda_{\phi}+6 \lph y_t^2 \nonumber \\
&~~~ +4 \lambda_1 \lc + 2 \lambda_2 \lc + 2 \lph Tr[Y_\nu^\dagger Y_\nu] , \nonumber \\ 
\beta_{\lc} & = -\frac{9}{10} g_{1}^{2} \lc -\frac{9}{2} g_{2}^{2} \lc + 12 \lc \lambda_\Phi + 4\lambda_1 \lph+ 2\lambda_2 \lph + \lc \lambda_\Phi + 4\lc^2, \nonumber  \\
 \beta_{\lambda_{\phi}} & =  
3 \Big(4 \lc^{2}  + 4 \lph^{2}  + \lambda_{\phi}^{2}\Big). 
\end{align}

The above $\beta$ functions are generated using the model implementation in the code {\tt SARAH} \cite{Staub:2013tta}.
The running of $\lambda_H$ as in Eqn.(\ref{eq:RGScalars}) is influenced adversely mostly by top Yukawa coupling $y_t\sim \mathcal{O}(1)$ 
and right handed neutrino Yukawa coupling as ${\rm Tr}[Y_\nu^\dagger Y_\nu]$. 
On the other hand, multiparticle scalar DM couplings present in the model help in compensating the strong negative 
effect from $y_t$ and ${\rm Tr}[Y_\nu^\dagger Y_\nu]$. To  evaluate $\textrm{Tr}[Y_\nu^\dagger Y_\nu]$, we 
employ Eq. \ref{trCa}. As an example, let us consider magnitude of all the entries of  $\mathcal{A}$ to be equal, say 
$a$ with all diagonal entries to be zero. Then using the best fit values of neutrino mixing angles and cosnidering 
the mass of lightest neutrino zero, one can find for$M_R$ = 1 TeV, $\textrm{Tr}[Y_\nu^\dagger Y_\nu]$ can be as 
large as 1 with  $a =8.1 $\cite{Casas:2001sr,Guo:2006qa}. 
Equipped with all these $\beta$ functions and strategy to estimate the relevant couplings present in them, let us now 
analyse the SM Higgs vacuum stability at high energy scale. Below we provide the stability and metastability criteria.

\vspace{2mm}
$\bullet$ \textbf{Stability criteria:} The stability of Higgs vacuum can be 
ensured by the condition $\lambda_H>0$ at any scale. 
However, we have multiple scalars (SM Higgs doublet, one inert doublet and one gauge real singlet) in the model. 
Therefore we should ensure the boundedness or stability of the entire scalar potential in any field direction. 
This can be guaranteed by using the co-positivity criterion in Eqn.~\ref{eq:stability}. Note that once $\lambda_H$ becomes negative
the other copositivity conditions no longer remain valid.

\vspace{2mm}
$\bullet$ \textbf{Metastability criteria:} On the other hand, when $\lambda_H$ becomes negative, there may exist
another deeper minimum other than the EW one. Then the estimate of the
tunnelling probability $\mathcal{P}_T$ of the EW vacuum to the second minimum is essential 
to confirm the metastability of the Higgs vacuum. Obviously, the Universe can be in a metastable
state only, provided the decay time of the EW vacuum is longer than the age of the Universe. The tunnelling
probability is given by \cite{Buttazzo:2013uya,Isidori:2001bm},
\begin{align}
 \mathcal{P}_T=T_U^4\mu_B^4e^{-\frac{8\pi^2}{3|\lambda_H(\mu_B)|}},
\end{align}
where $T_U$ is the age of the Universe, $\mu_B$ is the scale at which tunnelling probability is maximized, determined
from $\beta_{\lambda_H} = 0$. Therefore, solving above equation, we see that metastable Universe requires \cite{Buttazzo:2013uya,Isidori:2001bm} :
\begin{align}
 \lambda_H(\mu_B) ~>~ \frac{-0.065}{1-\textrm{ln}\Big(\frac{v}{\mu_B}\Big)}.
\label{eq:Met}
\end{align}
As noted in \cite{Buttazzo:2013uya}, for $\mu_B > M_P$ (Planck Scale, $M_P = 1.22 \times 10^{19}$ GeV), one can safely consider $\lambda_H(\mu_B)=\lambda_H(M_P)$. One should
also note, that even with meatstability of Higgs vacuum, the instability in other field direction may also occur. The conditions
to avoid the possible instability along various field directions for $\lambda_H<0$ are listed below \cite{Khan:2015ipa}:

\vspace{1mm}
$(i)$ $\lambda_\Phi>0$ to avoid the unboundedness of the potential along $H^0$, $A^0$ and $H^\pm$ directions,

\vspace{1mm}
$(ii)$ $\lambda_1>0$ to ensure the stability along a direction between $H^\pm$ and $h$,

\vspace{1mm}
$(iii)$ $\lambda_L>0$ to ensure the stability along a direction between $H^0$ and $h$,

\vspace{1mm}
$(iv)$ $\lambda_S>0$, to avoid unboundedness of potential along a direction between $A^0$ and $h$,

\vspace{1mm}
$(v)$ $\lambda_\phi>0$, otherwise the potential will be unbounded along $\phi$ direction,

\vspace{1mm}
$(vi)$ $\lambda_{\phi h}>0$, to ensure the stability along a direction between $\phi$ and $h$.

\vspace{2mm}

Now to begin the vacuum stability analysis in the present multi-component DM scenario, we first choose two benchmark 
values of DM masses ($\phi$ and $H^0$) which satisfy both the relic density and direct detection bounds. These along 
with corresponding values of other relevant parameters are denoted by the set of two benchmark points, BP1 and BP2, 
as tabulated in Table \ref{tab1:BP}. These parameters are mentioned in the caption of Table \ref{tab1:BP} for both benchmark points. We also show the value of electroweak parameters and $\mu_{\gamma\gamma}$ for these two benchmark points in Table \ref{tab2:BP}.
\begin{table}[h]
\begin{center}
\resizebox{\textwidth}{!}{
\begin{tabular}{|c|c|c|c|c|c|c|c|c|c|}
\hline
BPs &$m_{H^0}$ & $m_{\phi}$  & $m_{A^0}$ & $m_{H^\pm}$& $\lambda_L$ & $\lph$ & $\lc$ & $\Omega_{H^0}h^2$ & $\Omega_{\phi}h^2$   \\
\hline
BP1       & 330  & 343  &  333 & 339 &  0.043         & 0.065      & 0.2  & 0.033 & 0.086   \\
\hline
BP2       & 427  & 449  &  438 &  440 & 0.086         & 0.017      & 0.3  & 0.027 & 0.088   \\
\hline
\end{tabular}
}
\end{center}
\caption{ Benchmark points used to analyse EW vacuum stability in our model. All masses are in GeVs. The other couplings used in these benchmark points 
play an important role; they are: BP1: $\{\lambda_1=0.285,\lambda_2=-0.17,\lambda_3=-0.033\}$, BP2: $\{\lambda_1=0.544,\lambda_2=-0.215,\lambda_3=-0.157\}$.}
\label{tab1:BP}
\end{table}
\begin{table}[h]
\begin{center}
\begin{tabular}{|c|c|c|c|}
\hline
BPs  & $\Delta S ~ (10^{-4})$ & $\Delta T~ (10^{-4})$ & $\mu_{\gamma\gamma}~ (10^{-5})$ \\
\hline
BP1  &-12 & 5.1  & 3  \\
\hline
BP2  &-9 & 2.5 & 3 \\
\hline
\end{tabular}
\end{center}
\caption{ Estimate of electroweak precision parameters and $\mu_{\gamma\gamma}$ for the two benchmark points 
as chosen in Table~\ref{tab1:BP}.}
\label{tab2:BP}
\end{table}
\begin{figure}[htb!]
$$
 \includegraphics[height=4.9cm]{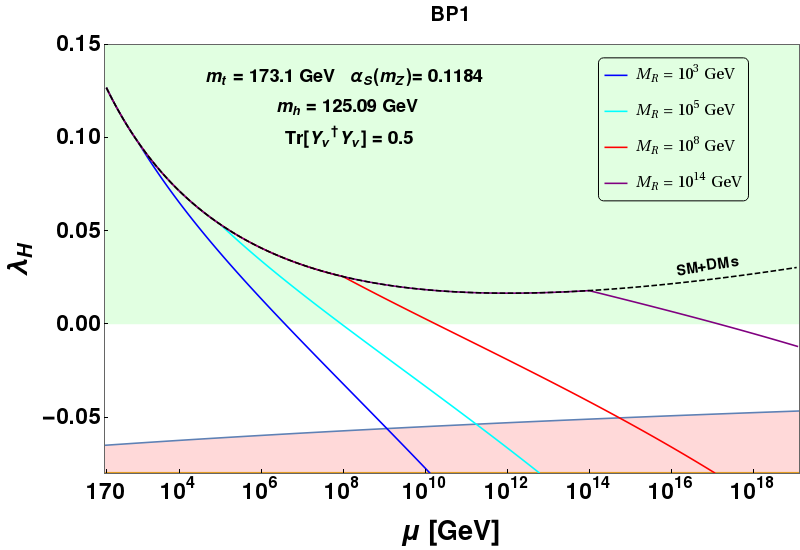}~~
 \includegraphics[height=4.9cm]{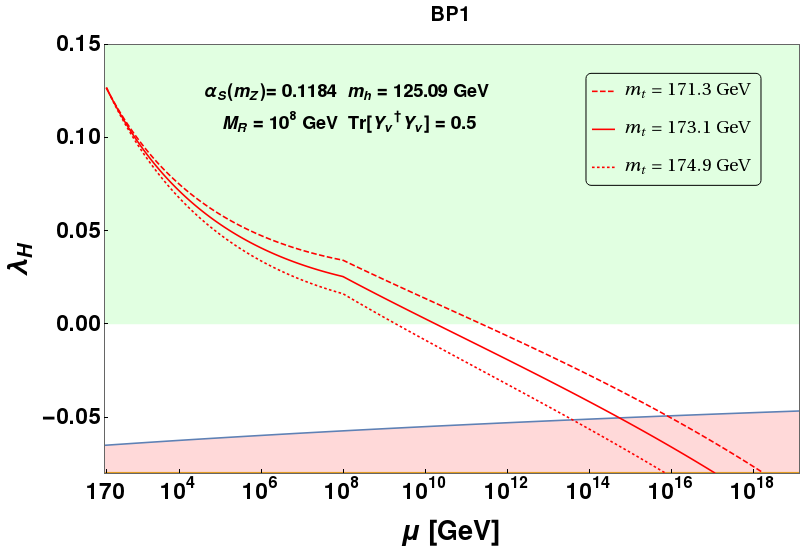}
 $$
 \caption{RG running of $\lambda_H$ with energy scale for BP1. In left panel, we have shown the effect of different right handed neutrinos masses, 
 $M_R$ (indicated by different colours and mentioned in figure inset) for a fixed top quark mass $m_t = 173.1$ GeV. The black dotted line corresponds to the case when right handed 
 neutrinos are absent in the scenario. In right panel, we choose a specific $M_R = 10^8$ GeV and consider top mass in $2\sigma$ range: $m_t = 173.1 \pm 0.9 $ GeV. 
 $\text{Tr}[Y_{\nu}^\dagger Y_{\nu}] = 0.5$ is kept constant in both plots.}
 \label{fig:EW_VS1}
\end{figure}
\begin{figure}[htb!]
$$
 \includegraphics[height=4.9cm]{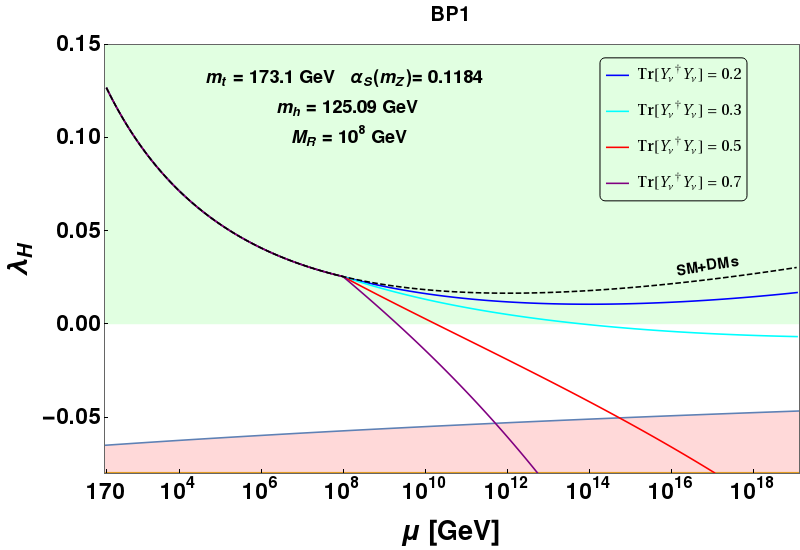}~~
 \includegraphics[height=4.9cm]{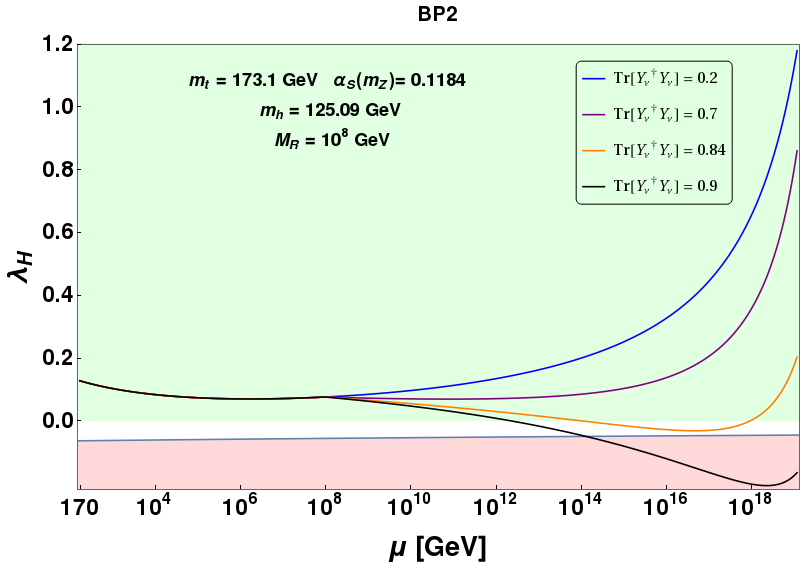}
 $$
 \caption{RG running of $\lambda_H$ with energy scale $\mu$ for different values of $\text{Tr}[Y_{\nu}^\dagger Y_{\nu}]$ (shown in different colours and values are mentioned in the figure inset) 
 for the benchmark points BP1 (left) and BP2 (right). Here we have chosen $M_R = 10^8$ GeV for illustration. 
 The black dotted line here corresponds to the case when right handed neutrinos are absent in this scenario.}
 \label{fig: EW_VS2}
\end{figure}
We run the two loops RG equations for all the SM couplings and
one loop RG equations for the other relevant couplings in the model from $\mu=m_t$ to $M_P$ energy scale. We use the inital boundary values of all the SM couplings as provided in \cite{Buttazzo:2013uya}. The boundary values have been
evaluated at $\mu=m_t$ in \cite{Buttazzo:2013uya} by taking various threshold corrections  and mismatch between top pole mass
and $\overline{MS}$ renormalised couplings into account. 

In Fig. \ref{fig:EW_VS1}, we show the running of $\lambda_H$ for BP1 as a function of energy scale $\mu$. The left panel
shows running of $\lambda_H$ for different values of RH neutrino mass $M_R$, considering top quark mass $m_t =173.1$ GeV, Higgs mass $m_h=125.09$ GeV and 
Tr$[Y_\nu^\dagger Y_\nu]=0.5$. As it is visible that for low value of $M_R\sim 10^4$ GeV, $\lambda_H$ enters into unstable region at
very early stage of its evolution (blue line in the figure). This has happened as the scalar couplings in 
$\beta_{\lambda_H}$ (Eqn. \ref{eq:RGScalars}) are not sufficiently large to
counter the strong negative impact of Tr$[Y_\nu^\dagger Y_\nu]$ which brings down the $\lambda_H$ shraply towards unstable region. On contrary, for large value of $M_R\sim 10^{14}$ GeV, the effect of $Y_\nu$ in $\beta_{\lambda_H}$ starts at a comparatively larger energy scale than the earlier case. As a consequence, although $\lambda_H$ becomes negative at some high energy scale,
it stays in metastable region till $M_P$ energy scale (violate line). Green region here describes stable, white region describes metastable (see Eqn.~\ref{eq:Met}) and the red region 
indicates unstable solution for the potential. For comparison, we also display the evolution of $\lambda_H$ (black dotted curve)
in absence of RH neutrinos in the theory with the inclusion of scalar couplings (related to DM) only. This clearly shows that in absence of RH neutrinos, EW vacuum could be absolutely stable till $M_P$ energy scale.
In right panel of Fig. \ref{fig:EW_VS1}, we study the evolution of $\lambda_H$ considering $2\sigma$
uncertainty of measured top mass $m_t$, keeping $m_h=125.09$ GeV, $M_R=10^8$ GeV and Tr$[Y_\nu^\dagger Y_\nu]=0.5$ fixed. It is trivial to find that
with the increase of top mass, $\lambda_H$ crosses zero at earlier stage in its evolution. \\
\begin{figure}[htb!]
$$
 \includegraphics[height=4.9cm]{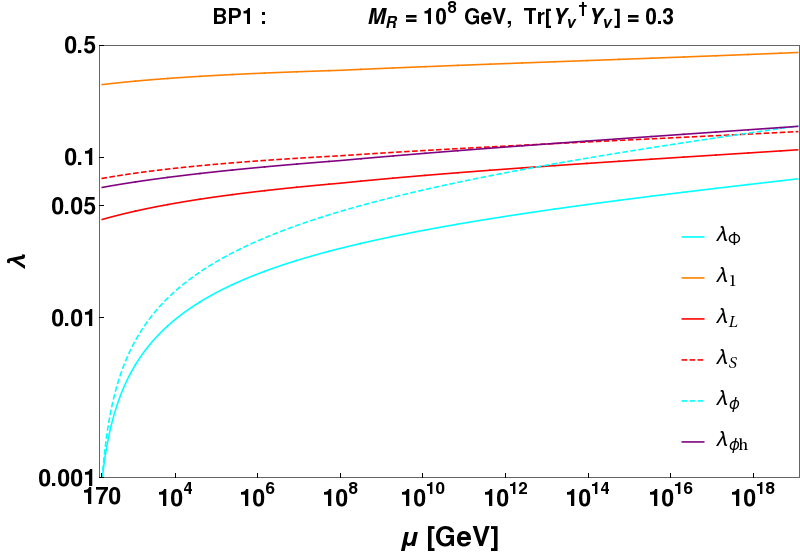}~~
 \includegraphics[height=4.9cm]{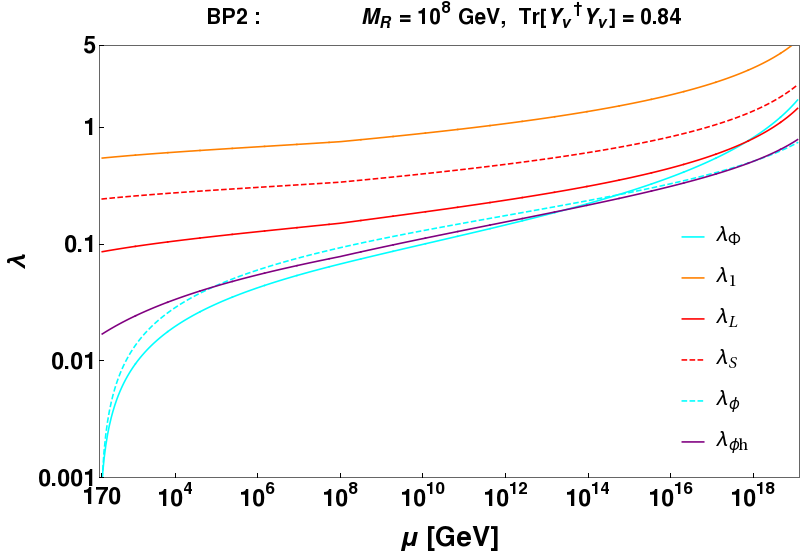}
 $$
 \caption{RG running of all the quartic couplings in metastability criteria for BP1 (left) and BP2 (right) to ensure the boundedness
 of the scalar potential in various field directions with energy scale $\mu$ for $\text{Tr}[Y_{\nu}^\dagger Y_{\nu}]=0.3$ (left)
 and 0.84 (right). The choices of $\text{Tr}[Y_{\nu}^\dagger Y_{\nu}]$ are demonstrated in cyan (0.3) and in orange (0.84) in left and right panels of Fig. \ref{fig: EW_VS2} to yield metastability.
}
 \label{fig: EW_VS2_met}
\end{figure}
\begin{figure}[htb!]
$$
 \includegraphics[height=4.8cm]{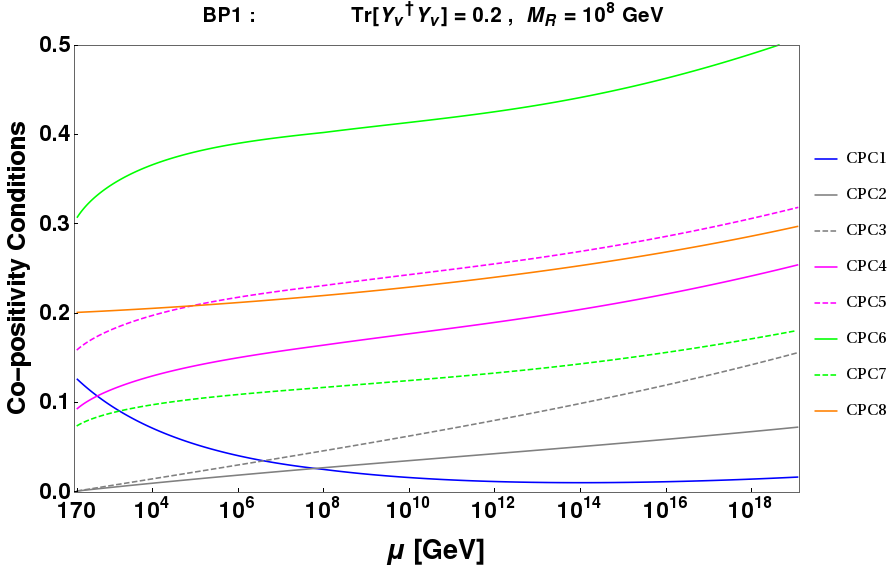}~~
 \includegraphics[height=4.8cm]{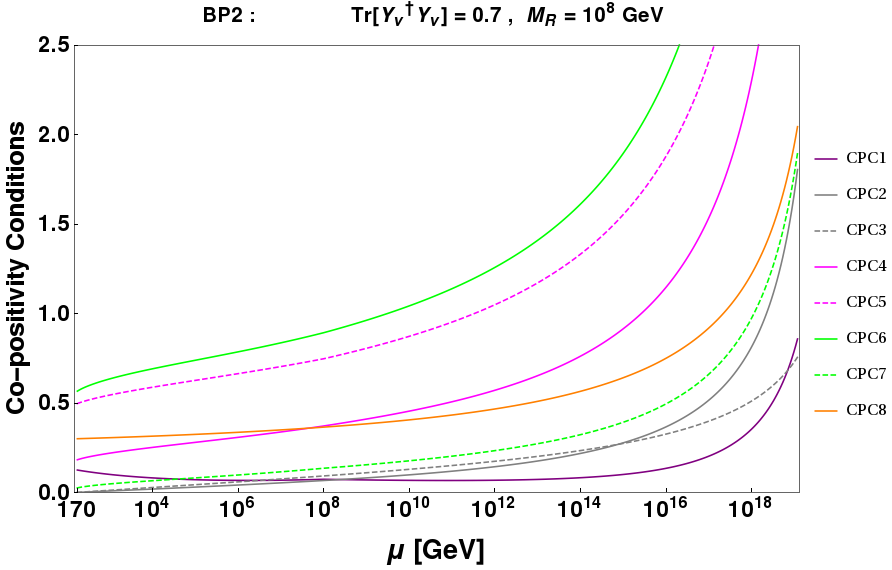}
 $$
 \caption{RG running of all copositivity criteria in Eqn.~\ref{eq:stability} for BP1 (left) and BP2 (right) to ensure the boundedness
 of the scalar potential in any field direction with energy scale $\mu$ for $\text{Tr}[Y_{\nu}^\dagger Y_{\nu}]=0.2$ (left) and 0.7 (right)
 The choices of $\text{Tr}[Y_{\nu}^\dagger Y_{\nu}]$ are shown in dark blue and purple colours in left and right panels of Fig. \ref{fig: EW_VS2} respectively to yield stability. 
}
 \label{fig: EW_VS2_copo}
\end{figure}
Next we show the effect of Tr$[Y_\nu^\dagger Y_\nu]$ in the RG evolution of $\lambda_H$ in Fig. \ref{fig: EW_VS2} for BP1 (left) and BP2 (right).
For the purpose we fix the RH neutrino mass scale $M_R=10^8$ GeV.
It can be seen from left panel that large value of Tr$[Y_\nu^\dagger Y_\nu]\sim 0.7$ brings down $\lambda_H$ towards the negative values at earlier energy scale.
This is obvious from 
the $\beta$ function of $\lambda_H$ as the amount of scalar couplings for BP1 are not that effective in presence of such large value of Tr$[Y_\nu^\dagger Y_\nu]$.
\begin{figure}[htb!]
$$
 \includegraphics[height=4.9cm]{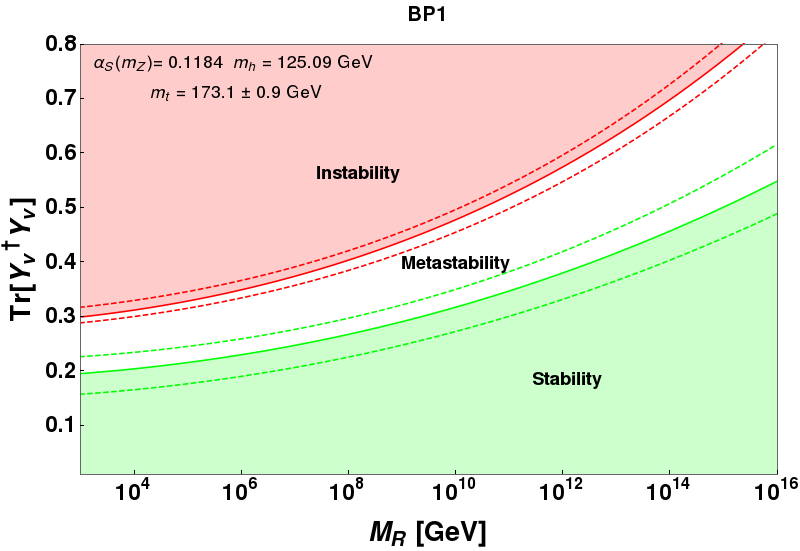}~~
 \includegraphics[height=4.9cm]{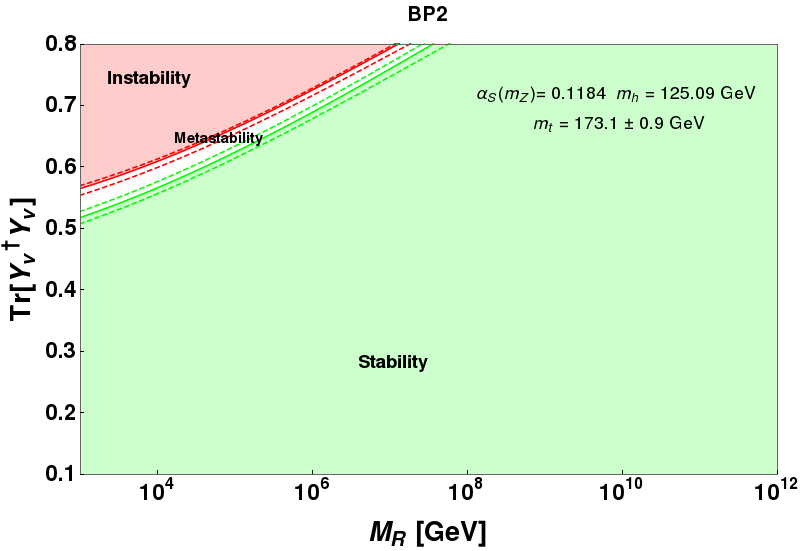}
 $$
 \caption{ Stability, metastability and instability regions plot on $M_R - \text{Tr}[Y_{\nu}^\dagger Y_{\nu}] $ plane for the benchmark point BP1 (left panel) and BP2 (right panel). 
 For illustration we have considered top mass variation in $1 \sigma$ range : $173.1 \pm 0.9$ GeV. }
 \label{fig: EW_VS3}
\end{figure}
With comparatively lesser value of Tr$[Y_\nu^\dagger Y_\nu]\sim 0.3$, the EW vacuum might be in metastable state
provided other conditions ($(i)-(vi)$) are satisfied as shown in left panel of Fig. \ref{fig: EW_VS2_met}. 
 If we further reduce the value of Tr$[Y_\nu^\dagger Y_\nu]\sim 0.2$ the EW vacuum might be absolutely stable. For the stability case we also show the satisfaction of all the copositivity criterias (Eqn.~\ref{eq:stability}) in
left panel of Fig. \ref{fig: EW_VS2_copo}. This ensures the boundness of the scalar potential
in any arbitary field direction. The analysis for BP2 turns out to be similar as observed from right
panels of Fig. \ref{fig: EW_VS2}-\ref{fig: EW_VS2_copo}. Note that due to comparatively larger values of the scalar couplings in BP2, we achieve a better results in stability point of view than BP1.

Based on the inputs from above analysis, now we constrain Tr$[Y_\nu^\dagger Y_\nu]-M_R$ parameter space using the stability, metastability and instability criteria (green, white and red regions respectively) 
for BP1 (left panel) and BP2 (right panel) in Fig. \ref{fig: EW_VS3}. The criteria has been set at Planck scale ($M_P$). 
We use $\alpha_S=0.1184$ and $m_{h}=125.09$ GeV for both the plots. The solid lines indicate the contour for $m_t=173.1$ GeV while 
the dotted lines denote 2$\sigma$ uncertainty of the measured value of $m_t$. It can be concluded from Fig. \ref{fig: EW_VS3},
that to have a stable/metastable EW vacuum, smaller values of $M_R$ requires low Tr$[Y_\nu^\dagger Y_\nu]$ and vice versa. 
One important distinction between the left and right panel figures, corresponding to two different benchmark points, is that BP2 has significantly larger parameter space available from high scale stability. This is because of the larger values of $\lambda_{1,2,3}$ 
parameters used in BP2 compared to BP1 (see Table \ref{tab1} for details). Therefore, it can be concluded, that larger is the mass splitting in IDM sector, the more favourable it is from the stability point of view. 
However there is an important catch to this statement, which we will illustrate next. 
\begin{figure}[htb!]
$$
 \includegraphics[height=4.9cm]{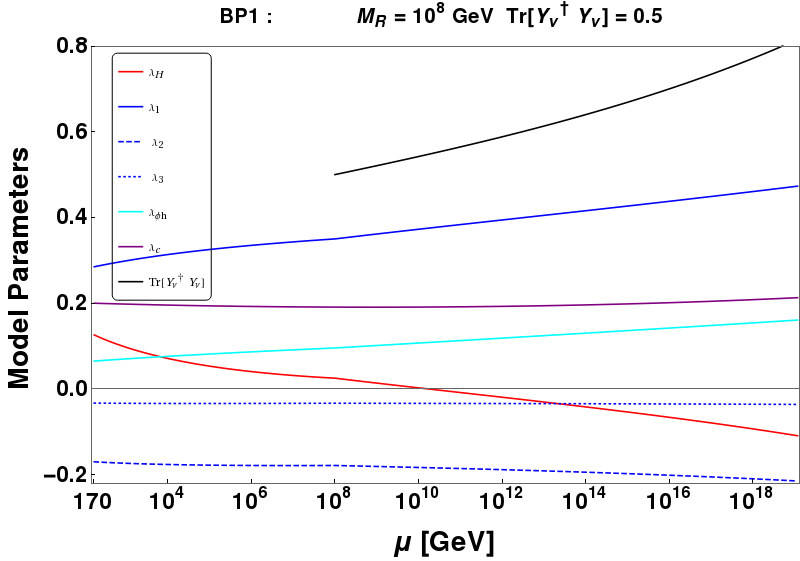}~~
 \includegraphics[height=4.9cm]{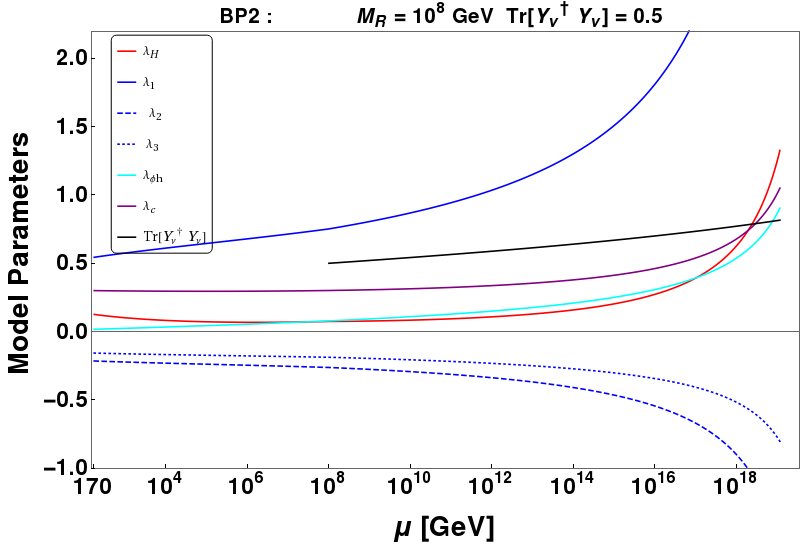}
 $$
 \caption{RG running of relevant coupling parameters for BP1 (left panel) and BP2 (right panel). $M_R = 10^8$ GeV, $\text{Tr}[Y_{\nu}^\dagger Y_{\nu}] = 0.5$ and $m_t = 173.1 $ GeV have been kept fixed in both plots. }
 \label{fig: EW_VS4}
\end{figure}

In Fig. \ref{fig: EW_VS4} we plot the running of all the individual couplings present in the set up.
We see that (fortunately) for the two benchmark points used in the analysis, we 
are still okay with the perturbative limit at the high scale, {\it  i.e.} all the couplings obey the perturbative limit, $|\lambda_i| < 4\pi, ~|\text{Tr}[Y_{\nu}^\dagger Y_{\nu}]|<4\pi$ at Planck scale. However, for BP1, with 
$M_R=10^8$ GeV, and $\text{Tr}[Y_{\nu}^\dagger Y_{\nu}]=0.5$ as shown in the left panel yields unstable solution to EW vacuum with $\lambda_H$ running negative. On the other hand, BP2 with  
same choices of right handed neutrino mass and Yukawa yields a stable vacuum. Therefore, once again we find that larger splitting in IDM sector is more favourable for EW vacuum stability, 
as we have also inferred before from Fig.~\ref{fig: EW_VS3}. But, it turns out that as larger splitting in the IDM sector also uses larger values of $\lambda_{1,2,3}$, it is possible, that many of those points 
become non-perturbative i.e. $|\lambda_i| >4\pi$ when run upto Planck scale. We will show in the next section, that this very phenomena disallows many of the 
relic density and direct search allowed parameter space of the model. Another point to end this section is to note that our available parameter space from DM constraints heavily depend on large DM-DM 
conversion, which naturally comes from large choices of the conversion coupling $\lc$. It is natural, that some of those cases will also be discarded by the perturbative limit $|\lc|<4\pi$, when we evaluate 
the validity of the model at high scale.
\subsection{Allowed parameter space of the model from high scale validity}  
Finally, we come to the point where we can assimilate all the constraints together, from DM constraints to high scale validity. 
In order to do that, we choose relic density and direct search at NLO allowed parameter space of the 
model as discussed in Section \ref{sec:RD-ps} and impose the high scale validity of the model till some energy scale $\mu$ by demanding: \\
$\bullet$  Stability of the scalar potential (Eqn.(\ref{eq:stability})) determined by satisfying copositivity conditions for any
energy scale $\mu$.\\
$\bullet$  Non-violation of perturbativity and unitarity of all the relevant couplings present in the model
as defined in Eqns. \ref{eq:perturbativity} and \ref{eq:unitarity}.\\
\begin{figure}[htb!]
$$
 \includegraphics[height=5.0cm]{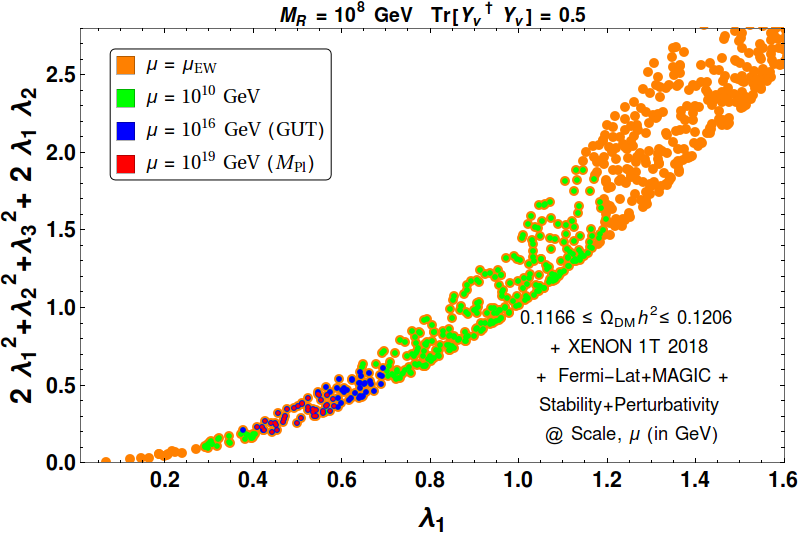}~~
  \includegraphics[height=5.0cm]{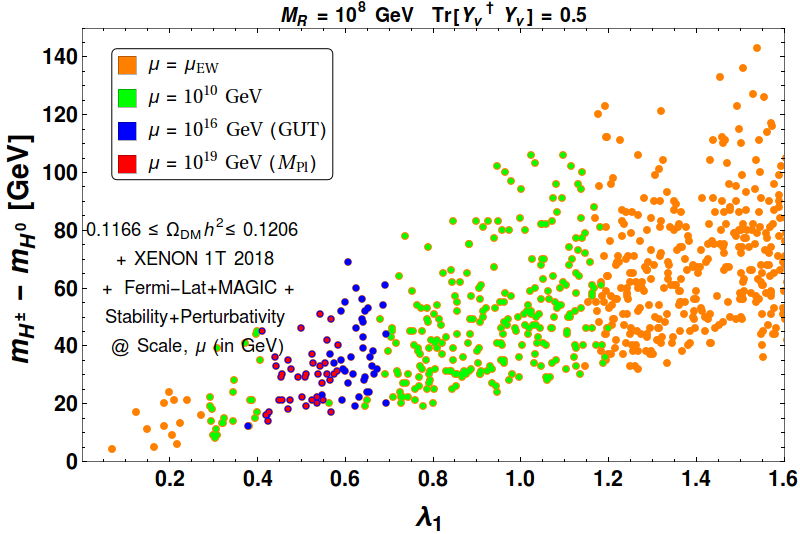}~~~
 $$
 \caption{Relic density, direct search and indirect search allowed points in $2\lambda_1^2+\lambda_2^2+\lambda_3^2+2\lambda_1\lambda_2$ versus $\lambda_1$ plane (left figure) and $m_{H^\pm}-m_{H^0}$ versus $\lambda_1$ plane (right figure) at EW scale (orange points). We also find
 high scale validity of the model following Eqns.~\ref{eq:stability},~\ref{eq:perturbativity} and \ref{eq:unitarity} by
considering the high scale to be  $10^{10}$ GeV (green), $10^{16}$ GeV (blue) and $10^{19}$ GeV (red).
 Right handed neutrino mass and Yukawa couplings are kept fixed at $M_R = 10^8$ GeV, $\text{Tr}[Y_{\nu}^\dagger Y_{\nu}] = 0.5$ with $m_t = 173.1 $ GeV.}
 \label{fig:EWVS5}
\end{figure}
\noindent Note that the high
scale validity of the models does not depend on the structure of mass hierarchy of the DM candidates ({\it i.e.} on the 
sign of mass difference $m_{H^0}-m_\phi$).
To study the EW vacuum stability we demand the positivity of $\lambda_H$ at each energy scale from EW to high scale $\mu$. 
As evident from $\beta_{\lambda_H}$ in Eqn.~\ref{eq:RGScalars}, a particular combination of the scalar couplings in the form of $2\lambda_1^2+\lambda_2^2+\lambda_3^2+2\lambda_1\lambda_2+\frac{1}{2}\lph^2$
determines the positivity of $\lambda_H$ during its running. However the factor $\lph\lesssim 0.1$ is strongly constrained
from the SI DD cross section bound for $m_{\phi}<500$ GeV. Hence, we can assume safely that the 
factor $2\lambda_1^2+\lambda_2^2+\lambda_3^2+2\lambda_1\lambda_2$ without $\lph$ effectively determines the
stability of Higgs vacuum in relic density direct search and indirect search allowed parameter space. It turns out that when $\lambda_H>0$.
all other copositivity conditions for all relic and DD cross section satisfied points in our model stays positive from $\mu=m_t$ to $\mu=M_P$.  \\
\begin{figure}[htb!]
 $$
 \includegraphics[height=5.0cm]{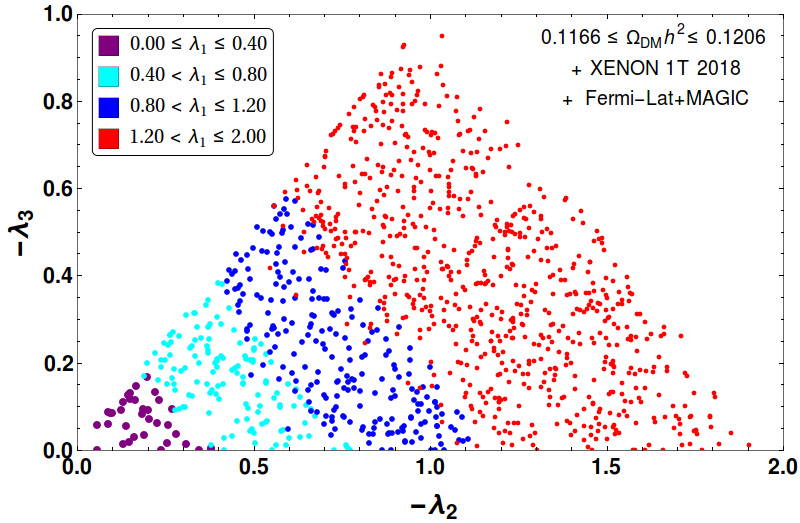}~~
 \includegraphics[height=5.0cm]{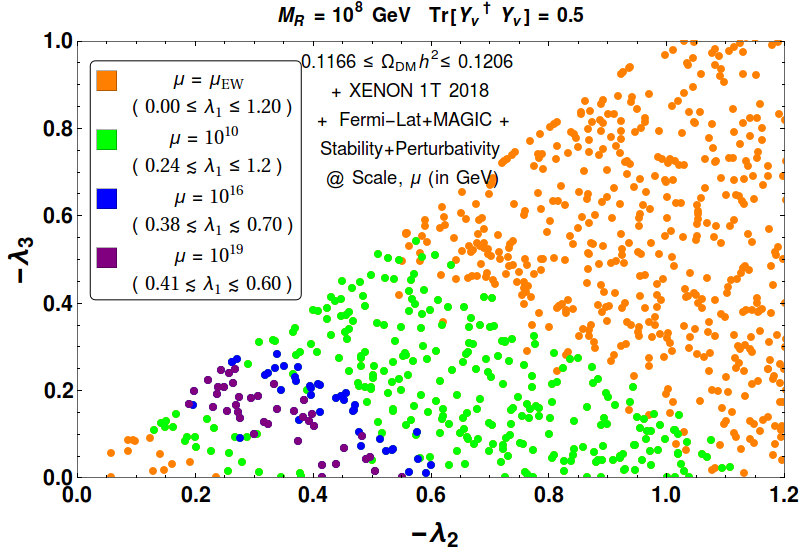}
  $$
 \caption{Relic density direct search and indirect search allowed points in $\lambda_2-\lambda_3$ plane for different values of $\lambda_1$ (left). Allowed points in the same parameter space from DM constraints (orange),  stability 
 and perturbativity conditions following Eqns.~\ref{eq:stability}-\ref{eq:unitarity} considering the high
scale $\mu=\mu_{\rm EW},~10^{10}$ GeV (green), $10^{16}$ GeV (blue) and $10^{19}$ GeV (purple) (right).}
  \label{fig:EWVS6}
\end{figure}
\begin{figure}[htb!]
$$
 \includegraphics[height=5.0cm]{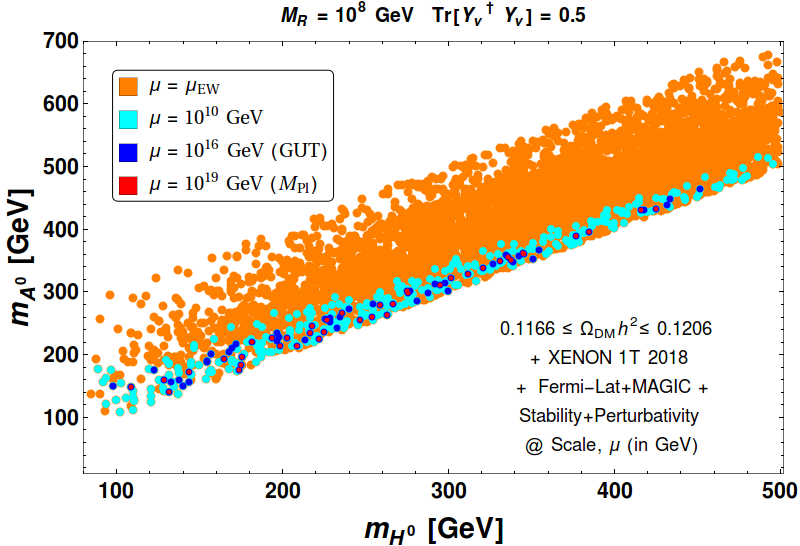}~~
 \includegraphics[height=5.0cm]{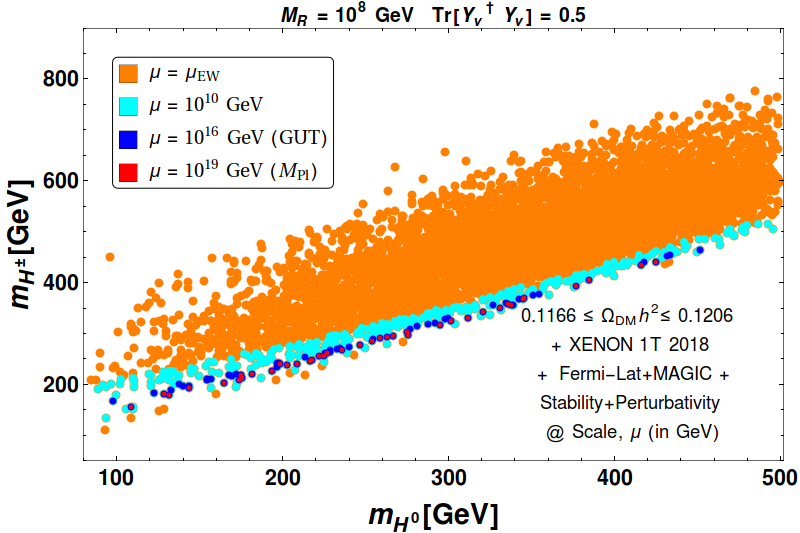}
 $$
 $$
  \includegraphics[height=5.0cm]{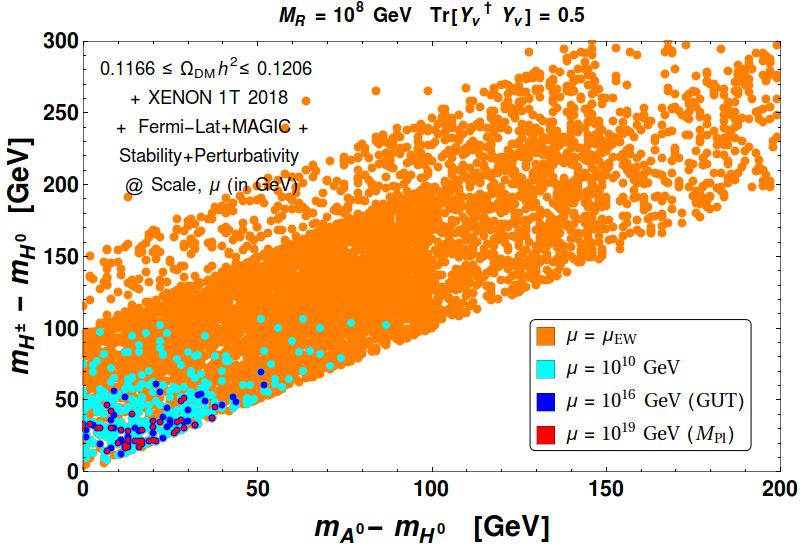}
 $$
 \caption{Relic density direct search and indirect search allowed points (orange) in $m_{A^0}-m_{H^0}$ (top left), $m_{H^\pm}-m_{H^0}$ (top left)
and  $\Delta m (=\mA-\mH)- \Delta M  (=\mHp-\mH$) (bottom). We further apply stability 
 and perturbativity conditions following Eqns.~\ref{eq:stability}-\ref{eq:unitarity} at different energy scales $\mu=10^{10}$ GeV
(light blue), $10^{16}$ GeV (dark blue) and $10^{19}$ GeV (red).}
  \label{fig: HSVA2}
\end{figure}
In Fig. \ref{fig:EWVS5}, we constrain relic density direct search and indirect search allowed points of the model in $2\lambda_1^2+\lambda_2^2+\lambda_3^2+2\lambda_1\lambda_2-\lambda_1$ plane 
to additionally satisfy perturbativity and vacuum stability conditions following Eqns.~\ref{eq:stability}-\ref{eq:unitarity}. Orange points satisfy relic density, DD bounds and indirect search bound, while the green, blue and red points 
on top of that satisfy perturbativity and vacuum stability conditions upto high energy scales $\mu=10^{10}$ GeV, $10^{16}$ GeV
and $10^{19}$ GeV respectively. This very figure essentially shows that all those points with either small values of $\lambda_1$ (i.e. small $m_{H^\pm} - m_{H^0}$)are discarded due to stability of EW vacuum, while 
those with large $\lambda_1$ (i.e. large $m_{H^\pm} - m_{H^0}$)are discarded by perturbative limits of the coupling at high scale.   \\

In Fig. \ref{fig:EWVS6} we study the correlation between the individual scalar couplings to satisfy
the DM constraints,  perturbativity limits and vacuum stability criteria. In  left panel of Fig. \ref{fig:EWVS6}, we 
show the DM relic density and DD cross section satisfied points in $\lambda_2-\lambda_3$ plane for
different values of $\lambda_1$. In right panel we first identify the relevant parameter space in the same plane which satisfy
the DM constraints, the perturbativity bound and vacuum stability criteria
till EW energy scale. Then we further impose perturbativity bound and vacuum stability conditions considering
the high scale as  $\mu=10^{10}$ GeV, $10^{16}$ GeV and $10^{19}$ GeV in addition to the DM constraints. 
It is seen that the lower 
portion of the available parameter space gets discarded by vacuum stability or high value of $\lambda_c$ while perturbativity bounds 
constrain the higher values of the couplings. In this plot also we kept $M_R = 10^8$ GeV, $\text{Tr}[Y_{\nu}^\dagger Y_{\nu}] = 0.5$ with $m_t = 173.1 $ GeV. 
We must also note that with larger $M_R$ and smaller Yukawa $\text{Tr}[Y_{\nu}^\dagger Y_{\nu}] $, we could obtain a larger available parameter space from high scale validity. 

In Fig.~\ref{fig: HSVA2}, we show the high scale validity of relic density and direct search allowed parameter space 
of the model in $m_{H^0}-m_{A^0}$ (top left), $\mH-\mHp$ 
 (top right) and  $\Delta m( = \mA-\mH) -\Delta M (= \mHp-\mH)$ planes with $M_R = 10^8$ GeV and $\text{Tr}[Y_{\nu}^\dagger Y_{\nu}] = 0.5 $ at different high energy 
 scales $\mu=\{10^{10}, 10^{16},10^{19}\}$ GeVs denoted by light blue,  dark blue and red points.
 The orange points are corresponding 
 to relic and direct search allowed parameter space at EW scale. We see that larger mass difference between inert higgs 
 components, $\Delta m-\Delta M$ which are related with quartic
 couplings, $\lambda_{1,2,3}$ (see Eqn.~\ref{eq:mass-coupling}), are discarded from perturbativity
 conditions mentioned in Eqn.~\ref{eq:perturbativity}. While the small mass differences between inert componets are
 also excluded from stability criteria of Higgs potential.\\
 
 Till now, while discussing the effect of stability and high scale validity of the proposed set up, we have considered  fixed right handed neutrino mass, $M_R = 10^8$ GeV and corresponding Yukawa coupling $\textrm{Tr}[Y_\nu^\dagger Y_\nu] = 0.5$. For sake of completeness we extend our study for few different 
 values of RH neutrino mass and $\textrm{Tr}[Y_\nu^\dagger Y_\nu]$
 and find out the allowed ranges of the relevant parameters considering both the hierarchies $m_\phi>m_{H^0}$ and $m_\phi \leq m_{H^0}$ separately. We note the corresponding results in Table \ref{TabSE1} and Table \ref{TabSE2} of Appendix \ref{sec:Allrange}.
 \section{Collider signature of Inert doublet DM at LHC}
 \label{sec:collider}
\begin{figure}[htb!]
$$
 \includegraphics[height=3.7cm]{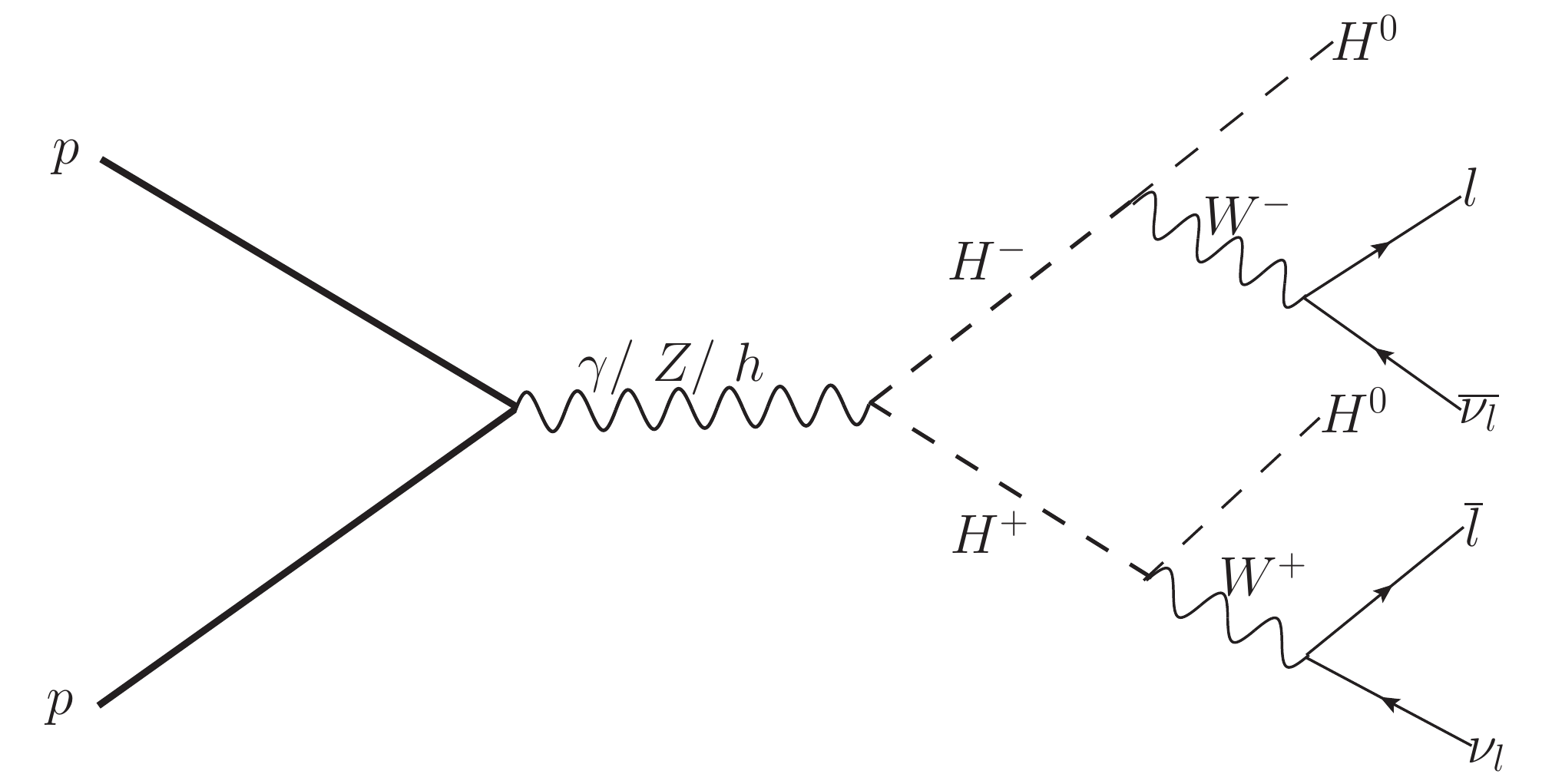}~
 \includegraphics[height=4.6cm]{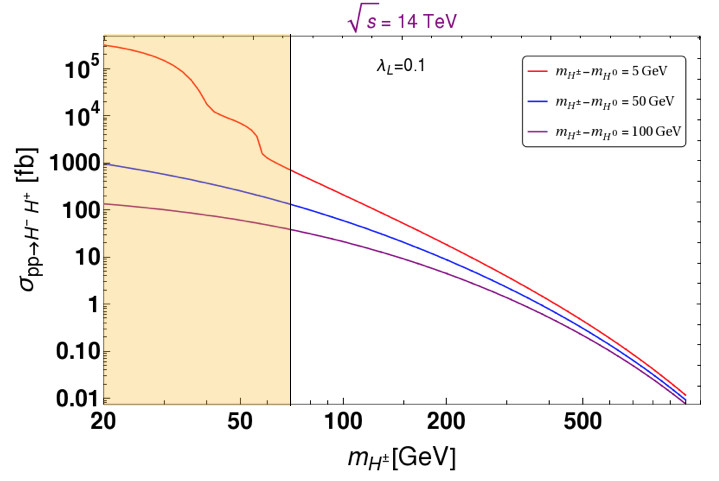}
 $$
 \caption{[Left] Feynman graph for OSDL+$\slashed{E}_T$ signature of IDM at LHC. [Right] Variation of production cross-section 
 $\sigma_{p p \to H^+~H^-}$ (in fb) with $m_{H^\pm}$ ($= m_{H^0}+\Delta M$) in GeV for different choices of $m_{H^\pm}-m_{H^0}$
 for center-of-mass energy $\sqrt{s}=14$ TeV at LHC. LEP limit on charged scalar is shown by the shaded region.}
 \label{fig:feyn_diag-production}
\end{figure}

Inert doublet has been an attractive DM framework, due to the possibility of collider detection~\cite{Gustafsson:2012aj,Bhardwaj:2019mts}. Here, we relook into the possible collider search strategies of IDM at LHC 
in presence of a second scalar singlet DM component. It is also worth mentioning here that the real scalar singlet, which interacts with SM only through Higgs portal coupling, does not have any promising 
collider signature excepting mono-$X$ signature arising out of initial state radiation (ISR), where $X$ stands for $W,Z,~\textrm{jet}$. Such signals are heavily submerged in SM background due to weak production cross-section of DM in relic density and 
direct search allowed parameter space~\cite{Han:2016gyy}. The charge components $H^+,~H^-$ of inert DM can be produced at LHC via Drell-Yan $Z ~{\textrm{and}} ~\gamma$ mediation as well as through Higgs mediation. 
Further decay of $H^\pm$ to DM ($H^0$) and leptonic final states through on/off shell $W^\pm$ yields hadronically quiet opposite sign di-lepton plus missing energy (OSDL+$\slashed{E}_T$)\footnote{There are other possible signatures (for example, three lepton final state) of inert DM arising from the the combination of $H^\pm,A^0$ production and their subsequent decays, for a detailed list see \cite{Chakraborti:2018aae,Bhardwaj:2019mts}.}, as shown in 
left panel of Fig.~\ref{fig:feyn_diag-production}. In this study, we focus on this particular signal of inert dark matter as detailed below:     
 \bea
\nonumber
\textrm{Signal ::}~{\rm OSDL}+\slashed{E}_T \equiv~~ {\ell^+\ell^{-} +({\slashed E}_T)} : ~~~ p~p \rightarrow H^+ ~~ H^- , ~(H^- \rightarrow ~\ell^- ~\overline{\nu_\ell} ~ H^0), \nonumber \\ ~~~~~~~~~~~~~~~(H^+ \rightarrow \ell^+ ~{\nu_\ell} ~H^0); {\rm where} ~~\ell = \{e ,\mu\} ~~ .
\eea
\begin{table}[htb!]
\begin{center}
\resizebox{\textwidth}{!}{
\begin{tabular}{|c|c|c|c|c|c|}
\hline
BPs &$ \{~ m_{H^0},~ m_{\phi},~ m_{A^0},~  m_{H^\pm},~ \lambda_L,~  \lph,~  \lc ~\}$ & $\Omega_{H^0}h^2$ & $\Omega_{\phi}h^2$  &
$\sigma_{H^0}^{eff}$ ($\textrm{cm}^2$) & $\sigma_{\phi}^{eff}$ ($\textrm{cm}^2$)  \\
\hline
BPC1       &$\{~ 209,~ 483,~ 213,~ 239,~  0.006,~ 0.09,~ 0.13~\}$  & $0.0082$ & $0.1084$  & $ 3.7 \times 10^{-47}$   &  $2.8 \times 10^{-46}$   \\
\hline
BPC2       &$\{~ 129,~ 476,~ 158,~ 180,~  0.011,~ 0.069,~ 0.13~\}$  & $0.0032$ & $0.1175$  & $ 6.7 \times 10^{-47}$   &  $1.8 \times 10^{-46}$   \\
\hline
BPC3       &$\{~ 100,~ 96,~ 127,~ 178,~  0.010,~ 0.002,~ 0.56~\}$  & $0.0021$ & $0.1156$  & $ 6.8 \times 10^{-47}$   &  $3.7 \times 10^{-48}$   \\
\hline
\hline
\end{tabular}
}
\end{center}
\begin{center}
\resizebox{\textwidth}{!}{
\begin{tabular}{|c|c|c|c|c|}
\hline
BPs &$ ~ m_{H^\pm} -  m_{H^0}~$ & $\{~\lambda_1,~\lambda_2,~\lambda_3~\}$ & $\{~M_R,~\textrm{Tr}[Y_{\nu}^\dagger Y_{\nu}]~\}$  &
Validity Scale ($\mu$) \\
\hline
BPC1       &$30$ $( < m_W)$  & $\{~0.456,~-0.416,~-0.028~\}$ & $\{10^8,~0.5\}$  & $ 1.22\times 10^{19}$ ($M_{pl}$)     \\
\hline
BPC2       &$51$  $( < m_W)$  & $\{~0.543,~-0.383,~-0.137~\}$ & $\{10^8,~0.5\}$  & $ 1.22\times 10^{19}$ ($M_{pl}$)     \\
\hline
BPC3       &$78$  $( \sim m_W)$  & $\{~0.737,~-0.615,~-0.101~\}$ & $\{10^8,~0.5\}$  & $ \sim 10^{14}$     \\
\hline
\hline
\end{tabular}
}
\end{center}
\caption{DM masses, quartic couplings, relic densities and spin independent effective DM-neucleon cross-section of selected benchmark points for collider study. All benchmark points chosen here have 
$m_{H^\pm}-m_{H^0} < m_{W^\pm}$ for off-shell production of $W^\pm$. The maximum scale ($\mu$) of Higgs vacuum satability and peturbativity in presence of right handed neutrinos are also noted. All masses and scales are in GeV.}
\label{tab1}
\end{table}
\begin{table}[h]
\begin{center}
\resizebox{\textwidth}{!}{
\begin{tabular}{|c|c|c|c|c|c|}
\hline
BPs &$ \{~ m_{H^0},~ m_{\phi},~ m_{A^0},~  m_{H^\pm},~ \lambda_L,~  \lph,~  \lc ~\}$ & $\Omega_{H^0}h^2$ & $\Omega_{\phi}h^2$  &
$\sigma_{H^0}^{eff}$ ($\textrm{cm}^2$) & $\sigma_{\phi}^{eff}$ ($\textrm{cm}^2$)  \\
\hline
BPD1       &$\{~ 92,~ 386,~ 155,~ 198,~  0.004,~ 0.063,~ 0.10~\}$  & $0.0032$ & $0.1167$  & $ 4.5 \times 10^{-47}$   &  $2.2 \times 10^{-46}$   \\
\hline
BPD2       &$\{~ 90,~ 437,~ 136,~ 233,~  0.006,~ 0.077,~ 0.11~\}$  & $0.0030$ & $0.1150$  & $ 6.9 \times 10^{-47}$   &  $2.6 \times 10^{-46}$   \\ 
\hline
\hline
\end{tabular}
}
\end{center}
\begin{center}
\resizebox{\textwidth}{!}{
\begin{tabular}{|c|c|c|c|c|}
\hline
BPs &$ ~ m_{H^\pm} -  m_{H^0}~$ & $\{~\lambda_1,~\lambda_2,~\lambda_3~\}$ & $\{~M_R,~\textrm{Tr}[Y_{\nu}^\dagger Y_{\nu}]~\}$  &
Validity Scale ($\mu$) \\
\hline
BPD1       &$106$  $( > m_W)$  & $\{~1.024,~-0.7588,~-0.2571~\}$ & $\{10^8,~0.5\}$  & $  10^{10}$      \\
\hline
BPD2       &$143$ $( > m_W)$ & $\{~1.538,~-1.354,~-0.1718~\}$ & $\{10^6,~0.5\}$  & $  10^{8}$      \\
\hline
\hline
\end{tabular}
}
\end{center}
\caption{ DM masses, quartic couplings, relic densities and spin independent effective DM-neucleon cross-section of selected benchmark points for collider study. All benchmark points chosen here have 
$m_{H^\pm}-m_{H^0} > m_{W^\pm}$ for on-shell production of $W^\pm$. The maximum scale ($\mu$) of Higgs vacuum satability and peturbativity in presence of right handed neutrinos are also noted. All masses and scales are in GeV.}
\label{tab2}
\end{table}
In the right panel of Fig.~\ref{fig:feyn_diag-production}, we show variation of charged pair $(H^+~H^-)$ production cross-section at LHC for center-of-mass energy $\sqrt{s}=14$ TeV as a function of  
$m_{H^\pm}= m_{H^0}+\Delta M$, where $\Delta M$ indicates the mass difference with the inert DM and serves as a very important variable for the signal characteristics. The plot on RHS show that production cross-section 
is decreasing with larger charged scalar mass $m_{H^\pm}$, where we have demonstrated three fixed values of $m_{H^\pm}-m_{H^0}$ $= 5, ~50$ and $100$ GeV. Around $m_{H^\pm} \sim m_h/2$, there is a sharpe fall of production cross-section. 
This is because, for $m_{H^\pm} \le m_h/2$, there is a significant contribution arising from Higgs production and its subsequent decay to the charged scalar components, which otherwise turns into an off-shell propagator
to yield a subdued contribution to Drell-Yan production. Following \cite{Kalinowski:2018ylg}, a conservative bound on the charge scalars is applied here as $m_{H^\pm} \geq 70$ GeV, as indicated in the RHS plot of Fig.~\ref{fig:feyn_diag-production}.  

We next choose a set of benchmark points (BPs) allowed from DM relic, direct search constraints as well as from Higgs invisible decay constraints for performing collider simulation, shown in Table~\ref{tab1} and Table~\ref{tab2}. 
The BPs are also allowed from absolute Higgs vacuum stability and perturbativity limits in presence of right handed neutrinos, valid upto scale $\mu$ as mentioned in the tables. The benchmark points are divided into two categories: 
(BPC1-BPC3) in Table~\ref{tab1} correspond to $\Delta M = m_{H^\pm} - m_{H^0} \lesssim m_{W^\pm}$ where the charged scalar, $H^\pm$ decay through off-shell $W^\pm$. On the other hand, benchmark points (BPD1-BPD2) in 
Table~\ref{tab2} correspond to $\Delta M = m_{H^\pm} - m_{H^0} > m_{W^\pm}$, where the charged scalar $H^\pm$ decay through on-shell $W^\pm$. Each table (Table~\ref{tab1} and Table~\ref{tab2}) consists of two parts: the first part 
contains all the relevant dark sector masses, couplings, relic density and direct search cross-sections of both DM components. The second part demonstrates the mass difference ($\Delta M = m_{H^\pm} - m_{H^0}$), choice of 
right handed neutrino mass, neutrino Yukawa and the maximum scale of validity ($\mu$) of the Higgs vacuum. 

The simulation technique adopted here is as follows. We first implemented the model in {\tt {FeynRule}} \cite{Alloul:2013bka} to generate UFO file which is required to feed into event generator {\tt Madgraph}\cite{Alwall:2011uj}. 
Then these events are passed to 
{\tt Pythia} \cite{Sjostrand:2006za} for hadronization. All parton level leading order (LO) signal events and SM background events\footnote{There are several SM process which contribute to the chosen ${\ell^+\ell^{-} +({\slashed E}_T)}$ signal, 
dominant processes are: $t \overline{t}$, $W^+ W^-$, $ZZ$ and $W^+ W^- Z$.} are generated in {\tt Madgraph} at $\sqrt{s}=14$ TeV using {\tt cteq6l1} \cite{Placakyte:2011az} parton distribution. 
Leptons ($\ell=e,\mu$) isolation, jet and unclustered event formation to mimic to the actual collider environment are performed as follows: \\
 (i) \underline{Lepton isolation}: The minimum transverse momentum required to identify a lepton $(\ell=e,\mu)$ has been kept as $p_T > 20$ GeV and we also require the lepton to be produced in the central region of detector followed by 
 pseudorapidity selection as $|\eta| < 2.5$. Two leptons are separated from each other with minimum distance $\Delta R \geq 0.2$ in $\eta-\phi$ plane. To separate leptons from jets we further imposed $\Delta R \geq 0.4$.\\
 (ii) \underline{Jet formation}: For jet formation, we used cone algorithm {\tt PYCELL} in built in {\tt Pythia}. All partons within a cone of $\Delta R \leq 0.4$ around a jet initiator with $p_T > 20$ GeV is identified to form a jet. It is important to identify 
 jets in our case because we require the final state signal to be hadronically quiet i.e. to have zero jets. \\
 (iii) \underline{Unclustered Objects}: All final state objects with $0.5<p_T <20$ GeV and $2.5< |\eta| < 5$ are considered as unclustered objects. Those objects neither form jets nor identified as isolated leptons and they only contribute to missing energy.
~\\
The main idea is to see if the signal events rise over SM background. For that there are three key kinematic variables where the signal and background show different sensitivity. They are: 
\begin{itemize}
 \item {\it Missing Energy ($\slashed{E_T}$):} The most important signature of DM being produced at collider. This is defined by a {\it vector sum} of transverse momentum of all the missing particles 
 (those are not registered in the detector); this in turn can be estimated form the momentum imbalance in the transverse direction associated to the visible particles. Thus missing energy (MET) is defined as: 
 \bea
 \slashed{E_T} = -\sqrt{(\sum_{\ell,j} p_x)^2+(\sum_{\ell,j} p_y)^2},
 \eea
 where the sum runs over all visible objects that include the leptons, jets and the unclustered components. 
 \item {\it Transverse Mass ($H_T$):} Transverse mass of an event is identified with the {\it scalar sum} of the transverse momentum of objects reconstructed in a collider event, namely lepton and jets as defined above. 
 \bea
 {H_T} = \sum_{\ell,j} \sqrt{(p_x)^2+ (p_y)^2}.
 \eea
 \item {\it Invariant mass ($m_{\ell\ell}$):} Invariant mass of opposite sign dilepton hints to the parent particle, from which the leptons have been produced and thus helps segregating signal from background. This is defined as:
  \bea
 {m_{\ell^+\ell^-}} = \sqrt{ (\sum_{\ell^+\ell^-} p_x)^2+ (\sum_{\ell^+\ell^-} p_y)^2+(\sum_{\ell^+\ell^-} p_z)^2}.
 \eea
\end{itemize}
\begin{figure}[htb!]
$$
 \includegraphics[height=4.8cm]{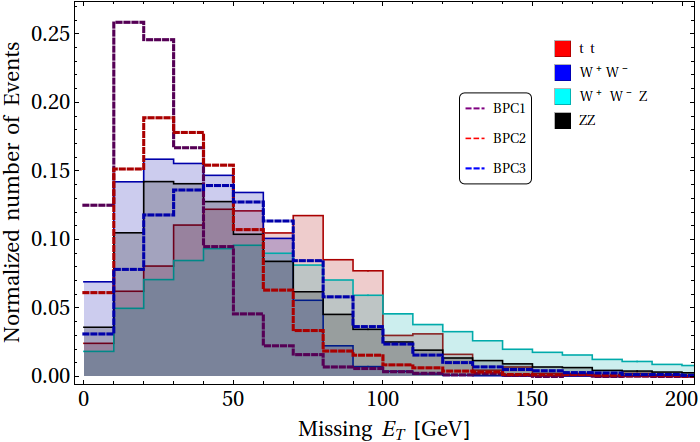}~
 \includegraphics[height=4.8cm]{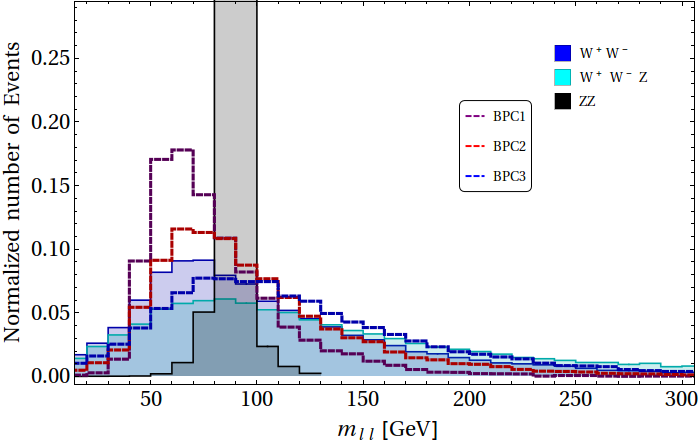}                             
$$
$$
\includegraphics[height=4.8cm]{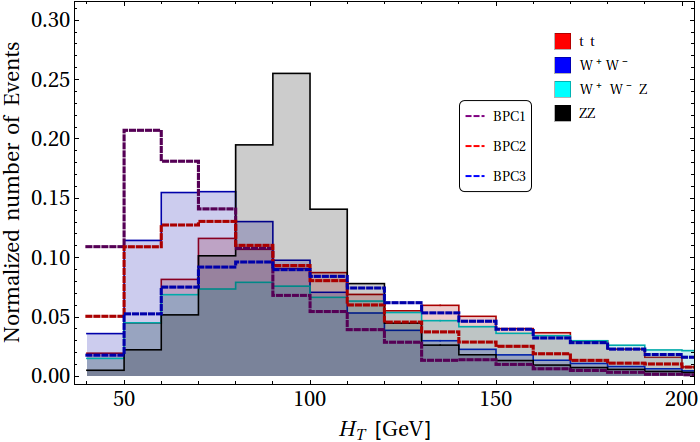} 
$$
 \caption{Distribution of missing energy ($\slashed{E}_T$), invariant mass of OSDL ($m_{ \ell^+ \ell^-}$) and transverse mass ($H_T$) for signal events $\ell^+\ell^{-} +({\slashed E}_T)$ and 
 dominant SM background events at LHC with $\sqrt{s} = 14 \rm ~TeV~$.}
 \label{fig:signal+backg-dist-1}
\end{figure}
\begin{figure}[htb!]
$$
 \includegraphics[height=4.8cm]{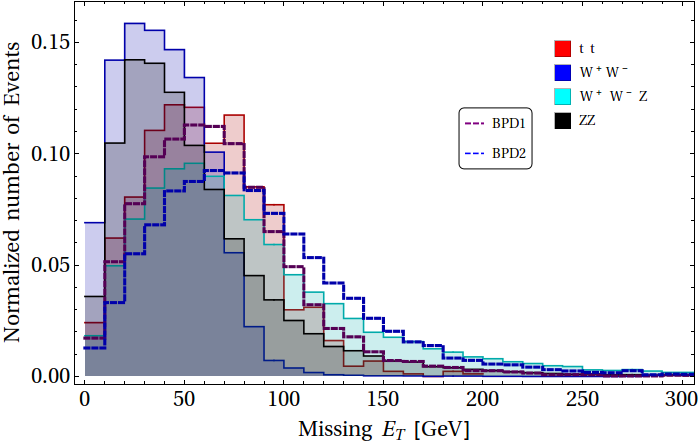}~
 \includegraphics[height=4.8cm]{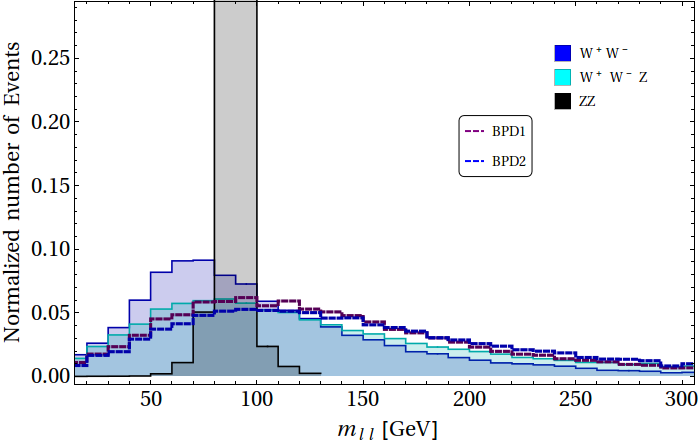}                             
$$
$$
\includegraphics[height=4.8cm]{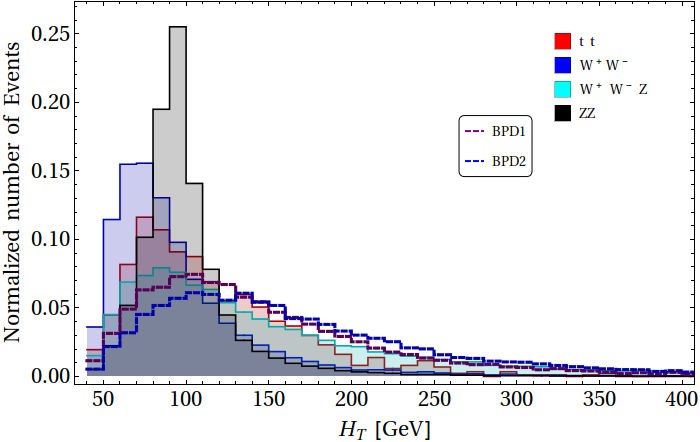} 
$$
 \caption{Distribution of missing energy ($\slashed{E}_T$), invariant mass of OSDL ($m_{ \ell^+ \ell^-}$) and transverse mass ($H_T$) for signal events $\ell^+\ell^{-} +({\slashed E}_T)$ and 
 dominant SM background events at LHC with $\sqrt{s} = 14 \rm ~TeV~$.}
 \label{fig:signal+backg-dist-2}
\end{figure}
The distribution of missing energy ($\slashed{E}_T$), invariant mass of opposite sign dilepton ($m_{ \ell^+ \ell^-}$) and transverse mass ($H_T$) for the BPs along with dominant SM background events are shown in 
Fig.~\ref{fig:signal+backg-dist-1} top left, top right and bottom panel respectively. All BPs depicted in Fig.~\ref{fig:signal+backg-dist-1} correspond to $m_{H^\pm}-m_{H^0}\equiv \Delta M < m_{W^\pm}$ where opposite sign 
di-lepton are produced from off-shell $W^\pm$ mediator. All the distributions are normalised to one event. Missing energy (as well as $H_T$) distributions of BPs (BPC1-BPC3) show that the peak of distribution for the signal is on the left 
of SM background. This is because the benchmark points are characterised by small $\Delta M$, where the charged scalars and inert DM have small mass splitting. Therefore, such situations are visibly segregated from SM background 
by MET and $H_T$ distribution. Clearly, when $\Delta M$ becomes $~m_{W^\pm}$ (for example, BPC3) the distribution closely mimic SM background. Therefore the signal events for this class of benchmark points can survive 
for a suitable upper $\slashed{E}_T$ and $H_T$ cut while reducing SM backgrounds. It is important to take a note that OSDL events coming from $ZZ$ background naturally peaks at $m_Z$ in $m_{\ell\ell}$ distribution. Therefore, 
we use invariant mass cut in the $Z$ mass window to get rid of this background. \\

\begin{figure}[htb!]
$$
 \includegraphics[height=5.0cm]{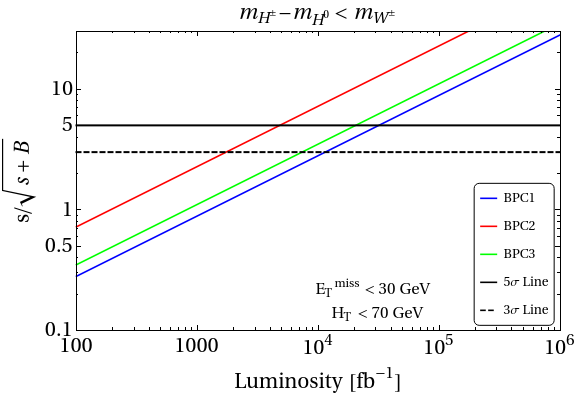}~
 \includegraphics[height=5.0cm]{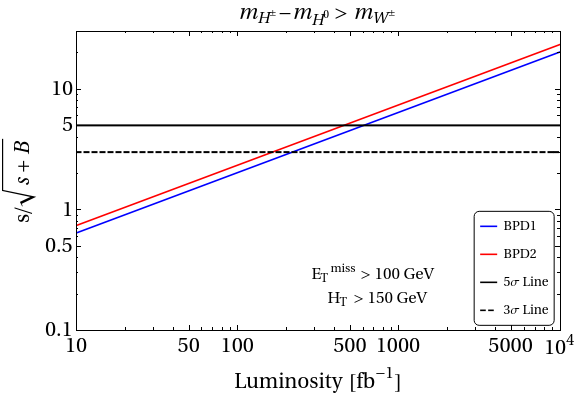}
 $$
 \caption{Signal significance of some select benchmark points at LHC for $\sqrt s=14$ TeV, in terms of Luminosity (fb$^{-1}$). 3$\sigma$ and 5$\sigma$ lines are shown. 
 Left: Points with $\Delta M < m_W$; Right: Points with $\Delta M > m_W$.}
 \label{fig:significance1}
\end{figure}

The situation is reversed for larger splitting between the charged scalar component with DM, i.e $\Delta M \equiv m_{H^\pm}-m_{H^0}> m_{W^\pm}$ corresponding to BPs (BPD1-BPD2) as in Table~\ref{tab2}. 
The distributions of $\slashed{E}_T,~m_{\ell\ell}$ and $H_T$ therefore become flatter and peak of the distribution shifts to higher value as shown in Fig.~\ref{fig:signal+backg-dist-2}. In such cases the signal events for 
large $\Delta M$ can be separated from SM background at a suitable lower end cut of $\slashed{E}_T$ and $H_T$.   \\
Therefore the selection cuts used in this analysis are summarised as follows: 
\begin{itemize}
 \item \underline{Invariant mass ($m_{\ell\ell}$) cuts}:  $m_{\ell\ell} < (m_Z -15)$ GeV and $m_{\ell\ell} > (m_Z +15)$ GeV. 
 \item \underline{$H_{T}$ cuts}: 
 \begin{itemize}
  \item $H_T < {70}$ when  $m_{H^\pm}-m_{H^0} < m_{W^\pm}$ .
  \item $H_T > {150, 200}$ when  $m_{H^\pm}-m_{H^0} > m_{W^\pm}$ .
 \end{itemize}
\item \underline{$\slashed{E}_{T}$ cuts}: 
 \begin{itemize}
  \item $\slashed{E}_T < {30,40}$ when  $m_{H^\pm}-m_{H^0} < m_{W^\pm}$ .
  \item $\slashed{E}_T > {100,150}$ when  $m_{H^\pm}-m_{H^0} > m_{W^\pm}$ .
 \end{itemize}
\end{itemize}
\begin{table}[htb!]
\begin{center}
\resizebox{\textwidth}{!}{
\begin{tabular}{|c|c|c|c|c|c|c|c|}
\hline
BPs & $\sigma^{\text{OSD}}$ (fb) & $\slashed{E}_T$ (GeV)  & $H_T$ (GeV) & $\sigma^{\text{OSD}}_{\text{eff}}$(fb) & $N^{\text{OSD}}_{\text{eff}} @  \mathcal{L} =10^2 \textrm{fb}^{-1}$ \\
\hline\hline 
\textrm{BPC1}& 9.16 &  $< 30$  &  $< 70 $ & $ 0.28 $ & 28 \\  
& &  $< 40$  &    & $0.33 $ & 33\\
\hline 
\textrm{BPC2}& 27.14&  $< 30$  &  $< 70 $ & $0.72 $ & 72 \\  
& &  $< 40$  &    & $0.97 $ & 97  \\
\hline 
\textrm{BPC3}& 28.65&  $< 30$  &  $< 70 $ & $ 0.35$ & 35 \\  
& &  $< 40$  &    & $ 0.55$ & 55 \\
\hline 
\hline
\end{tabular}
}
\end{center}
\caption {Signal cross-section for BPC1-BPC3 after the selection cuts are employed.} 
\label{tab:signal1}
\end{table}  
We next turn to signal and background events that survive after the selection cuts are employed. The signal events are listed in Table \ref{tab:signal1} 
for BPC1-BPC3.\\
\begin{table}[htb!]
\begin{center}
\resizebox{\textwidth}{!}{
\begin{tabular}{|c|c|c|c|c|c|c|c|}
\hline
BPs & $\sigma^{\text{OSD}}$ (fb) & $\slashed{E}_T$ (GeV)  & $H_T$ (GeV) & $\sigma^{\text{OSD}}_{\text{eff}}$(fb) & $N^{\text{OSD}}_{\text{eff}}@  \mathcal{L} =10^2 \textrm{fb}^{-1}$ \\
\hline\hline 
$\textrm{BPD1}$ & $19.91$  & $>$100 & $>$150  & $ 0.75$ &  $ 75$ \\
&&& $>$200 & $0.48 $ &  $ 48$\\
\cline{3-6}
 &  & $>$150  & $>$ 150    &  $ 0.23$ & $ 23$ \\
&&& $>$200 & $ 0.21$ & $ 21$ \\   
\cline{3-6}
\hline 
BPD2 & $ 11.16$& $>$100 & $>$150  & $0.87 $ & 87 \\
&&& $>$200 & $0.59 $ & 59 \\
\cline{3-6}
 &  & $>$150  & $>$ 150    & $0.34 $ & $34 $ \\
&&& $>$200 & $ 0.29$ & $29 $ \\   
\cline{3-6}
\hline 
\hline
\end{tabular}
}
\end{center}
\caption {Signal cross-section for BPD1-BPD2 after the selection cuts are employed.} 
\label{tab:signal2}
\end{table}  
\begin{table}[htb!]
\begin{center}
\resizebox{\textwidth}{!}{
\begin{tabular}{|c|c|c|c|c|c|c|c|}
\hline
SM Bkg. & $\sigma^{\text{OSD}}$ (fb) & $\slashed{E}_T$ (GeV)  & $H_T$ (GeV) & $\sigma^{\text{OSD}}_{\text{eff}}$(fb) & $N^{\text{OSD}}_{\text{eff}} @  \mathcal{L} =10^2 \textrm{fb}^{-1} $ \\
\hline\hline 
& &  $< 30$  &  $< 70 $ & 6.23 & 623\\  
& &  $< 40$  &    & 10.27 & 1027 \\
\cline{3-6}
$t~\bar {t}$ & $36.69 \times 10^3$& $>$100 & $>$150  & 10.64 & 1064 \\
&&& $>$200 & 4.40 &  440\\
\cline{3-6}
 &  & $>$150  & $>$ 150    &  1.47 & 147 \\
&&& $>$200 &  1.10 & 110  \\   
\cline{3-6}
\hline 
& &  $< 30$  &  $< 70 $ & 131.18 & 13118 \\  
& &  $< 40$  &    & 187.89 & 18789 \\
\cline{3-6}
$W^+~W^-$ & $4.74 \times 10^3$& $>$100 & $>$150  &  7.72 &  772 \\
&&& $>$200 & 4.97 & 4.97 \\
\cline{3-6}
 &  & $>$150  & $>$ 150    & 1.23 & 123 \\
&&& $>$200 & 1.18 & 118 \\   
\cline{3-6}
\hline 
& &  $< 30$  &  $< 70 $ & 0.53 & 53 \\  
& &  $< 40$  &    & 0.69  & 69  \\
\cline{3-6}
$Z~Z$ & $0.25 \times 10^3$& $>$100 & $>$150  &0.18  &  18 \\
&&& $>$200 & 0.05 & 5 \\
\cline{3-6}
 &  & $>$150  & $>$ 150    &  0.09 & 9\\
&&& $>$200 & 0.04 & 4 \\   
\cline{3-6}
\hline 
& &  $< 30$  &  $< 70 $ & 0.01 & 1 \\  
& &  $< 40$  &    & 0.02 & 2 \\
\cline{3-6}
$W^+~W^- Z$ & $1.00$& $>$100 & $>$150  & 0.06 & 6  \\
&&& $>$200 & 0.04 & 4 \\
\cline{3-6}
 &  & $>$150  & $>$ 150    &  0.03 & 3 \\
&&& $>$200 & 0.02 & 2\\   
\cline{3-6}
\hline 
\hline
\end{tabular}
}
\end{center}
 \caption {Dominant SM background contribution to $\ell^+\ell^{-} +({\slashed E}_T)$ signal events for $\sqrt{s}$ = 14 TeV at LHC. The effective number of 
 final state background events with different $\slashed{E}_T$, $H_T$ and $m_{\ell\ell}$ cuts are tabulated for luminosity $\mathcal{L} = 100~fb^{-1}$. 
 To incorporate the Next-to-Leading order (NLO) cross section of SM background we have used appropriate K-factors \cite{Alwall:2014hca}.} 
\label{tab:background}
\end{table}  
Similarly, signal events for BPD1-BPD2 are listed in Table~\ref{tab:signal2} using a lower cut on $\slashed{E}_T$ and $H_T$. The signal cross-section 
and event numbers for this class of points are much smaller due to the fact that the charged scalar masses are on higher side as for all of the cases 
$\Delta m >m_W$ independent of DM masses.  However, the SM background 
events get even more suppressed with the set of $\slashed{E}_T$, $H_T$ cuts. The SM background cross-section and event numbers after cut flow is mentioned 
in Table~\ref{tab:background}. Therefore, in spite of smaller signal events in this region of parameter space with BPD benchmark points, the discovery potential of the signal requires similar luminosity to that of BPC cases. \\~\\
Finally we present the discovery reach of the signal events in terms of significance $\sigma=\frac{S}{\sqrt{S+B}}$, where $S$ denotes signal events and $B$ denotes 
SM background events in terms of luminosity. This is shown in Fig.~\ref{fig:significance1}. This shows that the benchmark points that characterise the two component 
DM framework, can yield a visible signature  at high luminosity with $\mathcal{L}\sim 500 ~\rm {fb}^{-1}$  
depending on the charged scalar mass and its splitting with DM for the case $m_{H^\pm}-m_{H^0} > m_{W^\pm}$.
\section{Summary and Conclusions}
\label{sec:conclusion}
 We have studied a two component scalar DM model in presence of right handed neutrinos that address neutrino mass generation through type I seesaw. The DM components are (i) a singlet scalar and (ii) an inert scalar doublet, both studied 
 extensively as single component DM in literature. We show that the presence of second component enlarges the available parameter space significantly considering relic density, direct search and indirect search constraints. In particular, the inert scalar DM 
 will now be allowed in the so called `desert region': $\{m_W-550\}$ GeV. Also for singlet scalar, we can now revive it below TeV, which is otherwise discarded (except Higgs resonance) from direct search in single component framework. The
 results obtained for DM analysis crucially depends on DM-DM conversion, which have been demonstrated in details.  
 
 We also study the high scale perturbativity and vacuum stability of the Higgs potential by analysing two loop RGE $\beta$ functions. This in turn puts further constraints on the available DM parameter space of the model. One of the important conclusions 
 obtained are that the mass splitting of the charged scalar component to the corresponding DM component of inert doublet is crucially tamed depending on the absolute stability scale of the scalar potential, coming from the perturbativity 
 constraint on the quartic and Yukawa coupling. For example, we find :   
  \begin{itemize}
  \item  Validity scale ($\mu$) $\sim$ intermediate scale ($10^{10}$ GeV): $\Delta M= m_{H^\pm}-m_{H^0}$ $\sim \{8-106 \}$ GeV, 
 \item Validity scale ($\mu$) $\sim$ Planck scale ($10^{19}$ GeV):  $\Delta M=m_{H^\pm}-m_{H^0}$ $\sim \{14-51 \}$ GeV,
 \end{itemize}
 with RH neutrino mass $M_R=10^8$ GeV and Yukawa $\textrm{Tr}[Y_\nu^\dagger Y_\nu]=0.5$. The presence of RHNs in the model not only helps us addressing the neutrino masses but also controls the high scale validity of the model parameters, for example, low $\Delta M$ regions. This is how the neutrino and dark sector constraints 
 affect each other.
 
 Inert Higgs having charged scalars have collider detectability. We point out that the collider search prospect of the charged components 
 are not only limited to low DM masses ($<m_W$), but is open to a larger mass range in presence of the 
 second DM component, even after taking the high scale validity constraints. 
 We exemplified this at LHC for hadronically quiet dilepton channel with missing energy, where $\Delta M$, turns out to be a crucial kinematic 
 parameter, constrained from DM, high scale validity and neutrino sector. At LHC, due to its $t\bar{t}$ background, high $\Delta M$ 
 regions can be segregated from SM background more efficiently at the cost of small production cross section. 
 Low $\Delta M$ regions are more affected by SM background, although the signal cross-section is high. 
  Therefore the model can be probed within High Luminosity reach of LHC.
  On the other hand, $e^+e^-$ annihilation have better possibility to explore low 
 $\Delta M$ regions absent $t \bar{t}$ background. Here, we would like to comment that there are several studies that 
 have been done in this direction, but the high scale validity constraint may alter the conclusion significantly as we demonstrate. 
 
 Finally, we would like to mention that the analysis performed here, although focus on a specific model set up, but there are some 
 generic conclusions that can be borrowed. For example, if the two DM components 
 have sufficient interaction in between, the available parameter space will be enlarged significantly from both relic density and direct search. 
 The conversion of one DM into the other may also affect the collider outcome of the DM significantly. 
 It is obvious that richer signal is obtained when we have larger multiplets in dark sector (as scalar doublet produces two lepton 
 final state in the analysis). It is also possible that the dark sector and neutrino sector although may not 
 inherit a common origin, the high scale validity of the model can bring them together.
\paragraph*{Acknowledgments\,:} 
PG would like to thank Basabendu Barman and Rishav Roshan for useful discussions and also acknowledges MHRD, 
Government of India for research fellowship. SB would like to acknowledge the DST-INSPIRE research grant IFA13-PH-57 at IIT Guwahati. 
\appendix
\vspace{0.2cm}
\section{Tree Level Unitarity Constraints}\label{treeuni}
In this section, we perform the analysis to find the tree level unitarity limits
on quartic couplings present in our model at high energy. The scattering
amplitude for any 2 $\rightarrow$ 2 process can be expressed in terms
of the Legendre polynomial as \cite{Lee:1977eg}
\begin{align}
 \mathcal{M}^{2\rightarrow 2}= 16\pi
\sum_{l=0}^\infty
a_l(2l+1)P_l(\cos\theta),
\end{align}
where $\theta$ is the scattering angle and $P_l(\cos\theta)$ is the
Legendre polynomial of order $l$. In the high-energy limit,
only the s-wave ($l= 0$) partial amplitude $a_0$ will determine
the leading energy dependence of the scattering processes. The unitarity constraint  turns 
out to be \cite{Lee:1977eg,Horejsi:2005da,Bhattacharyya:2015nca}
\begin{align}
 {\rm Re}~|a_0|<\frac{1}{2}.
 \label{eq:pertT}
\end{align}
The constraint in Eqn.(\ref{eq:pertT}) can be further converted to a bound on
the scattering amplitude $\mathcal{M}$ \cite{Lee:1977eg,Horejsi:2005da,Bhattacharyya:2015nca}:
\begin{align}
 |\mathcal{M}| < 8\pi.
  \label{eq:pertT1}
\end{align}
In the present set up, we have multiple possible 2 $\rightarrow$ 2
scattering processes. Therefore, we need to construct
a matrix (${M}^{2\rightarrow 2}_{i;j}= \mathcal{M}_{i\rightarrow j}$) by considering all possible two particle
states. Finally, we  calculate the eigenvalues
of $\mathcal{M}$ and employ the bound as in Eqn. (\ref{eq:pertT1}).
In the high-energy limit, we express the SM Higgs
doublet as $H^T=\Big(w^+ ~ \frac{h+iz}{\sqrt2}\Big)$. Then, the scalar potential 
in Eqn.(\ref{eq:totP}) gives rise to 19 neutral combinations of two particle
states:
\begin{align}
w^+ w^-,~H^+ H^-,~\frac{h h}{\sqrt2},~\frac{z z}{\sqrt2},~\frac{H^0 H^0}{\sqrt2},~\frac{A^0 A^0}{\sqrt2},~\frac{\phi \phi }{\sqrt2},~ h ~z,~H^0~ A^0,~w^+ H^-,~H^+w^-, \nonumber \\
~h ~H^0,~h~ A^0,~ z~ H^0, ~z ~A^0,~h ~\phi,~ z~ \phi, ~H^0 ~\phi, ~A^0 ~\phi ~.
\end{align}
and 10 singly charged two-particle states:
\begin{align}
h ~w^+,~ z ~w^+,~ H^0 H^+,~A^0 H^+,~h ~H^+,~z~ H^+,~H^0~w^+,~A^0 ~w^+,~\phi~ w^+,~\phi ~H^+ ~.
\end{align}
Therefore, we can write the scattering amplitude matrix (M) in block-diagonal form by decomposing it into a neutral ($NC$) and singly
charged($SC$) sector as
\begin{align}
{{M}}=\left(
\begin{array}{cc}
  {M}^{NC}_{19\times 19}&  0  \\
 0 & {M}^{SC}_{10\times 10}  \\
\end{array}
\right).
 \label{eq:PU-Eigen}
\end{align}
where the sub-matrices are given by
\bea
{{{M}}^{NC}}_{19\times19}=\left(
\begin{array}{cccc}
 ({{M}}_{1}^{NC})_{7\times 7} & 0 & 0 & 0 \\
 0 & ({{M}}_{2}^{NC})_{2\times 2} & 0 & 0 \\
 0 & 0 & ({{M}}_{3}^{NC})_{6\times 6} & 0 \\
 0 & 0 & 0 & ({{M}}_{4}^{NC})_{4\times 4} \\
\end{array}
\right)
\eea
with
\bea
{{M}}_{1}^{NC}=\tiny{\left(
\begin{array}{ccccccc}
 4 \lambda_H & \lambda_1+\lambda_2+\lambda_3 & \sqrt{2} \lambda_H & \sqrt{2} \lambda_H & \frac{\lambda_1}{\sqrt{2}} & \frac{\lambda_1}{\sqrt{2}} & \frac{\lambda_{\phi h}}{\sqrt{2}} \\
 \lambda_1+\lambda_2+\lambda_3 & 4 \lambda_{\Phi}  & \frac{\lambda_1}{\sqrt{2}} & \frac{\lambda_1}{\sqrt{2}} & \sqrt{2} \lambda_{\Phi}  & \sqrt{2} \lambda_{\Phi}  & \frac{\lambda_{c}}{\sqrt{2}} \\
 \sqrt{2} \lambda_H & \frac{\lambda_1}{\sqrt{2}} & 3 \lambda_H & \lambda_H & 2 \left(\frac{\lambda_1}{4}+\frac{\lambda_2}{4}+\frac{\lambda_3}{4}\right) & 2 \left(\frac{\lambda_1}{4}+\frac{\lambda_2}{4}+\frac{\lambda_3}{4}\right) & \frac{\lambda_{\phi h}}{2} \\
 \sqrt{2} \lambda_H & \frac{\lambda_1}{\sqrt{2}} & \lambda_H & 3 \lambda_H & 2 \left(\frac{\lambda_1}{4}+\frac{\lambda_2}{4}+\frac{\lambda_3}{4}\right) & 2 \left(\frac{\lambda_1}{4}+\frac{\lambda_2}{4}+\frac{\lambda_3}{4}\right) & \frac{\lambda_{\phi h}}{2} \\
 \frac{\lambda_1}{\sqrt{2}} & \sqrt{2} \lambda_{\Phi}  & 2 \left(\frac{\lambda_1}{4}+\frac{\lambda_2}{4}+\frac{\lambda_3}{4}\right) & 2 \left(\frac{\lambda_1}{4}+\frac{\lambda_2}{4}+\frac{\lambda_3}{4}\right) & 3 \lambda_{\Phi}  & \lambda_{\Phi}  & \frac{\lambda_{c}}{2} \\
 \frac{\lambda_1}{\sqrt{2}} & \sqrt{2} \lambda_{\Phi}  & 2 \left(\frac{\lambda_1}{4}+\frac{\lambda_2}{4}+\frac{\lambda_3}{4}\right) & 2 \left(\frac{\lambda_1}{4}+\frac{\lambda_2}{4}+\frac{\lambda_3}{4}\right) & \lambda_{\Phi}  & 3 \lambda_{\Phi}  & \frac{\lambda_{c}}{2} \\
 \frac{\lambda_{\phi h}}{\sqrt{2}} & \frac{\lambda_{c}}{\sqrt{2}} & \frac{\lambda_{\phi h}}{2} & \frac{\lambda_{\phi h}}{2} & \frac{\lambda_{c}}{2} & \frac{\lambda_{c}}{2} & \frac{\lambda_{\phi} }{2} \\
\end{array}
\right)
}, \nonumber \\
\eea
\bea
{{M}}_{3}^{NC}=\tiny{\left(
\begin{array}{cccccc}
 \lambda_{1}+\lambda_{2}+\lambda_{3} & 0 & \frac{\lambda_{2}}{2}+\frac{\lambda_{3}}{2} & \frac{i \lambda_{2}}{2}+\frac{i \lambda_{3}}{2} & -\frac{i \lambda_{2}}{2}-\frac{i \lambda_{3}}{2} & \frac{\lambda_{2}}{2}+\frac{\lambda_{3}}{2} \\
 0 & \lambda_{1}+\lambda_{2}+\lambda_{3} & \frac{\lambda_{2}}{2}+\frac{\lambda_{3}}{2} & -\frac{i \lambda_{2}}{2}-\frac{i \lambda_{3}}{2} & \frac{i \lambda_{2}}{2}+\frac{i \lambda_{3}}{2} & \frac{\lambda_{2}}{2}+\frac{\lambda_{3}}{2} \\
 \frac{\lambda_{2}}{2}+\frac{\lambda_{3}}{2} & \frac{\lambda_{2}}{2}+\frac{\lambda_{3}}{2} & 4 \left(\frac{\lambda_{1}}{4}+\frac{\lambda_{2}}{4}+\frac{\lambda_{3}}{4}\right) & 0 & 0 & 0 \\
 -\frac{i \lambda_{2}}{2}-\frac{i \lambda_{3}}{2} & \frac{i \lambda_{2}}{2}+\frac{i \lambda_{3}}{2} & 0 & 4 \left(\frac{\lambda_{1}}{4}+\frac{\lambda_{2}}{4}+\frac{\lambda_{3}}{4}\right) & 0 & 0 \\
 \frac{i \lambda_{2}}{2}+\frac{i \lambda_{3}}{2} & -\frac{i \lambda_{2}}{2}-\frac{i \lambda_{3}}{2} & 0 & 0 & 4 \left(\frac{\lambda_{1}}{4}+\frac{\lambda_{2}}{4}+\frac{\lambda_{3}}{4}\right) & 0 \\
 \frac{\lambda_{2}}{2}+\frac{\lambda_{3}}{2} & \frac{\lambda_{2}}{2}+\frac{\lambda_{3}}{2} & 0 & 0 & 0 & 4 \left(\frac{\lambda_{1}}{4}+\frac{\lambda_{2}}{4}+\frac{\lambda_{3}}{4}\right) \\
\end{array}
\right)}, \nonumber \\
\eea
\bea
{{M}}_{2}^{NC}=\tiny{\left(
\begin{array}{cc}
 2 \lambda_H & 0 \\
 0 & 2 \lambda_\Phi  \\
\end{array}
\right)}, ~~~~~~~~
{{M}}_{4}^{NC}=\tiny{\left(
\begin{array}{cccc}
 \lph & 0 & 0 & 0 \\
 0 & \lph & 0 & 0 \\
 0 & 0 & \lc & 0 \\
 0 & 0 & 0 & \lc \\
\end{array}
\right)}.
\eea

and

\bea
{{{M}}^{SC}}=\tiny{\left(
\begin{array}{cccccccccc}
 2 \lambda_{H} & 0 & \frac{\lambda_{2}}{2}+\frac{\lambda_{3}}{2} & \frac{i \lambda_{2}}{2}+\frac{i \lambda_{3}}{2} & 0 & 0 & 0 & 0 & 0 & 0 \\
 0 & 2 \lambda_{H} & -\frac{i \lambda_{2}}{2}-\frac{i \lambda_{3}}{2} & \frac{\lambda_{2}}{2}+\frac{\lambda_{3}}{2} & 0 & 0 & 0 & 0 & 0 & 0 \\
 \frac{\lambda_{2}}{2}+\frac{\lambda_{3}}{2} & \frac{i \lambda_{2}}{2}+\frac{i \lambda_{3}}{2} & 2 \lambda_{\Phi}  & 0 & 0 & 0 & 0 & 0 & 0 & 0 \\
 -\frac{i \lambda_{2}}{2}-\frac{i \lambda_{3}}{2} & \frac{\lambda_{2}}{2}+\frac{\lambda_{3}}{2} & 0 & 2 \lambda_{\Phi}  & 0 & 0 & 0 & 0 & 0 & 0 \\
 0 & 0 & 0 & 0 & \lambda_{1} & 0 & \frac{\lambda_{2}}{2}+\frac{\lambda_{3}}{2} & -\frac{i \lambda_{2}}{2}-\frac{i \lambda_{3}}{2} & 0 & 0 \\
 0 & 0 & 0 & 0 & 0 & \lambda_{1} & \frac{i \lambda_{2}}{2}+\frac{i \lambda_{3}}{2} & \frac{\lambda_{2}}{2}+\frac{\lambda_{3}}{2} & 0 & 0 \\
 0 & 0 & 0 & 0 & \frac{\lambda_{2}}{2}+\frac{\lambda_{3}}{2} & -\frac{i \lambda_{2}}{2}-\frac{i \lambda_{3}}{2} & \lambda_{1} & 0 & 0 & 0 \\
 0 & 0 & 0 & 0 & \frac{i \lambda_{2}}{2}+\frac{i \lambda_{3}}{2} & \frac{\lambda_{2}}{2}+\frac{\lambda_{3}}{2} & 0 & \lambda_{1} & 0 & 0 \\
 0 & 0 & 0 & 0 & 0 & 0 & 0 & 0 & \lambda_{\phi h} & 0 \\
 0 & 0 & 0 & 0 & 0 & 0 & 0 & 0 & 0 & \lambda_{c} \\
\end{array}
\right)} \nonumber \\
\eea
After determining the eigenvalues of Eqn.(\ref{eq:PU-Eigen}) we conclude that tree level unitarity constraints in this set up are following:
\begin{align}
  |\lambda_H|<4\pi,~~~|\lambda_\Phi|< 4 \pi, \nonumber \\
 |\lc|<8\pi,~~~|\lph|< 8 \pi,\nonumber \\
 |\lambda_1|< 8 \pi,~~~|\lambda_1+2(\lambda_2+\lambda_3)|< 8 \pi \nonumber \\
  |\lambda_1+\lambda_2+\lambda_3| < 8\pi, ~~|\lambda_1-\lambda_2-\lambda_3| < 8\pi,\nonumber \\ 
 |(\lambda_\Phi + \lambda_H ) \pm \sqrt{(\lambda_2+\lambda_3)^2 + (\lambda_H-\lambda_\Phi)^2}| < 8 \pi, \nonumber \\
 ~\text{and}~~~ |x_{1,2,3}| < 16 \pi 
\end{align}
where $x_{1,2,3}$ be the roots of following cubic equation
\begin{eqnarray}\label{unitarity_eqn}
&& x^3+x^2 (-12 \lambda_{H}-12 \lambda_{\Phi} -\lambda_{\phi} )+x (-16 \lambda_{1}^2-16 \lambda_{1} \lambda_{2}-16 \lambda_{1} \lambda_{3}-4 \lambda_{2}^2-8 \lambda_{2} \lambda_{3} \nonumber \\
 &&~~~-4 \lambda_{3}^2-4 \lambda_{c}^2+144 \lambda_{H} \lambda_{\Phi} +12 \lambda_{H} \lambda_{\phi} +12 \lambda_{\Phi}  \lambda_{\phi} -4 \lambda_{\phi h}^2 )+16 \lambda_{1}^2 \lambda_{\phi} +16 \lambda_{1} \lambda_{2} \lambda_{\phi} \nonumber \\
 &&~~ +16 \lambda_{1} \lambda_{3} \lambda_{\phi} -32 \lambda_{1} \lambda_{c} \lambda_{\phi h}+4 \lambda_{2}^2 \lambda_{\phi} +8 \lambda_{2} \lambda_{3} \lambda_{\phi} -16 \lambda_{2} \lambda_{c} \lambda_{\phi h}+4 \lambda_{3}^2 \lambda_{\phi} \nonumber \\
 &&~~~-16 \lambda_{3} \lambda_{c} \lambda_{\phi h}+48 \lambda_{c}^2 \lambda_{H}-144 \lambda_{H} \lambda_{\Phi}  \lambda_{\phi} +48 \lambda_{\Phi}  \lambda_{\phi h}^2 ~= 0
\end{eqnarray}
\section{High Scale Validity of Single component DM models}\label{sec:HSV_single}
\begin{figure}[htb!]
$$
 \includegraphics[height=4.5cm]{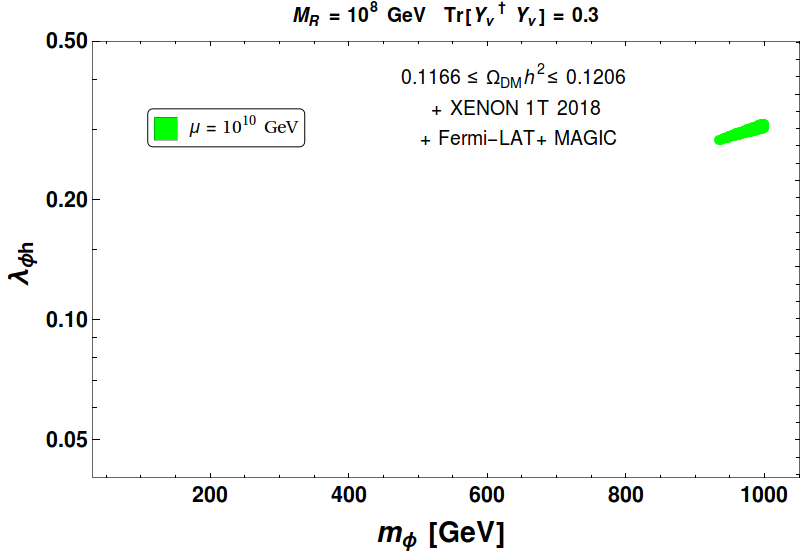}~
 \includegraphics[height=4.5cm]{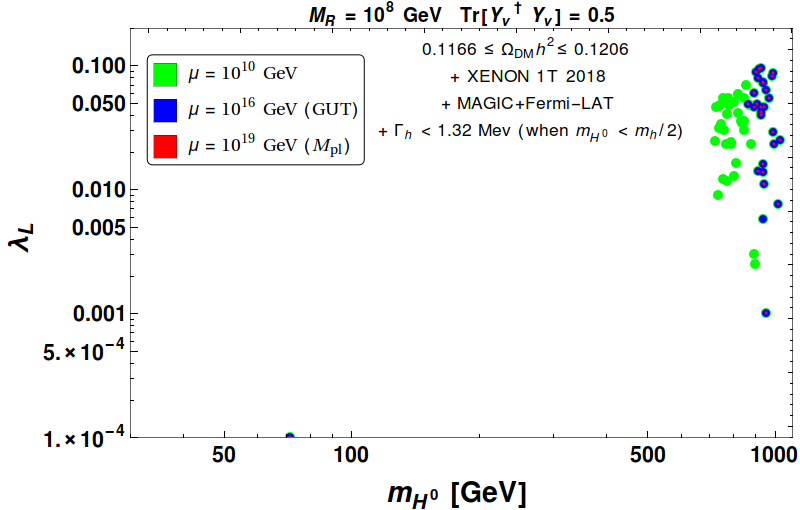}
 $$
 \caption{Parameter space scan for (left) scalar singlet DM and (right) IDM cosnidering satisfaction of relic density bound, direct detection cross section limit, indirect cross-section limit and high scale validity in presence of RH neutrinos. }
 \label{fig:Sepcases}
\end{figure}

Here we show the allowed parameter space for single component DM  when the vacuum stability conditions are taken into account in presence of RH neutrinos. In left panel of Fig. \ref{fig:Sepcases} the parameter space for scalar singlet DM ($\phi$) is shown considering $M_R=10^8$ GeV and Tr$[Y_\nu^\dagger Y_\nu]=0.3$, while in right panel we present the same for IDM ($H^0$) with $M_R=10^8$ GeV and Tr$[Y_\nu^\dagger Y_\nu]=0.5$. We see that for $\phi$, the DM mass is allowed beyond 900 GeV considering the absolute stability of the EW vacuum upto $10^{10}$ GeV \cite{Ghosh:2017fmr}. However for IDM, we notice that the absolute EW vacuum stability can be extended even upto Planck scale (with similar Yukawa) due to the presence of several scalar degrees of freedom.
\newpage
\section{Available parameters of the model from high scale validity for different choices of RH Neutrino mass and Yukawa coupling}
\label{sec:Allrange}
Here, we would like to show the available parameter space of the two component DM model with RH neutrinos, viable from high scale validity constraint after choosing different possible RH neutrino mass and 
Neutrino Yukawa coupling together with relic density, direct search and indirect search bound. In the main text, we elaborated only the case of $\textrm{Tr}[Y_\nu^\dagger Y_\nu]=0.5$ with RH neutrino mass of $10^8$ GeV. 
We show different possibilities in two tables, in Table \ref{TabSE1}, we depict the case of $m_\phi > \mH$ and in Table \ref{TabSE2}, the reverse hierarchy $m_\phi \le \mH$ is presented. 

\vspace{4mm}
\begin{table}[H]
 \resizebox{\linewidth}{!}{
 \begin{tabular}{|c|c|c|c|c|c|}
  \hline
  \multicolumn{1}{|c}{$~m_{\phi} > ~m_{H^0}$ (in GeV)}& \multicolumn{2}{|c}{RH Neutrinos} & \multicolumn{3}{|c|}{ Relic + DD (XENON 1T)+ ID (Fermi LAT+MAGIC) + Stability + Perturbativity}  \\ \hline
  Relic + DD (XENON 1T)+ID (Fermi LAT+MAGIC) & $M_R$ & $\textrm{Tr}[Y_\nu^\dagger Y_\nu]$ &  ~~$\mu = 10^{10}$ GeV ~~&  $\mu = 10^{16}$ GeV (GUT) & $ \mu = {10^{19}} {\rm GeV}  (M_{\rm pl})$        \\ \hline
  \multirow{9}{*}{ \makecell{ $m_{H^0} \sim \{85~-~500\} $\\
                              $m_{\phi} \sim \{170~-~500\} $\\ 
                             $\Delta m \equiv  m_{A^0}-m_{H^0} \sim \{0~ - ~200\} $  \\  
                              $\Delta M \equiv m_{H^\pm}-m_{H^0} \sim \{4~-~360\} $  \\ 
                              $\lambda_{c} \sim \{0.08~-~1.00\}$ \\
                              $\lambda_{L} \sim \{0.001~-~0.28\}$ \\
                              $\lambda_{\phi h} \sim \{0.001~-~0.14\} $ } }
                        & \multirow{2}{*}{\makecell{$10^4$ \\ GeV}}
                                &       0.1                          &  \makecell{$m_{H^0} \sim \{88~-~500\} $\\
                                                                                   $m_{\phi} \sim \{170~-~500\} $ \\
                                                                                    $\Delta m \sim \{0~-~90\} $   
                                                                                                    \\   $\Delta M  \sim \{5~-~106\} $ 
                                                                                                    \\ $\lambda_{c} \sim \{0.086~-~0.70\}$ 
                                                                                                     \\$\lambda_{L} \sim \{0.001~-~0.09\}$ 
                                                                                                     \\$\lambda_{\phi h} \sim \{0.001~-~0.14\} $ }   
                                                                                    & \makecell{$m_{H^0} \sim \{94~-~496\} $\\
                                                                                   $m_{\phi} \sim \{220~-~500\} $ \\
                                                                                    $\Delta m \sim \{0~-~52\} $   
                                                                                                    \\   $\Delta M  \sim \{6~-~70\} $ 
                                                                                                    \\ $\lambda_{c} \sim \{0.10~-~0.52\}$ 
                                                                                                     \\$\lambda_{L} \sim \{0.001~-~0.054\}$ 
                                                                                                     \\$\lambda_{\phi h} \sim \{0.001~-~0.14\} $ }   
                                                                                               &\makecell{$m_{H^0} \sim \{94~-~496\} $\\
                                                                                   $m_{\phi} \sim \{220~-~500\} $ \\
                                                                                    $\Delta m \sim \{0~-~38\} $   
                                                                                                    \\   $\Delta M  \sim \{6~-~55\} $ 
                                                                                                    \\ $\lambda_{c} \sim \{0.10~-~0.52\}$ 
                                                                                                        \\$\lambda_{L} \sim \{0.001~-~0.046\}$ 
                                                                                                     \\$\lambda_{\phi h} \sim \{0.001~-~0.14\} $ }   
                                                                                                                                      \\ \cline{3-6}
                        &       &       0.9                          & \makecell{$m_{H^0} \sim \{89~-~493\} $\\
                                                                                   $m_{\phi} \sim \{258~-~500\} $ \\
                                                                                    $\Delta m \sim \{0~-~87\} $   
                                                                                                    \\   $\Delta M  \sim \{21~-~106\} $ 
                                                                                                    \\ $\lambda_{c} \sim \{0.086~-~0.69\}$ 
                                                                                                     \\$\lambda_{L} \sim \{0.001~-~0.09\}$ 
                                                                                                     \\$\lambda_{\phi h} \sim \{0.001~-~0.11\} $ }  
                                                                                 &    \makecell{No parameter space \\ avialable}        &      \makecell{No parameter space \\ avialable}                             \\ \cline{2-6}
                        & \multirow{2}{*}{\makecell{$10^8$ \\ GeV}}
                                &       0.1                          & \makecell{$m_{H^0} \sim \{85~-~496\} $\\
                                                                                   $m_{\phi} \sim \{171~-~500\} $ \\
                                                                                    $\Delta m \sim \{0~-~87\} $   
                                                                                                    \\   $\Delta M  \sim \{4~-~123\} $ 
                                                                                                    \\ $\lambda_{c} \sim \{0.086~-~0.69\}$ 
                                                                                                     \\$\lambda_{L} \sim \{0.001~-~0.09\}$ 
                                                                                                     \\$\lambda_{\phi h} \sim \{0.001~-~0.14\} $ }   
                                                                                  &  \makecell{$m_{H^0} \sim \{94~-~496\} $\\
                                                                                   $m_{\phi} \sim \{221~-~500\} $ \\
                                                                                    $\Delta m \sim \{0~-~52\} $   
                                                                                                    \\   $\Delta M  \sim \{6~-~69\} $ 
                                                                                                    \\ $\lambda_{c} \sim \{0.10~-~0.52\}$ 
                                                                                                     \\$\lambda_{L} \sim \{0.001~-~0.06\}$ 
                                                                                                     \\$\lambda_{\phi h} \sim \{0.001~-~0.14\} $ }   
                                                                                                &  \makecell{$m_{H^0} \sim \{94~-~496\} $\\
                                                                                   $m_{\phi} \sim \{221~-~500\} $ \\
                                                                                    $\Delta m \sim \{0~-~38\} $   
                                                                                                    \\   $\Delta M  \sim \{6~-~55\} $ 
                                                                                                    \\ $\lambda_{c} \sim \{0.10~-~0.52\}$ 
                                                                                                     \\$\lambda_{L} \sim \{0.001~-~0.046\}$ 
                                                                                                     \\$\lambda_{\phi h} \sim \{0.001~-~0.14\} $ }             
                                                                                                                               \\ \cline{3-6}
                        &       &       0.9                          &  \makecell{$m_{H^0} \sim \{89~-~493\} $\\
                                                                                   $m_{\phi} \sim \{171~-~500\} $ \\
                                                                                    $\Delta m \sim \{0~-~87\} $   
                                                                                                    \\   $\Delta M  \sim \{17~-~106\} $ 
                                                                                                    \\ $\lambda_{c} \sim \{0.086~-~0.69\}$ 
                                                                                                     \\$\lambda_{L} \sim \{0.001~-~0.09\}$ 
                                                                                                     \\$\lambda_{\phi h} \sim \{0.001~-~0.10\} $ }     
                                                                                     &  \makecell{$m_{H^0} \sim \{98~-~432\} $\\
                                                                                   $m_{\phi} \sim \{317~-~498\} $ \\
                                                                                    $\Delta m \sim \{1~-~52\} $   
                                                                                                    \\   $\Delta M  \sim \{19~-~69\} $ 
                                                                                                    \\ $\lambda_{c} \sim \{0.10~-~0.40\}$ 
                                                                                                     \\$\lambda_{L} \sim \{0.001~-~0.046\}$ 
                                                                                                     \\$\lambda_{\phi h} \sim \{0.001~-~0.096\} $ }     
                                                                                                &         \makecell{No parameter space \\ avialable}   \\
                                                                                                                             \cline{2-6}
                        & \multirow{2}{*}{\makecell{$10^{12}$ \\ GeV}}
                                &       0.1                     &  \makecell{No effective contribution \\from \\RH Neutrinos}                    &  \makecell{$m_{H^0} \sim \{94~-~496\} $\\
                                                                                   $m_{\phi} \sim \{221~-~500\} $ \\
                                                                                    $\Delta m \sim \{0~-~52\} $   
                                                                                                    \\   $\Delta M  \sim \{6~-~69\} $ 
                                                                                                    \\ $\lambda_{c} \sim \{0.10~-~0.52\}$ 
                                                                                                     \\$\lambda_{L} \sim \{0.001~-~0.06\}$ 
                                                                                                     \\$\lambda_{\phi h} \sim \{0.001~-~0.14\} $ }      
                                                                                                     &    \makecell{$m_{H^0} \sim \{94~-~496\} $\\
                                                                                   $m_{\phi} \sim \{221~-~500\} $ \\
                                                                                    $\Delta m \sim \{0~-~38\} $   
                                                                                                    \\   $\Delta M  \sim \{6~-~55\} $ 
                                                                                                    \\ $\lambda_{c} \sim \{0.10~-~0.52\}$ 
                                                                                                     \\$\lambda_{L} \sim \{0.001~-~0.046\}$ 
                                                                                                     \\$\lambda_{\phi h} \sim \{0.001~-~0.14\} $ }      
        
                                                                                                                   \\ \cline{3-6}
                        &       &       0.9                  &   \makecell{No effective contribution \\from \\RH Neutrinos}                    &   \makecell{$m_{H^0} \sim \{98~-~493\} $\\
                                                                                   $m_{\phi} \sim \{248~-~498\} $ \\
                                                                                    $\Delta m \sim \{0~-~52\} $   
                                                                                                    \\   $\Delta M  \sim \{17~-~69\} $ 
                                                                                                    \\ $\lambda_{c} \sim \{0.10~-~0.52\}$ 
                                                                                                     \\$\lambda_{L} \sim \{0.001~-~0.054\}$ 
                                                                                                     \\$\lambda_{\phi h} \sim \{0.001~-~0.096\} $ } 
                                                                                                 &    \makecell{$m_{H^0} \sim \{129~-~493\} $\\
                                                                                   $m_{\phi} \sim \{271~-~495\} $ \\
                                                                                    $\Delta m \sim \{0~-~37\} $   
                                                                                                    \\   $\Delta M  \sim \{17~-~55\} $ 
                                                                                                    \\ $\lambda_{c} \sim \{0.10~-~0.25\}$ 
                                                                                                     \\$\lambda_{L} \sim \{0.003~-~0.046\}$ 
                                                                                                     \\$\lambda_{\phi h} \sim \{0.007~-~0.092\} $ }           \\ \hline                                                                                        
 \end{tabular}
 }
 \caption{Allowed ranges of relevant parameters considering $m_\phi>m_{H^0}$ for different values of RH neutrino mass and Tr[$Y_\nu^\dagger Y_\nu]$.}
 \label{TabSE1}
\end{table}
\newpage
\begin{table}[H]
 \resizebox{\linewidth}{!}{
 \begin{tabular}{|c|c|c|c|c|c|}
  \hline
  \multicolumn{1}{|c}{$~m_{\phi} \leq ~m_{H^0}$ (in GeV)}& \multicolumn{2}{|c}{RH Neutrinos} & \multicolumn{3}{|c|}{ Relic + DD (XENON 1T) +ID (Fermi LAT+MAGIC)+ Stability + Perturbativity} \\ \hline
  Relic + DD (XENON 1T)+ID (Fermi LAT+MAGIC) & $M_R$ & $\textrm{Tr}[Y_\nu^\dagger Y_\nu]$ &  ~~$\mu = 10^{10}$ GeV ~~&  $\mu = 10^{16}$ GeV (GUT) & $ \mu = {10^{19}} {\rm GeV}  (M_{\rm pl})$        \\ \hline
  \multirow{9}{*}{ \makecell{ $m_{H^0} \sim \{100~-~500\} $\\
                              $m_{\phi} \sim \{96~-~500\} $\\ 
                             $\Delta m \equiv  m_{A^0}-m_{H^0} \sim \{0~ - ~200\} $  \\  
                              $\Delta M \equiv m_{H^\pm}-m_{H^0} \sim \{12~-~300\} $  \\ 
                              $\lambda_{c} \sim \{0.30~-~1.00\}$ \\
                              $\lambda_{L} \sim \{0.001~-~0.30\}$ \\
                              $\lambda_{\phi h} \sim \{0.001~-~0.11\} $ } }
                        & \multirow{2}{*}{\makecell{$10^4$\\ GeV}}
                                &       0.1                          &  \makecell{$m_{H^0} \sim \{ 100~-~479\} $\\
                                                                                   $m_{\phi} \sim \{96~-~471\} $ \\
                                                                                    $\Delta m \sim \{0~-~57\} $   
                                                                                                    \\   $\Delta M  \sim \{12~-~78\} $ 
                                                                                                    \\ $\lambda_{c} \sim \{0.34~-~0.98\}$ 
                                                                                                     \\$\lambda_{L} \sim \{0.001~-~0.127\}$ 
                                                                                                     \\$\lambda_{\phi h} \sim \{0.001~-~0.096\} $ }   
                                                                                    & \makecell{$m_{H^0} \sim \{205~-~452\} $\\
                                                                                   $m_{\phi} \sim \{201~-~451\} $ \\
                                                                                    $\Delta m \sim \{5~-~11\} $   
                                                                                                    \\   $\Delta M  \sim \{12~-~29\} $ 
                                                                                                    \\ $\lambda_{c} \sim \{0.43~-~0.81\}$ 
                                                                                                     \\$\lambda_{L} \sim \{0.008~-~0.043\}$ 
                                                                                                     \\$\lambda_{\phi h} \sim \{0.002~-~0.043\} $ }   
                                                                                               &\makecell{$m_{H^0} \sim \{205~-~225\} $\\
                                                                                   $m_{\phi} \sim \{201~-~224\} $ \\
                                                                                    $\Delta m \sim \{5~-~11\} $   
                                                                                                    \\   $\Delta M  \sim \{16~-~29\} $ 
                                                                                                    \\ $\lambda_{c} \sim \{0.43~-~0.59\}$ 
                                                                                                     \\$\lambda_{L} \sim \{0.024~-~0.025\}$ 
                                                                                                     \\$\lambda_{\phi h} \sim \{0.002~-~0.016\} $ }   
                                                                                                                                      \\ \cline{3-6}
                        &       &       0.9                          & \makecell{$m_{H^0} \sim \{170~-~455\} $\\
                                                                                   $m_{\phi} \sim \{168~-~450\} $ \\
                                                                                    $\Delta m \sim \{0~-~57\} $   
                                                                                                    \\   $\Delta M  \sim \{26~-~67\} $ 
                                                                                                    \\ $\lambda_{c} \sim \{0.34~-~0.98\}$ 
                                                                                                     \\$\lambda_{L} \sim \{0.001~-~0.054\}$ 
                                                                                                     \\$\lambda_{\phi h} \sim \{0.004~-~0.086\} $ }  
                                                                                 &    \makecell{No parameter space \\ available}        &      \makecell{No parameter space \\ avialable}                             \\ \cline{2-6}
                        & \multirow{2}{*}{\makecell{$10^8$\\GeV}}
                                &       0.1                          & \makecell{$m_{H^0} \sim \{100~-~479\} $\\
                                                                                   $m_{\phi} \sim \{96~-~471\} $ \\
                                                                                    $\Delta m \sim \{0~-~57\} $   
                                                                                                    \\   $\Delta M  \sim \{12~-~78\} $ 
                                                                                                    \\ $\lambda_{c} \sim \{0.34~-~0.98\}$ 
                                                                                                     \\$\lambda_{L} \sim \{0.001~-~0.127\}$ 
                                                                                                     \\$\lambda_{\phi h} \sim \{0.001~-~0.096\} $ }   
                                                                                  &  \makecell{$m_{H^0} \sim \{126~-~452\} $\\
                                                                                   $m_{\phi} \sim \{122~-~451\} $ \\
                                                                                    $\Delta m \sim \{0~-~19\} $   
                                                                                                    \\   $\Delta M  \sim \{12~-~29\} $ 
                                                                                                    \\ $\lambda_{c} \sim \{0.43~-~0.81\}$ 
                                                                                                     \\$\lambda_{L} \sim \{0.004~-~0.043\}$ 
                                                                                                     \\$\lambda_{\phi h} \sim \{0.002~-~0.054\} $ }   
                                                                                    &  \makecell{$m_{H^0} \sim \{126~-~224\} $\\
                                                                                   $m_{\phi} \sim \{122~-~224\} $ \\
                                                                                    $\Delta m \sim \{0~-~19\} $   
                                                                                                    \\   $\Delta M  \sim \{16~-~29\} $ 
                                                                                                    \\ $\lambda_{c} \sim \{0.43~-~0.61\}$ 
                                                                                                     \\$\lambda_{L} \sim \{0.004~-~0.025\}$ 
                                                                                                     \\$\lambda_{\phi h} \sim \{0.002~-~0.016\} $ }             
                                                                                                                               \\ \cline{3-6}
                        &       &       0.9                          &  \makecell{$m_{H^0} \sim \{100~-~479\} $\\
                                                                                   $m_{\phi} \sim \{96~-~471\} $ \\
                                                                                    $\Delta m \sim \{0~-~57\} $   
                                                                                                    \\   $\Delta M  \sim \{19~-~78\} $ 
                                                                                                    \\ $\lambda_{c} \sim \{0.34~-~0.98\}$ 
                                                                                                     \\$\lambda_{L} \sim \{0.001~-~0.127\}$ 
                                                                                                     \\$\lambda_{\phi h} \sim \{0.001~-~0.096\} $ }     
                                                                                     &  \makecell{No parameter space \\ available }     
                                                                                                &         \makecell{No parameter space \\ available}   \\
                                                                                                                             \cline{2-6}
                        & \multirow{2}{*}{\makecell{$10^{12}$\\GeV}}
                                &       0.1                 &  \makecell{No effective contribution \\from \\RH Neutrinos}    &  \makecell{$m_{H^0} \sim \{126~-~452\} $\\
                                                                                   $m_{\phi} \sim \{122~-~451\} $ \\
                                                                                    $\Delta m \sim \{5~-~19\} $   
                                                                                                    \\   $\Delta M  \sim \{12~-~29\} $ 
                                                                                                    \\ $\lambda_{c} \sim \{0.43~-~0.81\}$ 
                                                                                                     \\$\lambda_{L} \sim \{0.004~-~0.043\}$ 
                                                                                                     \\$\lambda_{\phi h} \sim \{0.002~-~0.054\} $ }      
                                                                                        &    \makecell{$m_{H^0} \sim \{126~-~225\} $\\
                                                                                   $m_{\phi} \sim \{122~-~224\} $ \\
                                                                                    $\Delta m \sim \{5~-~19\} $   
                                                                                                    \\   $\Delta M  \sim \{16~-~29\} $ 
                                                                                                    \\ $\lambda_{c} \sim \{0.43~-~0.61\}$ 
                                                                                                     \\$\lambda_{L} \sim \{0.004~-~0.025\}$ 
                                                                                                     \\$\lambda_{\phi h} \sim \{0.002~-~0.016\} $ }      
        
                                                                                                                   \\ 
\hline
 \end{tabular}
 }
 \caption{Allowed ranges of relevant parameters considering $m_\phi \leq m_{H^0}$ for different values of RH neutrino mass and Tr[$Y_\nu^\dagger Y_\nu]$.}
 \label{TabSE2}
\end{table}

\bibliographystyle{JHEP}
\bibliography{ref.bib}

\providecommand{\href}[2]{#2}\begingroup\raggedright\begin{thebibliography}{100}

\bibitem{Chatrchyan:2012xdj}
{\scshape CMS} collaboration, \emph{{Observation of a new boson at a mass of
  125 GeV with the CMS experiment at the LHC}},
  \href{https://doi.org/10.1016/j.physletb.2012.08.021}{\emph{Phys. Lett.}
  {\bfseries B716} (2012) 30}
  [\href{https://arxiv.org/abs/1207.7235}{{\ttfamily 1207.7235}}].

\bibitem{Aad:2012tfa}
{\scshape ATLAS} collaboration, \emph{{Observation of a new particle in the
  search for the Standard Model Higgs boson with the ATLAS detector at the
  LHC}}, \href{https://doi.org/10.1016/j.physletb.2012.08.020}{\emph{Phys.
  Lett.} {\bfseries B716} (2012) 1}
  [\href{https://arxiv.org/abs/1207.7214}{{\ttfamily 1207.7214}}].

\bibitem{Alekhin:2012py}
S.~Alekhin, A.~Djouadi and S.~Moch, \emph{{The top quark and Higgs boson masses
  and the stability of the electroweak vacuum}},
  \href{https://doi.org/10.1016/j.physletb.2012.08.024}{\emph{Phys. Lett.}
  {\bfseries B716} (2012) 214}
  [\href{https://arxiv.org/abs/1207.0980}{{\ttfamily 1207.0980}}].

\bibitem{Buttazzo:2013uya}
D.~Buttazzo, G.~Degrassi, P.~P. Giardino, G.~F. Giudice, F.~Sala, A.~Salvio
  et~al., \emph{{Investigating the near-criticality of the Higgs boson}},
  \href{https://doi.org/10.1007/JHEP12(2013)089}{\emph{JHEP} {\bfseries 12}
  (2013) 089} [\href{https://arxiv.org/abs/1307.3536}{{\ttfamily 1307.3536}}].

\bibitem{Isidori:2001bm}
G.~Isidori, G.~Ridolfi and A.~Strumia, \emph{{On the metastability of the
  standard model vacuum}},
  \href{https://doi.org/10.1016/S0550-3213(01)00302-9}{\emph{Nucl. Phys.}
  {\bfseries B609} (2001) 387}
  [\href{https://arxiv.org/abs/hep-ph/0104016}{{\ttfamily hep-ph/0104016}}].

\bibitem{Anchordoqui:2012fq}
L.~A. Anchordoqui, I.~Antoniadis, H.~Goldberg, X.~Huang, D.~Lust, T.~R. Taylor
  et~al., \emph{{Vacuum Stability of Standard Model$^{++}$}},
  \href{https://doi.org/10.1007/JHEP02(2013)074}{\emph{JHEP} {\bfseries 02}
  (2013) 074} [\href{https://arxiv.org/abs/1208.2821}{{\ttfamily 1208.2821}}].

\bibitem{Tang:2013bz}
Y.~Tang, \emph{{Vacuum Stability in the Standard Model}},
  \href{https://doi.org/10.1142/S0217732313300024}{\emph{Mod. Phys. Lett.}
  {\bfseries A28} (2013) 1330002}
  [\href{https://arxiv.org/abs/1301.5812}{{\ttfamily 1301.5812}}].

\bibitem{Ellis:2009tp}
J.~Ellis, J.~R. Espinosa, G.~F. Giudice, A.~Hoecker and A.~Riotto, \emph{{The
  Probable Fate of the Standard Model}},
  \href{https://doi.org/10.1016/j.physletb.2009.07.054}{\emph{Phys. Lett.}
  {\bfseries B679} (2009) 369}
  [\href{https://arxiv.org/abs/0906.0954}{{\ttfamily 0906.0954}}].

\bibitem{EliasMiro:2011aa}
J.~Elias-Miro, J.~R. Espinosa, G.~F. Giudice, G.~Isidori, A.~Riotto and
  A.~Strumia, \emph{{Higgs mass implications on the stability of the
  electroweak vacuum}},
  \href{https://doi.org/10.1016/j.physletb.2012.02.013}{\emph{Phys. Lett.}
  {\bfseries B709} (2012) 222}
  [\href{https://arxiv.org/abs/1112.3022}{{\ttfamily 1112.3022}}].

\bibitem{PhysRevD.98.030001}
{\scshape Particle Data Group} collaboration, \emph{Review of particle
  physics}, \href{https://doi.org/10.1103/PhysRevD.98.030001}{\emph{Phys. Rev.
  D} {\bfseries 98} (2018) 030001}.

\bibitem{EliasMiro:2012ay}
J.~Elias-Miro, J.~R. Espinosa, G.~F. Giudice, H.~M. Lee and A.~Strumia,
  \emph{{Stabilization of the Electroweak Vacuum by a Scalar Threshold
  Effect}}, \href{https://doi.org/10.1007/JHEP06(2012)031}{\emph{JHEP}
  {\bfseries 06} (2012) 031} [\href{https://arxiv.org/abs/1203.0237}{{\ttfamily
  1203.0237}}].

\bibitem{Lebedev:2012zw}
O.~Lebedev, \emph{{On Stability of the Electroweak Vacuum and the Higgs
  Portal}}, \href{https://doi.org/10.1140/epjc/s10052-012-2058-2}{\emph{Eur.
  Phys. J.} {\bfseries C72} (2012) 2058}
  [\href{https://arxiv.org/abs/1203.0156}{{\ttfamily 1203.0156}}].

\bibitem{Spergel:2006hy}
{\scshape WMAP} collaboration, \emph{{Wilkinson Microwave Anisotropy Probe
  (WMAP) three year results: implications for cosmology}},
  \href{https://doi.org/10.1086/513700}{\emph{Astrophys. J. Suppl.} {\bfseries
  170} (2007) 377} [\href{https://arxiv.org/abs/astro-ph/0603449}{{\ttfamily
  astro-ph/0603449}}].

\bibitem{Aghanim:2018eyx}
{\scshape Planck} collaboration, \emph{{Planck 2018 results. VI. Cosmological
  parameters}},  \href{https://arxiv.org/abs/1807.06209}{{\ttfamily
  1807.06209}}.

\bibitem{Roszkowski:2017nbc}
L.~Roszkowski, E.~M. Sessolo and S.~Trojanowski, \emph{{WIMP dark matter
  candidates and searches—current status and future prospects}},
  \href{https://doi.org/10.1088/1361-6633/aab913}{\emph{Rept. Prog. Phys.}
  {\bfseries 81} (2018) 066201}
  [\href{https://arxiv.org/abs/1707.06277}{{\ttfamily 1707.06277}}].

\bibitem{Silveira:1985rk}
V.~Silveira and A.~Zee, \emph{{SCALAR PHANTOMS}},
  \href{https://doi.org/10.1016/0370-2693(85)90624-0}{\emph{Phys. Lett.}
  {\bfseries 161B} (1985) 136}.

\bibitem{McDonald:1993ex}
J.~McDonald, \emph{{Gauge singlet scalars as cold dark matter}},
  \href{https://doi.org/10.1103/PhysRevD.50.3637}{\emph{Phys. Rev.} {\bfseries
  D50} (1994) 3637} [\href{https://arxiv.org/abs/hep-ph/0702143}{{\ttfamily
  hep-ph/0702143}}].

\bibitem{Cline:2013gha}
J.~M. Cline, K.~Kainulainen, P.~Scott and C.~Weniger, \emph{{Update on scalar
  singlet dark matter}}, \href{https://doi.org/10.1103/PhysRevD.92.039906,
  10.1103/PhysRevD.88.055025}{\emph{Phys. Rev.} {\bfseries D88} (2013) 055025}
  [\href{https://arxiv.org/abs/1306.4710}{{\ttfamily 1306.4710}}].

\bibitem{Guo:2010hq}
W.-L. Guo and Y.-L. Wu, \emph{{The Real singlet scalar dark matter model}},
  \href{https://doi.org/10.1007/JHEP10(2010)083}{\emph{JHEP} {\bfseries 10}
  (2010) 083} [\href{https://arxiv.org/abs/1006.2518}{{\ttfamily 1006.2518}}].

\bibitem{Feng:2014vea}
L.~Feng, S.~Profumo and L.~Ubaldi, \emph{{Closing in on singlet scalar dark
  matter: LUX, invisible Higgs decays and gamma-ray lines}},
  \href{https://doi.org/10.1007/JHEP03(2015)045}{\emph{JHEP} {\bfseries 03}
  (2015) 045} [\href{https://arxiv.org/abs/1412.1105}{{\ttfamily 1412.1105}}].

\bibitem{Bhattacharya:2016qsg}
S.~Bhattacharya, S.~Jana and S.~Nandi, \emph{{Neutrino Masses and Scalar
  Singlet Dark Matter}},
  \href{https://doi.org/10.1103/PhysRevD.95.055003}{\emph{Phys. Rev.}
  {\bfseries D95} (2017) 055003}
  [\href{https://arxiv.org/abs/1609.03274}{{\ttfamily 1609.03274}}].

\bibitem{Casas:2017jjg}
J.~A. Casas, D.~G. Cerdeño, J.~M. Moreno and J.~Quilis, \emph{{Reopening the
  Higgs portal for single scalar dark matter}},
  \href{https://doi.org/10.1007/JHEP05(2017)036}{\emph{JHEP} {\bfseries 05}
  (2017) 036} [\href{https://arxiv.org/abs/1701.08134}{{\ttfamily
  1701.08134}}].

\bibitem{Bhattacharya:2017fid}
S.~Bhattacharya, P.~Ghosh, T.~N. Maity and T.~S. Ray, \emph{{Mitigating Direct
  Detection Bounds in Non-minimal Higgs Portal Scalar Dark Matter Models}},
  \href{https://doi.org/10.1007/JHEP10(2017)088}{\emph{JHEP} {\bfseries 10}
  (2017) 088} [\href{https://arxiv.org/abs/1706.04699}{{\ttfamily
  1706.04699}}].

\bibitem{Bhattacharya:2019mmy}
S.~Bhattacharya, P.~Ghosh and S.~Verma, \emph{{SIMPler realisation of Scalar
  Dark Matter}},  \href{https://arxiv.org/abs/1904.07562}{{\ttfamily
  1904.07562}}.

\bibitem{Akerib:2016vxi}
{\scshape LUX} collaboration, \emph{{Results from a search for dark matter in
  the complete LUX exposure}},
  \href{https://doi.org/10.1103/PhysRevLett.118.021303}{\emph{Phys. Rev. Lett.}
  {\bfseries 118} (2017) 021303}
  [\href{https://arxiv.org/abs/1608.07648}{{\ttfamily 1608.07648}}].

\bibitem{Aprile:2017iyp}
{\scshape XENON} collaboration, \emph{{First Dark Matter Search Results from
  the XENON1T Experiment}},
  \href{https://doi.org/10.1103/PhysRevLett.119.181301}{\emph{Phys. Rev. Lett.}
  {\bfseries 119} (2017) 181301}
  [\href{https://arxiv.org/abs/1705.06655}{{\ttfamily 1705.06655}}].

\bibitem{Messina:2018fmz}
{\scshape XENON} collaboration, \emph{{Latest results of 1 tonne x year Dark
  Matter Search with XENON1T}},
  \href{https://doi.org/10.22323/1.335.0017}{\emph{PoS} {\bfseries EDSU2018}
  (2018) 017}.

\bibitem{Zhang:2018xdp}
{\scshape PandaX} collaboration, \emph{{Dark matter direct search sensitivity
  of the PandaX-4T experiment}},
  \href{https://doi.org/10.1007/s11433-018-9259-0}{\emph{Sci. China Phys. Mech.
  Astron.} {\bfseries 62} (2019) 31011}
  [\href{https://arxiv.org/abs/1806.02229}{{\ttfamily 1806.02229}}].

\bibitem{Bhattacharya:2013hva}
S.~Bhattacharya, A.~Drozd, B.~Grzadkowski and J.~Wudka, \emph{{Two-Component
  Dark Matter}}, \href{https://doi.org/10.1007/JHEP10(2013)158}{\emph{JHEP}
  {\bfseries 10} (2013) 158} [\href{https://arxiv.org/abs/1309.2986}{{\ttfamily
  1309.2986}}].

\bibitem{Bian:2013wna}
L.~Bian, R.~Ding and B.~Zhu, \emph{{Two Component Higgs-Portal Dark Matter}},
  \href{https://doi.org/10.1016/j.physletb.2013.11.034}{\emph{Phys. Lett.}
  {\bfseries B728} (2014) 105}
  [\href{https://arxiv.org/abs/1308.3851}{{\ttfamily 1308.3851}}].

\bibitem{Esch:2014jpa}
S.~Esch, M.~Klasen and C.~E. Yaguna, \emph{{A minimal model for two-component
  dark matter}}, \href{https://doi.org/10.1007/JHEP09(2014)108}{\emph{JHEP}
  {\bfseries 09} (2014) 108} [\href{https://arxiv.org/abs/1406.0617}{{\ttfamily
  1406.0617}}].

\bibitem{Karam:2015jta}
A.~Karam and K.~Tamvakis, \emph{{Dark matter and neutrino masses from a
  scale-invariant multi-Higgs portal}},
  \href{https://doi.org/10.1103/PhysRevD.92.075010}{\emph{Phys. Rev.}
  {\bfseries D92} (2015) 075010}
  [\href{https://arxiv.org/abs/1508.03031}{{\ttfamily 1508.03031}}].

\bibitem{Karam:2016rsz}
A.~Karam and K.~Tamvakis, \emph{{Dark Matter from a Classically Scale-Invariant
  $SU(3)_X$}}, \href{https://doi.org/10.1103/PhysRevD.94.055004}{\emph{Phys.
  Rev.} {\bfseries D94} (2016) 055004}
  [\href{https://arxiv.org/abs/1607.01001}{{\ttfamily 1607.01001}}].

\bibitem{Ahmed:2017dbb}
A.~Ahmed, M.~Duch, B.~Grzadkowski and M.~Iglicki, \emph{{Multi-Component Dark
  Matter: the vector and fermion case}},
  \href{https://doi.org/10.1140/epjc/s10052-018-6371-2}{\emph{Eur. Phys. J.}
  {\bfseries C78} (2018) 905}
  [\href{https://arxiv.org/abs/1710.01853}{{\ttfamily 1710.01853}}].

\bibitem{Herrero-Garcia:2018qnz}
J.~Herrero-Garcia, A.~Scaffidi, M.~White and A.~G. Williams, \emph{{On the
  direct detection of multi-component dark matter: implications of the relic
  abundance}}, \href{https://doi.org/10.1088/1475-7516/2019/01/008}{\emph{JCAP}
  {\bfseries 1901} (2019) 008}
  [\href{https://arxiv.org/abs/1809.06881}{{\ttfamily 1809.06881}}].

\bibitem{Poulin:2018kap}
A.~Poulin and S.~Godfrey, \emph{{Multicomponent dark matter from a hidden
  gauged SU(3)}}, \href{https://doi.org/10.1103/PhysRevD.99.076008}{\emph{Phys.
  Rev.} {\bfseries D99} (2019) 076008}
  [\href{https://arxiv.org/abs/1808.04901}{{\ttfamily 1808.04901}}].

\bibitem{Aoki:2018gjf}
M.~Aoki and T.~Toma, \emph{{Boosted Self-interacting Dark Matter in a
  Multi-component Dark Matter Model}},
  \href{https://doi.org/10.1088/1475-7516/2018/10/020}{\emph{JCAP} {\bfseries
  1810} (2018) 020} [\href{https://arxiv.org/abs/1806.09154}{{\ttfamily
  1806.09154}}].

\bibitem{Aoki:2017eqn}
M.~Aoki, D.~Kaneko and J.~Kubo, \emph{{Multicomponent Dark Matter in Radiative
  Seesaw Models}},
  \href{https://doi.org/10.3389/fphy.2017.00053}{\emph{Front.in Phys.}
  {\bfseries 5} (2017) 53} [\href{https://arxiv.org/abs/1711.03765}{{\ttfamily
  1711.03765}}].

\bibitem{Bhattacharya:2016ysw}
S.~Bhattacharya, P.~Poulose and P.~Ghosh, \emph{{Multipartite Interacting
  Scalar Dark Matter in the light of updated LUX data}},
  \href{https://doi.org/10.1088/1475-7516/2017/04/043}{\emph{JCAP} {\bfseries
  1704} (2017) 043} [\href{https://arxiv.org/abs/1607.08461}{{\ttfamily
  1607.08461}}].

\bibitem{Bhattacharya:2018cgx}
S.~Bhattacharya, P.~Ghosh and N.~Sahu, \emph{{Multipartite Dark Matter with
  Scalars, Fermions and signatures at LHC}},
  \href{https://doi.org/10.1007/JHEP02(2019)059}{\emph{JHEP} {\bfseries 02}
  (2019) 059} [\href{https://arxiv.org/abs/1809.07474}{{\ttfamily
  1809.07474}}].

\bibitem{Biswas:2013nn}
A.~Biswas, D.~Majumdar, A.~Sil and P.~Bhattacharjee, \emph{{Two Component Dark
  Matter : A Possible Explanation of 130 GeV $\gamma-$ Ray Line from the
  Galactic Centre}},
  \href{https://doi.org/10.1088/1475-7516/2013/12/049}{\emph{JCAP} {\bfseries
  1312} (2013) 049} [\href{https://arxiv.org/abs/1301.3668}{{\ttfamily
  1301.3668}}].

\bibitem{Borah:2019aeq}
D.~Borah, R.~Roshan and A.~Sil, \emph{{Minimal Two-component Scalar Doublet
  Dark Matter with Radiative Neutrino Mass}},
  \href{https://arxiv.org/abs/1904.04837}{{\ttfamily 1904.04837}}.

\bibitem{Chakraborti:2018lso}
S.~Chakraborti and P.~Poulose, \emph{{Interplay of Scalar and Fermionic
  Components in a Multi-component Dark Matter Scenario}},
  \href{https://arxiv.org/abs/1808.01979}{{\ttfamily 1808.01979}}.

\bibitem{Chakraborti:2018aae}
S.~Chakraborti, A.~Dutta~Banik and R.~Islam, \emph{{Probing Multicomponent
  Extension of Inert Doublet Model with a Vector Dark Matter}},
  \href{https://arxiv.org/abs/1810.05595}{{\ttfamily 1810.05595}}.

\bibitem{Barman:2018esi}
B.~Barman, S.~Bhattacharya and M.~Zakeri, \emph{{Multipartite Dark Matter in
  $SU(2)_N$ extension of Standard Model and signatures at the LHC}},
  \href{https://doi.org/10.1088/1475-7516/2018/09/023}{\emph{JCAP} {\bfseries
  1809} (2018) 023} [\href{https://arxiv.org/abs/1806.01129}{{\ttfamily
  1806.01129}}].

\bibitem{DuttaBanik:2016jzv}
A.~Dutta~Banik, M.~Pandey, D.~Majumdar and A.~Biswas, \emph{{Two component
  WIMP–FImP dark matter model with singlet fermion, scalar and pseudo
  scalar}}, \href{https://doi.org/10.1140/epjc/s10052-017-5221-y}{\emph{Eur.
  Phys. J.} {\bfseries C77} (2017) 657}
  [\href{https://arxiv.org/abs/1612.08621}{{\ttfamily 1612.08621}}].

\bibitem{Bhattacharya:2018cqu}
S.~Bhattacharya, A.~K. Saha, A.~Sil and J.~Wudka, \emph{{Dark Matter as a
  remnant of SQCD Inflation}},
  \href{https://doi.org/10.1007/JHEP10(2018)124}{\emph{JHEP} {\bfseries 10}
  (2018) 124} [\href{https://arxiv.org/abs/1805.03621}{{\ttfamily
  1805.03621}}].

\bibitem{YaserAyazi:2018lrv}
S.~Yaser~Ayazi and A.~Mohamadnejad, \emph{{Scale-Invariant Two Component Dark
  Matter}}, \href{https://doi.org/10.1140/epjc/s10052-019-6651-5}{\emph{Eur.
  Phys. J.} {\bfseries C79} (2019) 140}
  [\href{https://arxiv.org/abs/1808.08706}{{\ttfamily 1808.08706}}].

\bibitem{Abdallah:2015ter}
J.~Abdallah et~al., \emph{{Simplified Models for Dark Matter Searches at the
  LHC}}, \href{https://doi.org/10.1016/j.dark.2015.08.001}{\emph{Phys. Dark
  Univ.} {\bfseries 9-10} (2015) 8}.

\bibitem{Abercrombie:2015wmb}
D.~Abercrombie et~al., \emph{{Dark Matter Benchmark Models for Early LHC Run-2
  Searches: Report of the ATLAS/CMS Dark Matter Forum}},
  \href{https://arxiv.org/abs/1507.00966}{{\ttfamily 1507.00966}}.

\bibitem{Han:2016gyy}
H.~Han, J.~M. Yang, Y.~Zhang and S.~Zheng, \emph{{Collider Signatures of
  Higgs-portal Scalar Dark Matter}},
  \href{https://doi.org/10.1016/j.physletb.2016.03.010}{\emph{Phys. Lett.}
  {\bfseries B756} (2016) 109}
  [\href{https://arxiv.org/abs/1601.06232}{{\ttfamily 1601.06232}}].

\bibitem{Gustafsson:2012aj}
M.~Gustafsson, S.~Rydbeck, L.~Lopez-Honorez and E.~Lundstrom, \emph{{Status of
  the Inert Doublet Model and the Role of multileptons at the LHC}},
  \href{https://doi.org/10.1103/PhysRevD.86.075019}{\emph{Phys. Rev.}
  {\bfseries D86} (2012) 075019}
  [\href{https://arxiv.org/abs/1206.6316}{{\ttfamily 1206.6316}}].

\bibitem{Bhardwaj:2019mts}
A.~Bhardwaj, P.~Konar, T.~Mandal and S.~Sadhukhan, \emph{{Probing Inert Doublet
  Model using jet substructure with multivariate analysis}},
  \href{https://arxiv.org/abs/1905.04195}{{\ttfamily 1905.04195}}.

\bibitem{Mohapatra:1979ia}
R.~N. Mohapatra and G.~Senjanovic, \emph{{Neutrino Mass and Spontaneous Parity
  Nonconservation}},
  \href{https://doi.org/10.1103/PhysRevLett.44.912}{\emph{Phys. Rev. Lett.}
  {\bfseries 44} (1980) 912}.

\bibitem{Schechter:1980gr}
J.~Schechter and J.~W.~F. Valle, \emph{{Neutrino Masses in SU(2) x U(1)
  Theories}}, \href{https://doi.org/10.1103/PhysRevD.22.2227}{\emph{Phys. Rev.}
  {\bfseries D22} (1980) 2227}.

\bibitem{Gabrielli:2013hma}
E.~Gabrielli, M.~Heikinheimo, K.~Kannike, A.~Racioppi, M.~Raidal and
  C.~Spethmann, \emph{{Towards Completing the Standard Model: Vacuum Stability,
  EWSB and Dark Matter}},
  \href{https://doi.org/10.1103/PhysRevD.89.015017}{\emph{Phys. Rev.}
  {\bfseries D89} (2014) 015017}
  [\href{https://arxiv.org/abs/1309.6632}{{\ttfamily 1309.6632}}].

\bibitem{Chen:2012faa}
C.-S. Chen and Y.~Tang, \emph{{Vacuum stability, neutrinos, and dark matter}},
  \href{https://doi.org/10.1007/JHEP04(2012)019}{\emph{JHEP} {\bfseries 04}
  (2012) 019} [\href{https://arxiv.org/abs/1202.5717}{{\ttfamily 1202.5717}}].

\bibitem{Rodejohann:2012px}
W.~Rodejohann and H.~Zhang, \emph{{Impact of massive neutrinos on the Higgs
  self-coupling and electroweak vacuum stability}},
  \href{https://doi.org/10.1007/JHEP06(2012)022}{\emph{JHEP} {\bfseries 06}
  (2012) 022} [\href{https://arxiv.org/abs/1203.3825}{{\ttfamily 1203.3825}}].

\bibitem{Rose:2015fua}
L.~Delle~Rose, C.~Marzo and A.~Urbano, \emph{{On the stability of the
  electroweak vacuum in the presence of low-scale seesaw models}},
  \href{https://doi.org/10.1007/JHEP12(2015)050}{\emph{JHEP} {\bfseries 12}
  (2015) 050} [\href{https://arxiv.org/abs/1506.03360}{{\ttfamily
  1506.03360}}].

\bibitem{Lindner:2015qva}
M.~Lindner, H.~H. Patel and B.~Radovčić, \emph{{Electroweak Absolute, Meta-,
  and Thermal Stability in Neutrino Mass Models}},
  \href{https://doi.org/10.1103/PhysRevD.93.073005}{\emph{Phys. Rev.}
  {\bfseries D93} (2016) 073005}
  [\href{https://arxiv.org/abs/1511.06215}{{\ttfamily 1511.06215}}].

\bibitem{Chakrabortty:2012np}
J.~Chakrabortty, M.~Das and S.~Mohanty, \emph{{Constraints on TeV scale
  Majorana neutrino phenomenology from the Vacuum Stability of the Higgs}},
  \href{https://doi.org/10.1142/S0217732313500326}{\emph{Mod. Phys. Lett.}
  {\bfseries A28} (2013) 1350032}
  [\href{https://arxiv.org/abs/1207.2027}{{\ttfamily 1207.2027}}].

\bibitem{Coriano:2014mpa}
C.~Coriano, L.~Delle~Rose and C.~Marzo, \emph{{Vacuum Stability in U(1)-Prime
  Extensions of the Standard Model with TeV Scale Right Handed Neutrinos}},
  \href{https://doi.org/10.1016/j.physletb.2014.09.001}{\emph{Phys. Lett.}
  {\bfseries B738} (2014) 13}
  [\href{https://arxiv.org/abs/1407.8539}{{\ttfamily 1407.8539}}].

\bibitem{Ng:2015eia}
J.~N. Ng and A.~de~la Puente, \emph{{Electroweak Vacuum Stability and the
  Seesaw Mechanism Revisited}},
  \href{https://doi.org/10.1140/epjc/s10052-016-3981-4}{\emph{Eur. Phys. J.}
  {\bfseries C76} (2016) 122}
  [\href{https://arxiv.org/abs/1510.00742}{{\ttfamily 1510.00742}}].

\bibitem{Bonilla:2015kna}
C.~Bonilla, R.~M. Fonseca and J.~W.~F. Valle, \emph{{Vacuum stability with
  spontaneous violation of lepton number}},
  \href{https://doi.org/10.1016/j.physletb.2016.03.037}{\emph{Phys. Lett.}
  {\bfseries B756} (2016) 345}
  [\href{https://arxiv.org/abs/1506.04031}{{\ttfamily 1506.04031}}].

\bibitem{Khan:2012zw}
S.~Khan, S.~Goswami and S.~Roy, \emph{{Vacuum Stability constraints on the
  minimal singlet TeV Seesaw Model}},
  \href{https://doi.org/10.1103/PhysRevD.89.073021}{\emph{Phys. Rev.}
  {\bfseries D89} (2014) 073021}
  [\href{https://arxiv.org/abs/1212.3694}{{\ttfamily 1212.3694}}].

\bibitem{Garg:2017iva}
I.~Garg, S.~Goswami, V.~K. N. and N.~Khan, \emph{{Electroweak vacuum stability
  in presence of singlet scalar dark matter in TeV scale seesaw models}},
  \href{https://doi.org/10.1103/PhysRevD.96.055020}{\emph{Phys. Rev.}
  {\bfseries D96} (2017) 055020}
  [\href{https://arxiv.org/abs/1706.08851}{{\ttfamily 1706.08851}}].

\bibitem{Chakrabarty:2015yia}
N.~Chakrabarty, D.~K. Ghosh, B.~Mukhopadhyaya and I.~Saha, \emph{{Dark matter,
  neutrino masses and high scale validity of an inert Higgs doublet model}},
  \href{https://doi.org/10.1103/PhysRevD.92.015002}{\emph{Phys. Rev.}
  {\bfseries D92} (2015) 015002}
  [\href{https://arxiv.org/abs/1501.03700}{{\ttfamily 1501.03700}}].

\bibitem{Ghosh:2017pxl}
D.~K. Ghosh, N.~Ghosh, I.~Saha and A.~Shaw, \emph{{Revisiting the high-scale
  validity of the type II seesaw model with novel LHC signature}},
  \href{https://doi.org/10.1103/PhysRevD.97.115022}{\emph{Phys. Rev.}
  {\bfseries D97} (2018) 115022}
  [\href{https://arxiv.org/abs/1711.06062}{{\ttfamily 1711.06062}}].

\bibitem{Ma:2006km}
E.~Ma, \emph{{Verifiable radiative seesaw mechanism of neutrino mass and dark
  matter}}, \href{https://doi.org/10.1103/PhysRevD.73.077301}{\emph{Phys. Rev.}
  {\bfseries D73} (2006) 077301}
  [\href{https://arxiv.org/abs/hep-ph/0601225}{{\ttfamily hep-ph/0601225}}].

\bibitem{Kannike:2012pe}
K.~Kannike, \emph{{Vacuum Stability Conditions From Copositivity Criteria}},
  \href{https://doi.org/10.1140/epjc/s10052-012-2093-z}{\emph{Eur. Phys. J.}
  {\bfseries C72} (2012) 2093}
  [\href{https://arxiv.org/abs/1205.3781}{{\ttfamily 1205.3781}}].

\bibitem{Chakrabortty:2013mha}
J.~Chakrabortty, P.~Konar and T.~Mondal.

\bibitem{Horejsi:2005da}
J.~Horejsi and M.~Kladiva, \emph{{Tree-unitarity bounds for THDM Higgs masses
  revisited}}, \href{https://doi.org/10.1140/epjc/s2006-02472-3}{\emph{Eur.
  Phys. J.} {\bfseries C46} (2006) 81}
  [\href{https://arxiv.org/abs/hep-ph/0510154}{{\ttfamily hep-ph/0510154}}].

\bibitem{Bhattacharyya:2015nca}
G.~Bhattacharyya and D.~Das, \emph{{Scalar sector of two-Higgs-doublet models:
  A minireview}},
  \href{https://doi.org/10.1007/s12043-016-1252-4}{\emph{Pramana} {\bfseries
  87} (2016) 40} [\href{https://arxiv.org/abs/1507.06424}{{\ttfamily
  1507.06424}}].

\bibitem{Peskin:1991sw}
M.~E. Peskin and T.~Takeuchi, \emph{{Estimation of oblique electroweak
  corrections}}, \href{https://doi.org/10.1103/PhysRevD.46.381}{\emph{Phys.
  Rev.} {\bfseries D46} (1992) 381}.

\bibitem{Arhrib:2012ia}
A.~Arhrib, R.~Benbrik and N.~Gaur, \emph{{$H\to \gamma \gamma$ in Inert Higgs
  Doublet Model}},
  \href{https://doi.org/10.1103/PhysRevD.85.095021}{\emph{Phys. Rev.}
  {\bfseries D85} (2012) 095021}
  [\href{https://arxiv.org/abs/1201.2644}{{\ttfamily 1201.2644}}].

\bibitem{Barbieri:2006dq}
R.~Barbieri, L.~J. Hall and V.~S. Rychkov, \emph{{Improved naturalness with a
  heavy Higgs: An Alternative road to LHC physics}},
  \href{https://doi.org/10.1103/PhysRevD.74.015007}{\emph{Phys. Rev.}
  {\bfseries D74} (2006) 015007}
  [\href{https://arxiv.org/abs/hep-ph/0603188}{{\ttfamily hep-ph/0603188}}].

\bibitem{Baak:2014ora}
{\scshape Gfitter Group} collaboration, \emph{{The global electroweak fit at
  NNLO and prospects for the LHC and ILC}},
  \href{https://doi.org/10.1140/epjc/s10052-014-3046-5}{\emph{Eur. Phys. J.}
  {\bfseries C74} (2014) 3046}
  [\href{https://arxiv.org/abs/1407.3792}{{\ttfamily 1407.3792}}].

\bibitem{Lundstrom:2008ai}
E.~Lundstrom, M.~Gustafsson and J.~Edsjo, \emph{{The Inert Doublet Model and
  LEP II Limits}},
  \href{https://doi.org/10.1103/PhysRevD.79.035013}{\emph{Phys. Rev.}
  {\bfseries D79} (2009) 035013}
  [\href{https://arxiv.org/abs/0810.3924}{{\ttfamily 0810.3924}}].

\bibitem{Pierce:2007ut}
A.~Pierce and J.~Thaler, \emph{{Natural Dark Matter from an Unnatural Higgs
  Boson and New Colored Particles at the TeV Scale}},
  \href{https://doi.org/10.1088/1126-6708/2007/08/026}{\emph{JHEP} {\bfseries
  08} (2007) 026} [\href{https://arxiv.org/abs/hep-ph/0703056}{{\ttfamily
  hep-ph/0703056}}].

\bibitem{Cao:2007rm}
Q.-H. Cao, E.~Ma and G.~Rajasekaran, \emph{{Observing the Dark Scalar Doublet
  and its Impact on the Standard-Model Higgs Boson at Colliders}},
  \href{https://doi.org/10.1103/PhysRevD.76.095011}{\emph{Phys. Rev.}
  {\bfseries D76} (2007) 095011}
  [\href{https://arxiv.org/abs/0708.2939}{{\ttfamily 0708.2939}}].

\bibitem{Djouadi:2005gj}
A.~Djouadi, \emph{{The Anatomy of electro-weak symmetry breaking. II. The Higgs
  bosons in the minimal supersymmetric model}},
  \href{https://doi.org/10.1016/j.physrep.2007.10.005}{\emph{Phys. Rept.}
  {\bfseries 459} (2008) 1}
  [\href{https://arxiv.org/abs/hep-ph/0503173}{{\ttfamily hep-ph/0503173}}].

\bibitem{Swiezewska:2012eh}
B.~Swiezewska and M.~Krawczyk, \emph{{Diphoton rate in the inert doublet model
  with a 125 GeV Higgs boson}},
  \href{https://doi.org/10.1103/PhysRevD.88.035019}{\emph{Phys. Rev.}
  {\bfseries D88} (2013) 035019}
  [\href{https://arxiv.org/abs/1212.4100}{{\ttfamily 1212.4100}}].

\bibitem{Krawczyk:2013jta}
M.~Krawczyk, D.~Sokolowska, P.~Swaczyna and B.~Swiezewska, \emph{{Constraining
  Inert Dark Matter by $R_{\gamma\gamma}$ and WMAP data}},
  \href{https://doi.org/10.1007/JHEP09(2013)055}{\emph{JHEP} {\bfseries 09}
  (2013) 055} [\href{https://arxiv.org/abs/1305.6266}{{\ttfamily 1305.6266}}].

\bibitem{Aad:2014eha}
{\scshape ATLAS} collaboration, \emph{{Measurement of Higgs boson production in
  the diphoton decay channel in pp collisions at center-of-mass energies of 7
  and 8 TeV with the ATLAS detector}},
  \href{https://doi.org/10.1103/PhysRevD.90.112015}{\emph{Phys. Rev.}
  {\bfseries D90} (2014) 112015}
  [\href{https://arxiv.org/abs/1408.7084}{{\ttfamily 1408.7084}}].

\bibitem{Khachatryan:2014ira}
{\scshape CMS} collaboration, \emph{{Observation of the diphoton decay of the
  Higgs boson and measurement of its properties}},
  \href{https://doi.org/10.1140/epjc/s10052-014-3076-z}{\emph{Eur. Phys. J.}
  {\bfseries C74} (2014) 3076}
  [\href{https://arxiv.org/abs/1407.0558}{{\ttfamily 1407.0558}}].

\bibitem{Vagnozzi:2017ovm}
S.~Vagnozzi, E.~Giusarma, O.~Mena, K.~Freese, M.~Gerbino, S.~Ho et~al.,
  \emph{{Unveiling $\nu$ secrets with cosmological data: neutrino masses and
  mass hierarchy}},
  \href{https://doi.org/10.1103/PhysRevD.96.123503}{\emph{Phys. Rev.}
  {\bfseries D96} (2017) 123503}
  [\href{https://arxiv.org/abs/1701.08172}{{\ttfamily 1701.08172}}].

\bibitem{deSalas:2017kay}
P.~F. de~Salas, D.~V. Forero, C.~A. Ternes, M.~Tortola and J.~W.~F. Valle,
  \emph{{Status of neutrino oscillations 2018: 3$\sigma$ hint for normal mass
  ordering and improved CP sensitivity}},
  \href{https://doi.org/10.1016/j.physletb.2018.06.019}{\emph{Phys. Lett.}
  {\bfseries B782} (2018) 633}
  [\href{https://arxiv.org/abs/1708.01186}{{\ttfamily 1708.01186}}].

\bibitem{Esteban:2016qun}
I.~Esteban, M.~C. Gonzalez-Garcia, M.~Maltoni, I.~Martinez-Soler and
  T.~Schwetz, \emph{{Updated fit to three neutrino mixing: exploring the
  accelerator-reactor complementarity}},
  \href{https://doi.org/10.1007/JHEP01(2017)087}{\emph{JHEP} {\bfseries 01}
  (2017) 087} [\href{https://arxiv.org/abs/1611.01514}{{\ttfamily
  1611.01514}}].

\bibitem{Ilakovac:1994kj}
A.~Ilakovac and A.~Pilaftsis, \emph{{Flavor violating charged lepton decays in
  seesaw-type models}},
  \href{https://doi.org/10.1016/0550-3213(94)00567-X}{\emph{Nucl. Phys.}
  {\bfseries B437} (1995) 491}
  [\href{https://arxiv.org/abs/hep-ph/9403398}{{\ttfamily hep-ph/9403398}}].

\bibitem{Tommasini:1995ii}
D.~Tommasini, G.~Barenboim, J.~Bernabeu and C.~Jarlskog, \emph{{Nondecoupling
  of heavy neutrinos and lepton flavor violation}},
  \href{https://doi.org/10.1016/0550-3213(95)00201-3}{\emph{Nucl. Phys.}
  {\bfseries B444} (1995) 451}
  [\href{https://arxiv.org/abs/hep-ph/9503228}{{\ttfamily hep-ph/9503228}}].

\bibitem{Dinh:2012bp}
D.~N. Dinh, A.~Ibarra, E.~Molinaro and S.~T. Petcov, \emph{{The $\mu - e$
  Conversion in Nuclei, $\mu \to e \gamma, \mu \to 3e$ Decays and TeV Scale
  See-Saw Scenarios of Neutrino Mass Generation}},
  \href{https://doi.org/10.1007/JHEP09(2013)023,
  10.1007/JHEP08(2012)125}{\emph{JHEP} {\bfseries 08} (2012) 125}
  [\href{https://arxiv.org/abs/1205.4671}{{\ttfamily 1205.4671}}].

\bibitem{Bambhaniya:2016rbb}
G.~Bambhaniya, P.~Bhupal~Dev, S.~Goswami, S.~Khan and W.~Rodejohann,
  \emph{{Naturalness, Vacuum Stability and Leptogenesis in the Minimal Seesaw
  Model}}, \href{https://doi.org/10.1103/PhysRevD.95.095016}{\emph{Phys. Rev.}
  {\bfseries D95} (2017) 095016}
  [\href{https://arxiv.org/abs/1611.03827}{{\ttfamily 1611.03827}}].

\bibitem{Ghosh:2017fmr}
P.~Ghosh, A.~K. Saha and A.~Sil, \emph{{Study of Electroweak Vacuum Stability
  from Extended Higgs Portal of Dark Matter and Neutrinos}},
  \href{https://doi.org/10.1103/PhysRevD.97.075034}{\emph{Phys. Rev.}
  {\bfseries D97} (2018) 075034}
  [\href{https://arxiv.org/abs/1706.04931}{{\ttfamily 1706.04931}}].

\bibitem{LopezHonorez:2006gr}
L.~Lopez~Honorez, E.~Nezri, J.~F. Oliver and M.~H.~G. Tytgat, \emph{{The Inert
  Doublet Model: An Archetype for Dark Matter}},
  \href{https://doi.org/10.1088/1475-7516/2007/02/028}{\emph{JCAP} {\bfseries
  0702} (2007) 028} [\href{https://arxiv.org/abs/hep-ph/0612275}{{\ttfamily
  hep-ph/0612275}}].

\bibitem{Alarcon:2011zs}
J.~M. Alarcon, J.~Martin~Camalich and J.~A. Oller, \emph{{The chiral
  representation of the $\pi N$ scattering amplitude and the pion-nucleon sigma
  term}}, \href{https://doi.org/10.1103/PhysRevD.85.051503}{\emph{Phys. Rev.}
  {\bfseries D85} (2012) 051503}
  [\href{https://arxiv.org/abs/1110.3797}{{\ttfamily 1110.3797}}].

\bibitem{Alarcon:2012nr}
J.~M. Alarcon, L.~S. Geng, J.~Martin~Camalich and J.~A. Oller, \emph{{The
  strangeness content of the nucleon from effective field theory and
  phenomenology}},
  \href{https://doi.org/10.1016/j.physletb.2014.01.065}{\emph{Phys. Lett.}
  {\bfseries B730} (2014) 342}
  [\href{https://arxiv.org/abs/1209.2870}{{\ttfamily 1209.2870}}].

\bibitem{Herrero-Garcia:2017vrl}
J.~Herrero-Garcia, A.~Scaffidi, M.~White and A.~G. Williams, \emph{{On the
  direct detection of multi-component dark matter: sensitivity studies and
  parameter estimation}},
  \href{https://doi.org/10.1088/1475-7516/2017/11/021}{\emph{JCAP} {\bfseries
  1711} (2017) 021} [\href{https://arxiv.org/abs/1709.01945}{{\ttfamily
  1709.01945}}].

\bibitem{Klasen:2013btp}
M.~Klasen, C.~E. Yaguna and J.~D. Ruiz-Alvarez, \emph{{Electroweak corrections
  to the direct detection cross section of inert higgs dark matter}},
  \href{https://doi.org/10.1103/PhysRevD.87.075025}{\emph{Phys. Rev.}
  {\bfseries D87} (2013) 075025}
  [\href{https://arxiv.org/abs/1302.1657}{{\ttfamily 1302.1657}}].

\bibitem{Abe:2015rja}
T.~Abe and R.~Sato, \emph{{Quantum corrections to the spin-independent cross
  section of the inert doublet dark matter}},
  \href{https://doi.org/10.1007/JHEP03(2015)109}{\emph{JHEP} {\bfseries 03}
  (2015) 109} [\href{https://arxiv.org/abs/1501.04161}{{\ttfamily
  1501.04161}}].

\bibitem{Barducci:2016pcb}
D.~Barducci, G.~Belanger, J.~Bernon, F.~Boudjema, J.~Da~Silva, S.~Kraml et~al.,
  \emph{{Collider limits on new physics within micrOMEGAs 4.3}},
  \href{https://doi.org/10.1016/j.cpc.2017.08.028}{\emph{Comput. Phys. Commun.}
  {\bfseries 222} (2018) 327}
  [\href{https://arxiv.org/abs/1606.03834}{{\ttfamily 1606.03834}}].

\bibitem{Semenov:2008jy}
A.~Semenov, \emph{{LanHEP: A Package for the automatic generation of Feynman
  rules in field theory. Version 3.0}},
  \href{https://doi.org/10.1016/j.cpc.2008.10.012}{\emph{Comput. Phys. Commun.}
  {\bfseries 180} (2009) 431}
  [\href{https://arxiv.org/abs/0805.0555}{{\ttfamily 0805.0555}}].

\bibitem{Kolb:1990vq}
E.~W. Kolb and M.~S. Turner, \emph{{The Early Universe}}, {\emph{Front. Phys.}
  {\bfseries 69} (1990) 1}.

\bibitem{Aprile:2018cxk}
{\scshape XENON} collaboration, \emph{{First results on the scalar WIMP-pion
  coupling, using the XENON1T experiment}},
  \href{https://doi.org/10.1103/PhysRevLett.122.071301}{\emph{Phys. Rev. Lett.}
  {\bfseries 122} (2019) 071301}
  [\href{https://arxiv.org/abs/1811.12482}{{\ttfamily 1811.12482}}].

\bibitem{Ahnen:2016qkx}
{\scshape MAGIC, Fermi-LAT} collaboration, \emph{{Limits to Dark Matter
  Annihilation Cross-Section from a Combined Analysis of MAGIC and Fermi-LAT
  Observations of Dwarf Satellite Galaxies}},
  \href{https://doi.org/10.1088/1475-7516/2016/02/039}{\emph{JCAP} {\bfseries
  1602} (2016) 039} [\href{https://arxiv.org/abs/1601.06590}{{\ttfamily
  1601.06590}}].

\bibitem{Gustafsson:2007pc}
M.~Gustafsson, E.~Lundstrom, L.~Bergstrom and J.~Edsjo, \emph{{Significant
  Gamma Lines from Inert Higgs Dark Matter}},
  \href{https://doi.org/10.1103/PhysRevLett.99.041301}{\emph{Phys. Rev. Lett.}
  {\bfseries 99} (2007) 041301}
  [\href{https://arxiv.org/abs/astro-ph/0703512}{{\ttfamily
  astro-ph/0703512}}].

\bibitem{Garcia-Cely:2015khw}
C.~Garcia-Cely, M.~Gustafsson and A.~Ibarra, \emph{{Probing the Inert Doublet
  Dark Matter Model with Cherenkov Telescopes}},
  \href{https://doi.org/10.1088/1475-7516/2016/02/043}{\emph{JCAP} {\bfseries
  1602} (2016) 043} [\href{https://arxiv.org/abs/1512.02801}{{\ttfamily
  1512.02801}}].

\bibitem{Eiteneuer:2017hoh}
B.~Eiteneuer, A.~Goudelis and J.~Heisig, \emph{{The inert doublet model in the
  light of Fermi-LAT gamma-ray data: a global fit analysis}},
  \href{https://doi.org/10.1140/epjc/s10052-017-5166-1}{\emph{Eur. Phys. J.}
  {\bfseries C77} (2017) 624}
  [\href{https://arxiv.org/abs/1705.01458}{{\ttfamily 1705.01458}}].

\bibitem{Queiroz:2015utg}
F.~S. Queiroz and C.~E. Yaguna, \emph{{The CTA aims at the Inert Doublet
  Model}}, \href{https://doi.org/10.1088/1475-7516/2016/02/038}{\emph{JCAP}
  {\bfseries 1602} (2016) 038}
  [\href{https://arxiv.org/abs/1511.05967}{{\ttfamily 1511.05967}}].

\bibitem{Garcia-Cely:2013zga}
C.~Garcia-Cely and A.~Ibarra, \emph{{Novel Gamma-ray Spectral Features in the
  Inert Doublet Model}},
  \href{https://doi.org/10.1088/1475-7516/2013/09/025}{\emph{JCAP} {\bfseries
  1309} (2013) 025} [\href{https://arxiv.org/abs/1306.4681}{{\ttfamily
  1306.4681}}].

\bibitem{Casas:2001sr}
J.~A. Casas and A.~Ibarra, \emph{{Oscillating neutrinos and muon ---> e,
  gamma}}, \href{https://doi.org/10.1016/S0550-3213(01)00475-8}{\emph{Nucl.
  Phys.} {\bfseries B618} (2001) 171}
  [\href{https://arxiv.org/abs/hep-ph/0103065}{{\ttfamily hep-ph/0103065}}].

\bibitem{Branco:2011iw}
G.~C. Branco, P.~M. Ferreira, L.~Lavoura, M.~N. Rebelo, M.~Sher and J.~P.
  Silva, \emph{{Theory and phenomenology of two-Higgs-doublet models}},
  \href{https://doi.org/10.1016/j.physrep.2012.02.002}{\emph{Phys. Rept.}
  {\bfseries 516} (2012) 1} [\href{https://arxiv.org/abs/1106.0034}{{\ttfamily
  1106.0034}}].

\bibitem{Pirogov:1998tj}
{\relax Yu}.~F. Pirogov and O.~V. Zenin, \emph{{Two loop renormalization group
  restrictions on the standard model and the fourth chiral family}},
  \href{https://doi.org/10.1007/s100520050602,
  10.1007/s100529900035}{\emph{Eur. Phys. J.} {\bfseries C10} (1999) 629}
  [\href{https://arxiv.org/abs/hep-ph/9808396}{{\ttfamily hep-ph/9808396}}].

\bibitem{Staub:2013tta}
F.~Staub, \emph{{SARAH 4 : A tool for (not only SUSY) model builders}},
  \href{https://doi.org/10.1016/j.cpc.2014.02.018}{\emph{Comput. Phys. Commun.}
  {\bfseries 185} (2014) 1773}
  [\href{https://arxiv.org/abs/1309.7223}{{\ttfamily 1309.7223}}].

\bibitem{Guo:2006qa}
W.-l. Guo, Z.-z. Xing and S.~Zhou, \emph{{Neutrino Masses, Lepton Flavor Mixing
  and Leptogenesis in the Minimal Seesaw Model}},
  \href{https://doi.org/10.1142/S0218301307004898}{\emph{Int. J. Mod. Phys.}
  {\bfseries E16} (2007) 1}
  [\href{https://arxiv.org/abs/hep-ph/0612033}{{\ttfamily hep-ph/0612033}}].

\bibitem{Khan:2015ipa}
N.~Khan and S.~Rakshit, \emph{{Constraints on inert dark matter from the
  metastability of the electroweak vacuum}},
  \href{https://doi.org/10.1103/PhysRevD.92.055006}{\emph{Phys. Rev.}
  {\bfseries D92} (2015) 055006}
  [\href{https://arxiv.org/abs/1503.03085}{{\ttfamily 1503.03085}}].

\bibitem{Kalinowski:2018ylg}
J.~Kalinowski, W.~Kotlarski, T.~Robens, D.~Sokolowska and A.~F. Zarnecki,
  \emph{{Benchmarking the Inert Doublet Model for $e^+ e^-$ colliders}},
  \href{https://doi.org/10.1007/JHEP12(2018)081}{\emph{JHEP} {\bfseries 12}
  (2018) 081} [\href{https://arxiv.org/abs/1809.07712}{{\ttfamily
  1809.07712}}].

\bibitem{Alloul:2013bka}
A.~Alloul, N.~D. Christensen, C.~Degrande, C.~Duhr and B.~Fuks,
  \emph{{FeynRules 2.0 - A complete toolbox for tree-level phenomenology}},
  \href{https://doi.org/10.1016/j.cpc.2014.04.012}{\emph{Comput. Phys. Commun.}
  {\bfseries 185} (2014) 2250}
  [\href{https://arxiv.org/abs/1310.1921}{{\ttfamily 1310.1921}}].

\bibitem{Alwall:2011uj}
J.~Alwall, M.~Herquet, F.~Maltoni, O.~Mattelaer and T.~Stelzer, \emph{{MadGraph
  5 : Going Beyond}},
  \href{https://doi.org/10.1007/JHEP06(2011)128}{\emph{JHEP} {\bfseries 06}
  (2011) 128} [\href{https://arxiv.org/abs/1106.0522}{{\ttfamily 1106.0522}}].

\bibitem{Sjostrand:2006za}
T.~Sjostrand, S.~Mrenna and P.~Z. Skands, \emph{{PYTHIA 6.4 Physics and
  Manual}}, \href{https://doi.org/10.1088/1126-6708/2006/05/026}{\emph{JHEP}
  {\bfseries 05} (2006) 026}
  [\href{https://arxiv.org/abs/hep-ph/0603175}{{\ttfamily hep-ph/0603175}}].

\bibitem{Placakyte:2011az}
R.~Placakyte, \emph{{Parton Distribution Functions}},  in \emph{{Proceedings,
  31st International Conference on Physics in collisions (PIC 2011): Vancouver,
  Canada, August 28-September 1, 2011}}, 2011,
  \href{https://arxiv.org/abs/1111.5452}{{\ttfamily 1111.5452}}.

\bibitem{Alwall:2014hca}
J.~Alwall, R.~Frederix, S.~Frixione, V.~Hirschi, F.~Maltoni, O.~Mattelaer
  et~al., \emph{{The automated computation of tree-level and next-to-leading
  order differential cross sections, and their matching to parton shower
  simulations}}, \href{https://doi.org/10.1007/JHEP07(2014)079}{\emph{JHEP}
  {\bfseries 07} (2014) 079} [\href{https://arxiv.org/abs/1405.0301}{{\ttfamily
  1405.0301}}].

\bibitem{Lee:1977eg}
B.~W. Lee, C.~Quigg and H.~B. Thacker, \emph{{Weak Interactions at Very
  High-Energies: The Role of the Higgs Boson Mass}},
  \href{https://doi.org/10.1103/PhysRevD.16.1519}{\emph{Phys. Rev.} {\bfseries
  D16} (1977) 1519}.

\end{thebibliography}\endgroup


\end{document}